\documentclass[twocolumn,secnumarabic,amssymb,nobibnotes,aps,pre,superscriptaddress]{revtex4-1}

\usepackage{subfiles}
\usepackage{float}
\usepackage{graphicx,subcaption}
\usepackage{amsmath,amssymb}
\usepackage{epstopdf}
\usepackage{setspace}
\usepackage[shortlabels]{enumitem}
\usepackage{mathrsfs}
\usepackage{color,soul}
\usepackage[dvipsnames]{xcolor}
\usepackage{dcolumn}
\usepackage{comment}
\usepackage{upgreek}
\usepackage{bm}
\usepackage{svg}

\usepackage{hyperref}
\hypersetup{colorlinks = true,citecolor = blue,urlcolor = black}

\usepackage{ragged2e}
\DeclareCaptionJustification{justified}{\justifying}

\newcommand{\nn}{\nonumber}
\newcommand{\eq}[1]{Eq.~\eqref{#1}}

\newcommand{\fig}[1]{Fig.~\ref{#1}}
\newcommand{\sctn}[1]{\S~\ref{#1}}

\newcommand{\appndx}[1]{SI~\ref{#1}}

\newcommand{\ue}{School of Physics and Astronomy, The University of Edinburgh, Peter Guthrie Tait Road, Edinburgh, EH9 3FD, United Kingdom}

\newcommand{\av}[1]{\left\langle #1 \right\rangle}

\renewcommand{\vec}[1]{\underline{#1}}

\newcommand{\tens}[1]{\underline{\underline{#1}}}

\newcommand{\grad}{\vec{\nabla}}

\newcommand{\kbt}{k_{\mathrm{B}}T}

\newcounter{siequation}
\newcounter{sifigure}
\newcounter{sisection}
\newcounter{simovie}
\newcommand{\correctText}[2]{#2}

\newcommand{\tocite}[1]{\textcolor{Plum}{[c]}}

\begin{document}
\title{Lock-Key Microfluidics: \\Simulating Nematic Colloid Advection along Wavy-Walled Channels}

\author{Karolina Wamsler}
\affiliation{\ue}
\author{Louise C. Head}
\affiliation{\ue}
\author{Tyler N. Shendruk}
\email{t.shendruk@ed.ac.uk}
\affiliation{\ue}

\begin{abstract}
    Liquid crystalline media mediate interactions between suspended particles and confining \correctText{geomtries}{geometries}, which not only has potential to guide patterning and bottom-up colloidal assembly, but can also control colloidal migration in microfluidic devices. 
    However, simulating such dynamics is challenging because nemato-elasticity, diffusivity and hydrodynamic interactions must all be accounted for within complex boundaries. 
    We model the advection of colloids dispersed in flowing and fluctuating nematic fluids confined within 2D wavy channels. 
    A lock-key mechanism between colloids and troughs is found to be stronger for planar anchoring compared to homeotropic anchoring due to the relative location of the colloid-associated defects. 
    Sufficiently large amplitudes result in stick-slip trajectories and even permanent locking of colloids in place. 
    These results demonstrate that wavy walls not only have potential to direct colloids to specific docking sites but also to control site-specific resting duration and intermittent elution. 
\end{abstract}

\maketitle

\section{Introduction}\label{sctn:intro}
\begin{figure*}[tb]
    \centering
    \includegraphics[width=0.95\textwidth]{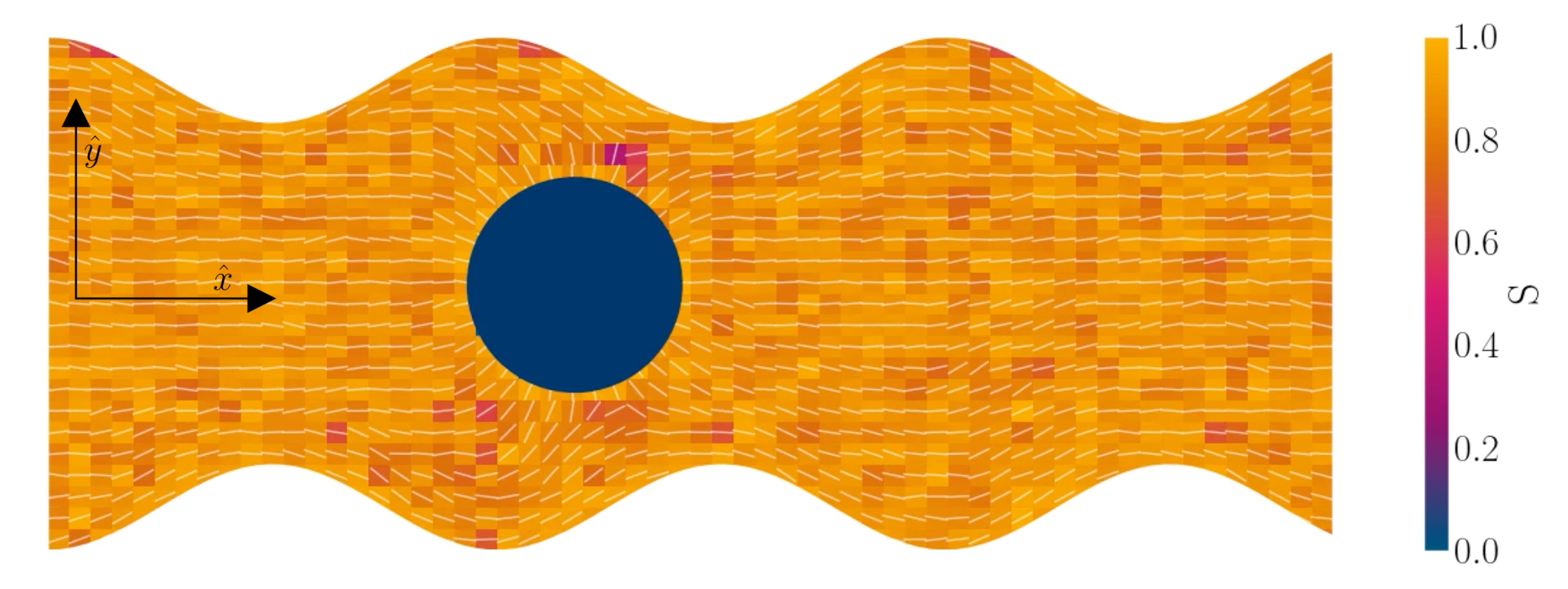}
    \caption{
        Simulation snapshot, with planar anchoring on the wavy channel walls, and homeotropic anchoring on the mobile colloid. 
        In this study, all colloids have homeotropic anchoring but the anchoring on the channel walls varies. 
        The colour shows local scalar order parameter \correctText{}{field} $S$ and white dashes show the director field $\vec{n}$. 
        Colloid radius $R=5a$; wall amplitude $B_0=2.5a$; wall wavelength $\lambda=20a$; average channel height $\av{h}_x=20a$; and pressure gradient $\vec{\nabla}P= 0.02 ma^{1-d}\tau^{-2} \hat{x}$. 
        Snapshot from $t=20\tau$, early in the simulation. 
    }
    \label{fig:setup}
\end{figure*}

Manipulating particle trajectories is an important feature of many microfluidic applications, including drug delivery~\cite{Liu2022}, cell sorting~\cite{Sivaramakrishnan2020} and medical diagnoses~\cite{Burklund2020}. 
Accordingly, a variety of approaches have been developed to control the trajectories of microparticles~\cite{Shendruk2013,Zhang2016,Zhang2018,Zhang2020}. 
One means of sculpting particle trajectories is by suspending them in complex carrier fluids, as in topological microfluidics~\cite{Sengupta2013b}, in which particles are guided by liquid crystalline carrier fluids~\cite{ Lavrentovich2020}. 
Liquid crystalline materials are alluring for microfluidic transport because they are highly responsive to flows~\cite{fedorowicz2023,dalby2022,Sengupta2020}, suspended inclusions~\cite{Smalyukh2018,stark2001} and confining surfaces~\cite{han2023,han2022,chen2021}. 
When colloidal particles are dispersed in liquid crystalline fluids, the anisotropic nature of liquid crystals give rise to emergent properties~\cite{Dogic2014,Smalyukh2020,Mundoor2018,paget2023} and imposed anchoring at colloidal surfaces results in topological defects in the vicinity of the colloids, to ensure topological charge neutrality. 
Strong homeotropic and planar anchoring endows the surface with a topological charge of $+1$ and necessitates the existence of an accompanying $-1$ charge in the bulk fluid, either as two $-1/2$ point defects in 2D, or in 3D as defect loops (Saturn rings), -1 point defect (hyperbolic hedgehog), or surface defects (boojum defects). 

Such defects mediate particle-particle~\cite{yuan2019,Senyuk2017,muvsevivc2019} and particle-wall interactions~\cite{yoshida2015,nikkhou2015}. 
Particle-wall elastic interactions allow micropatterned structures on walls to have significant, long-ranged effects on colloids~\cite{Hung2007,Lapointe2008}. 
For example, indentations in walls can govern colloid position: When the width of an indentation is comparable to colloid size there can be a strong attraction~\cite{Silvestre2004}, whereas this attraction is weak when the width is much greater or smaller than the diameter of the colloid~\cite{Cheung2008,Hung2007,Eskandari2014}. 
These ideas were advanced by investigating periodic wavy boundaries, in which periodic nematic deformations near the wavy walls causes colloids with Saturn ring defects to be attracted to the troughs, and dock via a `lock-and-key' mechanism, while docking location depends on the orientation of the defect for colloids with hedgehog defects~\cite{Luo2016}. 
Variations of this setup have included imposed twist~\cite{Boniello2019}, saw-tooth or crenellated substrates~\cite{RojasGomez2017}, and ellipsoidal~\cite{Luo2019} or four-armed colloids~\cite{Yao2022}. 
In addition to setting stable docking sites, structured boundaries can control dynamic trajectories. 
Trajectories can arise due to diffusion or external fields~\cite{Luo2018}, with docking or continued motion controlled by a balance between nemato-elasticity and driving force. 
Periodic structures, in particular, allow precise control of colloidal transport~\cite{Chen2018}. 
In these cases, nematic colloids must respond to both elastic deformations through free energy minimization and advection through drag. 

In this manuscript, we model \correctText{spherical}{discoidal} colloids advecting in fluctuating 2D nematics flowing through microfluidic wavy channels (\fig{fig:setup}) \correctText{}{and vary the amplitude of the undulations}. 
We investigate this using a Multi-Particle Collision Dynamics (MPCD) algorithm to simulate thermalized nematodynamics and the inclusion of colloidal particles, and we provide a method for implementing wavy boundaries. 
This approach includes colloidal diffusion, nemato-elasticity and hydrodynamics within complex boundaries. 
\correctText{We}{Inspired by experimental studies in which docking sites are of comparable size to colloids~\cite{Luo2016,Boniello2019,Luo2019}, we} focus on the effect of the amplitude of the wavy boundaries on the colloidal trajectories and velocities, for different anchoring conditions, and demonstrate that these systems have the potential to trap colloids at given points or allow colloidal conveyance. 
For intermediate amplitudes, elution follows stick-slip dynamics, with the ``sticking'' duration increasing with amplitude until the colloid locks in place. 
The amplitude at which the transition from stick-slip to locking occurs decreases with nematic elasticity for a given pressure gradient. 
\correctText{}{We consider colloids with strong homeotropic anchoring but allow the anchoring on the channel walls to vary.}
Our results show simple microfluidic systems can temporarily lock particles in place by combining surface structure and advective flow, allowing precise control over elution dynamics and the ability to lock colloids in place for specific times.

\section{Methods}\label{sctn:methods}
To simulate colloids in a flowing and thermally fluctuating nematic liquid crystal confined within a complex channel, the Multi-Particle Collision Dynamics (MPCD) algorithm is used~\cite{kapral2008,Gompper2009,Slater2009,Howard2019}. 
MPCD has been employed to simulate reaction-diffusion dynamics~\cite{Sayyidmousavi2018,Reigh2020}, electrophoresis~\cite{Hickey2012,Shendruk2015b}, thermophoresis~\cite{Burelbach2018}, swimmers~\cite{zottl2018,Kuhr2019,Clopes2020}, polymers~\cite{Lamura2021,Clopes2022,Choi2023,wang2023}, colloidal suspensions~\cite{Wani2022,Chen2019}, binary mixtures~\cite{Eisenstecken2018}, viscoelastic fluids~\cite{Toneian2019}, ferrofluids~\cite{Ilg2022,Ilg2022b}, and dense stellar systems~\cite{dicintio2020,dicintio2022a,dicintio2022b}. 
Most relevantly, MPCD has been used to simulate nematic liquid crystals~\cite{shendruk2015}. 
In this context, MPCD has simulated nematohydrodynamics~\cite{Mandal2019}, suspended colloids~\cite{ReyesArango2020,Hijar2020}, magnetic colloids~\cite{Armendariz2021b,Armendariz2021}, living liquid crystals~\cite{Mandal2021}, and active nematics~\cite{kozhukhov2022,kozhukhov2023,Keogh2023}.
For the present study, MPCD is chosen because it is ideal for moderate P\'{e}clet numbers, mobile solutes and complex \correctText{boundary}{boundaries}. 

The \correctText{MPCD}{nematic MPCD (N-MPCD)} algorithm discretizes the continuous fluid into \correctText{point $N$}{$N$ point} particles. 
Each such point particle $i$ is given a set mass $m_i = m \ \forall \ i$, possesses an instantaneous velocity $\vec{v}_i$ and unit-length orientation $\vec{u}_i$ to model the direction of the nematogens. 
The MPCD algorithm consists of two steps~\cite{Malevanets1999,Malevanets2000}: 
    \textit{(i)} streaming and 
    \textit{(ii)} collision. 
The streaming step (see \correctText{SI}{\appndx{app:methods}}) updates each particle position $\vec{r}_i$ assuming ballistic motion over a time step $\delta t$. 
The collision step (see \correctText{SI}{\appndx{app:methods}}) simulates the interactions between fluid particles via a coarse-grained collision operator. 
The collision operator stochastically updates the particles' velocities and orientations, while constrained to respect conservation laws. 
To do this, the simulation domain is split into cubic cells of size $a$ and index $c$ containing  $N_c(t)$ particles at any instant $t$. 
A random grid shift ensures Galilean invariance~\cite{Ihle2001}. 
Only particles in the same cell interact and the interactions involve all particles in the same cell. 
The average number of particles per cell sets the fluid density $\rho = m \av{N_c}/a^d$, where $\av{\cdot}$ is the average over all cells and $d=2$ for a two-dimensional fluid. 
The collision event can itself be broken into three stages (see \correctText{SI}{\appndx{app:methods}}): 

    {\bf 1. Stochastic momentum exchange:} Particle velocities are updated by a collision operator $\Xi_{i,c}(t)$ for particle $i$ in cell $c$ at time $t$. 
    An Andersen-thermostatted collision operator randomly generates velocities from a Boltzmann distribution with energy $\kbt$~\cite{Noguchi2007,Noguchi2008}. 
    This operator conserves momentum and thermostats the energy. 

    {\bf 2. External forces:} A pressure gradient is applied as $\grad P = \rho \vec{g}$, where the effective external acceleration $\vec{g}$ is included in the collision operator. 

    {\bf 3. Orientation exchange:} 
    \correctText{Each}{In N-MPCD, each} particle's orientation $\vec{u}_i$ is stochastically drawn from the Maier-Saupe distribution about each cell's director $\vec{n}_c$ as $p_c(\vec{u}_i) \propto \exp{\left(U S_c (\vec{n}_c \cdot \vec{u}_i)^2/\kbt \right)} $, where $S_c$ is the scalar order parameter \correctText{$S_c$}{of cell $c$} and $U$ is the mean field potential, which controls how strongly the orientations align. 
    Additionally, MPCD particle orientation responds to velocity gradients through Jeffery's equation, which is parameterized by a bare tumbling parameter $\xi$ and heuristic shear coupling coefficient $\chi$. 
    \correctText{}{The $\chi$ parameter acts as a relaxation parameter, effectively allowing Jeffrey's equation to be averaged over a small number of time steps of the fluctuating hydrodynamic field.}
    Finally, director dynamics are coupled back to the velocity field through local torques parameterized by a rotational mobility coefficient $\gamma_\text{R}$ (See \correctText{SI}{\appndx{app:methods}} for details)~\cite{shendruk2015,Armendariz2021}. \correctText{}{The small value chosen keeps backflow effects to a minimum.}
    Numerical analysis has shown that N-MPCD describes a linearized nematohydrodynamic model in which viscosity and elastic effects are isotropic~\cite{Hijar2019}.

The wavy 2D channel is modelled with plane-wavy solid boundaries (see \appndx{app:bc}). 
Each \correctText{boundary surface}{surface (both bounding walls and colloidal)} is labelled by index $b$ and represented as a surface on which $\mathcal{S}_b(\vec{r})=0$. 
A baseline surface is defined for planes and, to account for wavy boundaries, a second term is included, which allows waves with amplitude $B_{b,0}$ and \correctText{frequencies $B_{b,1}$ and $B_{b,2}$}{frequency $B_{b,1}$}. 
\correctText{In 3D, this allows us to simulate egg-cartoon (Fig.~2a) and corrugated (Fig.~2b) boundaries, as well as wavy colloids and cylinders (see SI)}{}. 
In this study, we focus on 2D \correctText{($B_{b,2} = 0$ and smooth spherical colloids ($B_{b,0}=B_{b,1}=0$)}{}\correctText{}{; however, this framework allows simulations of egg-cartoon and corrugated walls (\appndx{app:bc})}. 
Both the top and bottom wavy walls of the channel have the same amplitude $B_{b,0}=B_0$ and wavelength $\lambda$ given by \correctText{$B_{b,1} = 2\pi\mathcal{L}/\lambda \ \forall \ b$}{$B_{b,1} = B_{1} = 2\pi\mathcal{L}/\lambda$ for all wall boundaries}, where $\mathcal{L}$ is the system size in that direction. 
\correctText{}{The boundaries defining the discoidal colloids are smooth and so $B_{b,0}=B_{b,1}=0$ for all colloids, though wavy colloids and cylinders are possible in our framework (\appndx{app:wavyspheres}).} 
Bounce-back boundary conditions with phantom particles ensure no-slip~\cite{Lamura2001,Whitmer2010,Bolintineanu2012}. 
Periodic boundary conditions are \correctText{planar}{plane} surfaces at the channel extremities.

\subsection{Simulation Parameters}\label{sctn:params}
The simulation units are particle mass $m$, thermal energy $\kbt$ and cell size $a$. 
From these, the time unit is $\tau=a\sqrt{m/\kbt}$. 
Additional derived units include dynamic viscosity $\mu_0 \equiv \kbt \tau / a^{d}$~\cite{Noguchi2008}, stress $\kbt a^{-d}$ and Frank elasticity $\kbt a^{2-d}$. 
In these two dimensional ($d=2$) simulations, the time step is set to $\delta t=0.1\tau$ and  density is set to $\rho=20m/a^2$. 
The mean field potential is $U=10\kbt$ (unless otherwise stated), placing the fluid deep in the nematic phase~\cite{shendruk2015}. 
A mean field potential of $U=5\kbt$ is relatively close to the nematic-isotropic transition point, with a system-\correctText{averaged}{wide (averaged over all cells $c$)} scalar parameter $S\approx0.5$, while $U=10\kbt$ and $20\kbt$ have scalar parameters much closer to $S\approx1$ \cite{shendruk2015}. 
\correctText{}{These large values are an idealization that ensures the system is deep in the nematic phase and that elastic constants are large enough to ensure small Ericksen numbers (\appndx{app:planeChannel}).}
The Frank elastic constants were previously found to obey a one constant approximation and be linear with  the mean field potential as $K=\left(113\pm6\right) U$ for $\rho=20m/a^2$~\cite{shendruk2015}. 
The nematic is in the flow aligning regime with $\xi=2$, shear susceptibility $\chi=0.5$ and rotational friction $\gamma_\text{R}=0.01\tau\kbt$. \correctText{}{The choice of small rotational friction minimizes the amount of backflow and limits any backflow-related effects in this study.}
The gravitational acceleration is set to $\vec{g}=0.001a/\tau^2\hat{x}$ to generate a pressure gradient $\vec{\nabla}P= 0.02 \kbt/a^{1+d} \hat{x}$. 
Ten repeats \correctText{}{($n=10$)} of each configuration are run for $1.5\times10^4\tau$. 

The channel is modelled with plane-wavy solid boundaries along $\hat{x}$. 
Both walls have the same amplitude $B_0$ and frequency $B_1$. 
Planes set $60a$ apart define the channel walls with average normal $\pm\hat{y}$. 
The domain size is $60a \times (20a+2B_0)$ and the frequency of the wavy walls, $B_1$, is set to $\pi$, so that 3 full periods fit within the length of the channel, which gives a wavelength $\lambda=20a$. 
Positions of $x=\lambda/4$ and $x=3\lambda/4$ correspond to the center of the trough and crest, respectively. 
The amplitude of the waves $B_0$ is varied between $0$ and $4.5a$. 

The colloid is simulated as a moving circular boundary with a radius $R=5a$ and mass $\rho\pi r^2$. 
\correctText{}{The presence of the $-1/2$ companion defect pair (\fig{fig:setup}) indicates that the anchoring is strong (\appndx{app:bc}) and that the characteristic anchoring length is small compared to the colloid size~\cite{Louise}.}
Colloids are initialized with their center at $\vec{r}_c=(x,y)=(25a,10a+B_0)$, which is between the wavy walls and offset to the left between a crest and a trough on the walls (\fig{fig:setup}), unless otherwise stated. 
\correctText{}{Immediately after initialization, colloids are free to move in response to nematic and hydrodynamic forces.} \correctText{}{The colloid position $\vec{r}_c$ is a continuous variable that is not constrained to the MPCD collision lattice.}
These choices define a characteristic length scale $L$ for the geometry of our system. 
Here, $L=10a$ coincides with the diameter of the colloid $2R$, the average distance from the centerline to the wall $\av{h}/2$, and the distance from trough to crest $\lambda/2$. 
This length scale results in characteristic velocities. 
The velocity scale of the nematoelastic response is $\tilde{V} = K / \mu L^{d-1}$ for viscosity $\mu$ and the scale due to pressure gradients is $V=L^2 \rho g_x /\mu$. 
For the parameters used here, the nematoelastic speed $\tilde{V} \approx 12 a/\tau$ is much larger than the advective speed $V = 0.208(9) a/\tau$, which we identify to be the characteristic speed of interest in our lock-key dynamics. 

\correctText{}{In this study, all colloids have homeotropic anchoring but the anchoring on the channel walls varies.}

\section{Results}\label{sctn:results}

\begin{figure}[tb]
    \centering
    \begin{subfigure}[b]{0.23\textwidth}
        \centering
        \includegraphics[width=\textwidth]{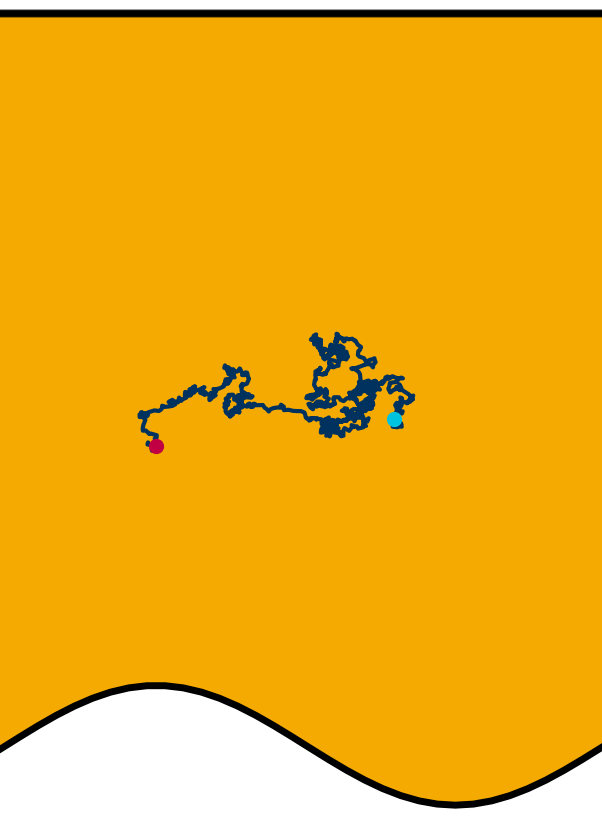}
        \caption{Homeotropic \correctText{}{anchoring} - crest.} 
        \label{fig:homeoHill}
    \end{subfigure}
    \hfill
    \begin{subfigure}[b]{0.23\textwidth}
        \centering
        \includegraphics[width=\textwidth]{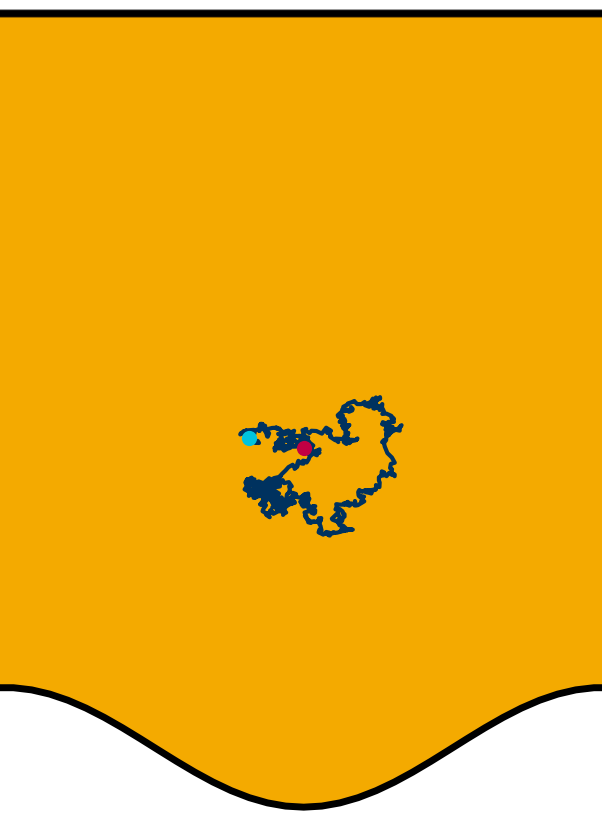}
        \caption{Homeotropic \correctText{}{anchoring} - trough.}
        \label{fig:homeoWell}
    \end{subfigure}
    \\
    \begin{subfigure}[b]{0.23\textwidth}
        \centering
        \includegraphics[width=\textwidth]{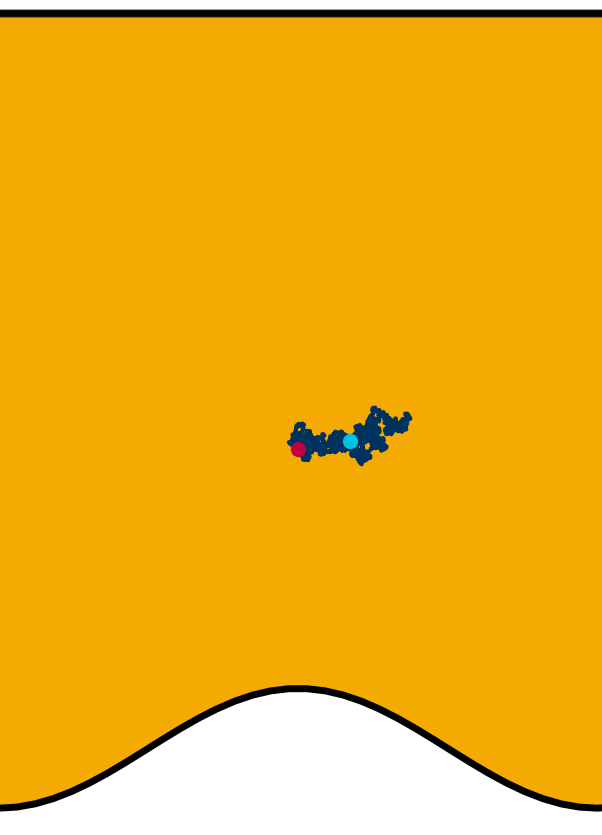}
        \caption{Planar \correctText{}{anchoring} - crest.}
        \label{fig:planarHill}
    \end{subfigure}
    \hfill
    \begin{subfigure}[b]{0.23\textwidth}
        \centering
        \includegraphics[width=\textwidth]{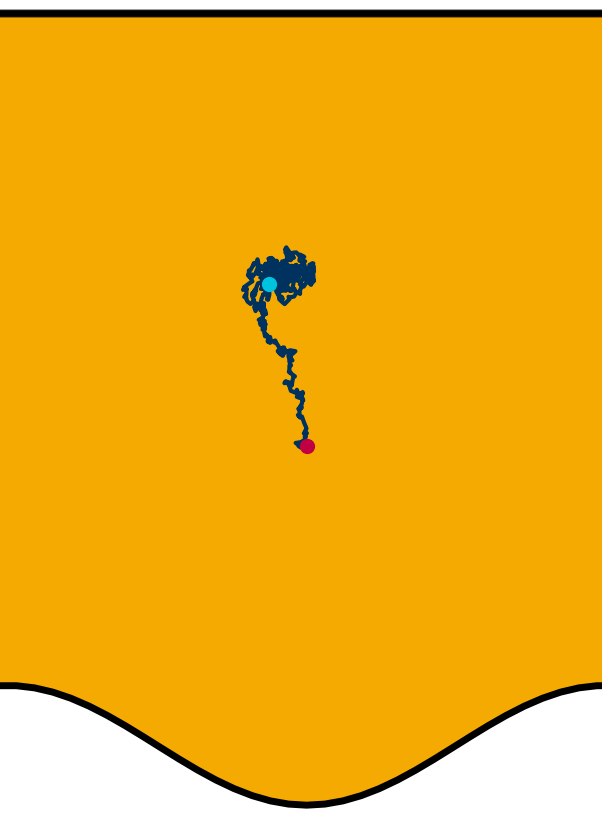}
        \caption{Planar \correctText{}{anchoring} - trough.}
        \label{fig:planarWell}
    \end{subfigure}
    \caption{
        Typical \correctText{trajectory}{trajectories} for a colloid with homeotropic anchoring confined between a plane wall above and wavy wall below. 
        \correctText{}{The start and end points are shown by magenta and cyan dots respectively.}
        \textbf{(top row)}
        Walls have strong homeotropic anchoring. 
        The colloid is either initiated above a crest (a) or above a trough (b). 
        \correctText{The start and end points are shown by magenta and cyan dots respectively.}{} 
        \textbf{(bottom row)} Walls have strong planar anchoring, while the anchoring on the colloid surface is still strong homeotropic.
        Colloid radius $R=5a$; wall amplitude $B_0=2.5a$; wall wavelength $\lambda=20a$; and average channel height $\av{h}_x=20a$.
    }
    \label{fig:trajsHomeo}
\end{figure}

Lock-key microfluidics result from two competing \correctText{process}{processes}: 
nematoelastic relaxation pushes the colloids to locations that minimize the distortion free energy, while the minimum dissipation theorem leads colloids to advect with a velocity that matches the fluid flow. 
For this reason, we begin by simulating homeotropic colloid kinetics in the absence of \correctText{}{externally driven} flows (\sctn{sctn:noFlow}), then proceed to consider nematohydrodynamic flows in wavy channels in the absence of the colloid (\sctn{sctn:noColloid}) before putting the two aspects together to study lock-key microfluidics (\sctn{sctn:colloidFlow}). 
\correctText{}{In all cases, each colloid has a topological charge of $+1$ and so necessitates the existence of companion defects in the vicinity of the colloid.
In 3D, both Saturn rings and hedgehog defects are possible~\cite{Grollau2003} with rings more likely in simulations~\cite{Andrienko2001,Sussman2019}. 
In 2D only the quadrapolar state is expected to be stable~\cite{fukuda2001,tasinkevych2002} and the results reported here are consistent with this since they exhibit a pair of $-1/2$ companion defects.}

\subsection{Colloid Trajectories With No Pressure Gradient} \label{sctn:noFlow}
\correctText{Colloid Trajectories With No Flow}{}

We place colloids anchoring into channels with one plane boundary \correctText{}{wall} and one wavy \correctText{boundary}{wall}, with an amplitude of $B_0=2.5a$. 
The nematic anchoring on the colloid is always homeotropic but both planar and homeotropic anchoring on the channel walls is considered. 
No pressure gradient is applied. 
The colloid is initially positioned either directly above a crest or trough (\fig{fig:trajsHomeo}; top row) for homeotropic anchoring on the boundary walls. 
These numerical trajectories are consistent with Luo \textit{et al.} \cite{Luo2018}. 
For homeotropic anchoring on the boundary walls, the colloid is attracted to the trough, traversing away from the crest where it settles (\fig{fig:homeoHill}). 
The principle motion is in the $\hat{x}$ direction with little only stochastic motion in the $\hat{y}$ direction. 
\correctText{}{Although the colloid does not move substantially in the $\hat{y}$ direction, careful inspection shows that it does drift upward since the repulsive forces from the different walls are not expected to balance at the same height for each $x$-position.} 
When the colloid is initiated above the trough ($x=\lambda/4$), it still diffuses, but remains within the trough in the vicinity of the free energy minimum, exhibiting no deterministic forcing away from this point (\fig{fig:homeoWell}). 

\correctText{Planar}{Homeotropic anchoring on the colloid and planar} anchoring on the walls was not investigated by Luo \textit{et al.} \cite{Luo2018}, but shows interesting features (\fig{fig:trajsHomeo}; bottom row). 
In the case of planar anchoring on the walls, the colloid is stable above the crest (\fig{fig:planarHill}), rather than the trough. 
\correctText{}{This is consistent with experimental images of the analogous case of a planar colloid being attracted to the crests of a homeotropic wall~\cite{Luo2018} and with the trapping of colloidal particles at the tips of sharp protrusions whenever the colloid and surface have opposing anchoring conditions~\cite{Silvestre2014}.}
In this case, the colloid simply diffuses as though it is in a well within the free energy landscape. 
Likewise, a colloid initiated above the trough does not move left or right; but rather, moves directly away from the wavy wall towards the upper plane wall (\fig{fig:planarWell}). 
While the wall above the colloid is also repulsive, the repulsion is smaller than from the wavy wall. 
Thus, the homeotropic colloid's equilibrium position is closer to the \correctText{planar}{plane} wall than the wavy wall. 

\begin{figure}[tb]
    \centering
    \includegraphics[width=0.5\textwidth]{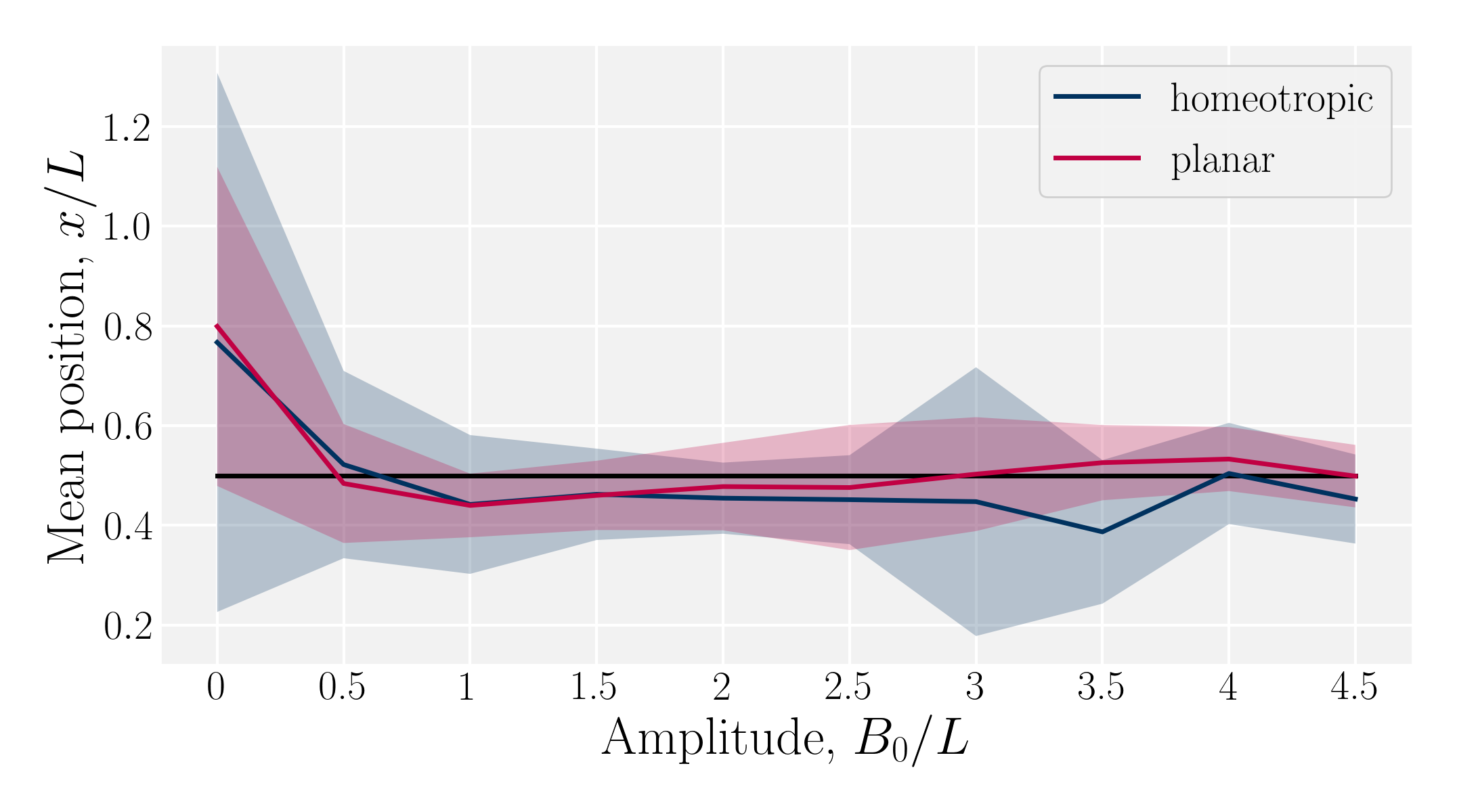}
    \caption{Mean $x$-position of the colloid \correctText{}{with homeotropic anchoring} in relation to the trough for planar and homeotropic anchoring on wavy boundaries. 
    The equilibrium position is above the troughs ($x=\lambda/4$) for all anchoring conditions an non-zero amplitudes $B_0$. 
    Center of the trough shown as the black line and errorbars correspond to the standard deviation of the position. 
    Colloid radius $R=5a$; wall amplitude $B_0=2.5a$; wall wavelength $\lambda=20a$; and average channel height $\av{h}_x=20a$.
    }
    \label{fig:mean_trajs}
\end{figure}

Having considered the dynamics of a colloid confined by one wavy wall, we further consider how homeotropic colloids move in response to wavy boundaries on both the lower and upper wall. 
The colloid is initialized in the center of the channel, and at the inflection point halfway between a trough and a crest. 
For both homeotropic and planar anchoring on the walls, colloids are observed to move towards the troughs (\fig{fig:mean_trajs}) and generally remains centered in the trough for all amplitudes. 

\begin{figure}[tb]
    \centering
    \begin{subfigure}[b]{0.225\textwidth}
        \centering
        \includegraphics[width=\textwidth]{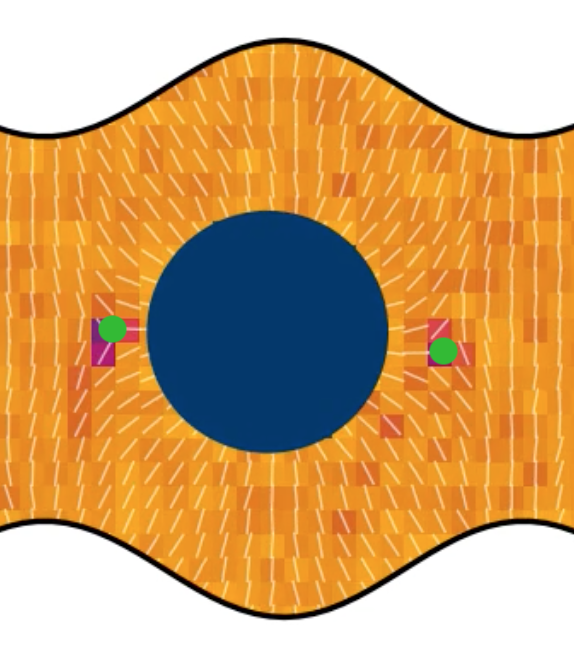}
        \caption{Homeotropic \correctText{}{anchoring}.} 
        \label{fig:def_homeo}
    \end{subfigure}
    \begin{subfigure}[b]{0.235\textwidth}
        \centering
        \includegraphics[width=\textwidth]{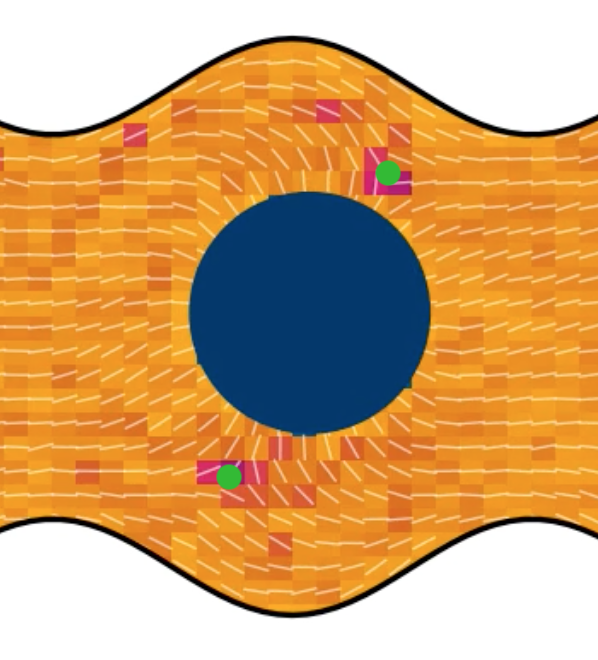}
        \caption{Planar \correctText{}{anchoring} - centered.}
        \label{fig:def_planar_c}
    \end{subfigure}
    \begin{subfigure}[b]{0.235\textwidth}
        \centering
        \includegraphics[width=\textwidth]{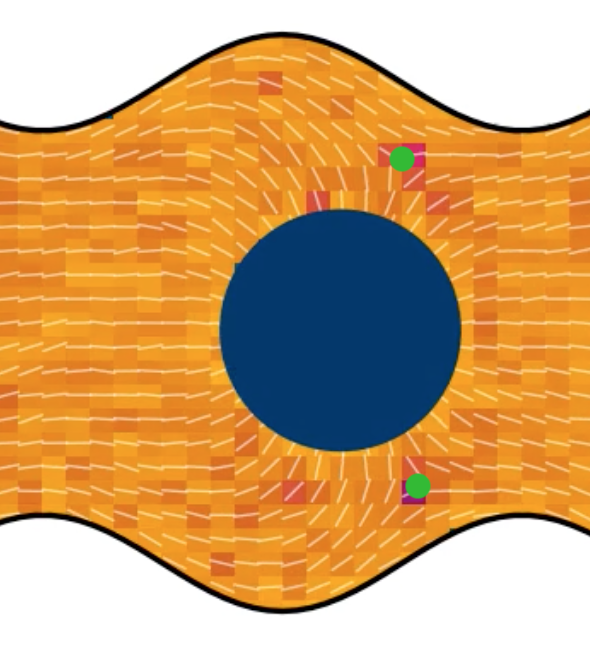}
        \caption{Planar \correctText{}{anchoring} - offset.}
        \label{fig:def_planar_o}
    \end{subfigure}
    \begin{subfigure}[b]{0.2\textwidth}
        \centering
        \includegraphics[width=\textwidth]{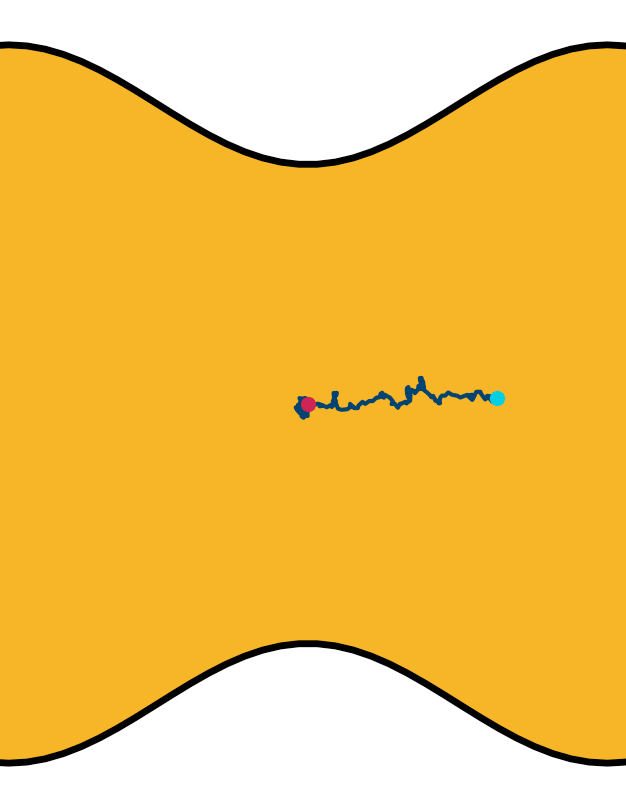}
        \caption{Planar \correctText{}{anchoring -} crest.}
        \label{fig:traj_planar_h}
    \end{subfigure}
    \caption{
        Colloid \correctText{}{with homeotropic anchoring} between two wavy walls \correctText{}{with various anchoring conditions}. 
        (a-c) The defect positions \correctText{}{(green dots)} for homeotropic and planar anchoring on the wavy walls. 
        The green dots show the defect positions. 
        (a) Homeotropic anchoring. 
        (b) Planar anchoring with the colloid is centered within the trough.
        (c) Planar anchoring with the colloid offset to the right. 
        (d) Trajectory of a colloid initiated above a crest. 
        The magenta and cyan dots show the start and end positions respectively. 
        Colloid radius $R=5a$; wall amplitude $B_0=2.5a$; wall wavelength $\lambda=20a$; and average channel height $\av{h}_x=20a$. 
    }
    \label{fig:defects_res}
\end{figure}

While the colloid's equilibrium position is centered on the troughs in all cases, the locations of the colloid-associated defects differ (\fig{fig:defects_res}). 
When the colloid with strong homeotropic anchoring, possessing two satellite $-1/2$ defects, is confined between two wavy walls with strong homeotropic anchoring the defects are found immediately to the left and right of the colloid along the centerline (\fig{fig:def_homeo}). 
However, for the same colloid initialized in the trough of a wavy channel with planar anchoring, the defects take a diagonal configuration (\fig{fig:def_planar_c}), rotating off the centerline to the points where the walls are closest to the colloid. 
By residing in the smallest space between the \correctText{planar walls}{walls, with their planar anchoring} and \correctText{homeotropic colloid surfaces}{the colloid, with its homeotropic surface,} the defect pair can reduce the nematoelastic free energy. 
Since there are four equal points for which the distance between surfaces are minimal, the orientation of the diagonal configuration is spontaneous. 
When the colloid is initialized off center from at the crest between planar\correctText{}{-anchored} walls (\fig{fig:def_planar_o}), the diagonal pairing of defects is no longer the preferred configuration. 
Instead the system prefers to have the two defects shifted towards the closest boundary. 
This configuration lowers the instantaneous deformation free energy by placing the pair of defects at the two minimal separation points between surfaces but the state is not stable. 
In this state, the colloid moves away from the crests towards the center of the trough. 
In this non-diagonal configuration, the defect separation is not as large as when they are found at polar opposite locations. 
Once the colloid approaches the center of the troughs one of the defects is able to escape the deformation barrier cost of crossing to the opposite side. 
Unlike the case of one wavy and one \correctText{planar}{plane} wall (\fig{fig:trajsHomeo}), the crest is not a stable point (\fig{fig:traj_planar_h}). 

\subsection{Flowing Nematodynamics}\label{sctn:noColloid}

The previous section investigated the diffusive dynamics of colloids through the distortion free energy landscape of the nematic in the absence of \correctText{advection}{externally driven flows} (\sctn{sctn:noFlow}). 
We now consider \correctText{the}{pressure-driven} flow of a nematic liquid crystal through a wavy microfluidic channel. 

\begin{figure}[tb]
    \centering
    \begin{subfigure}[b]{0.495\textwidth}
        \includegraphics[width=\textwidth]{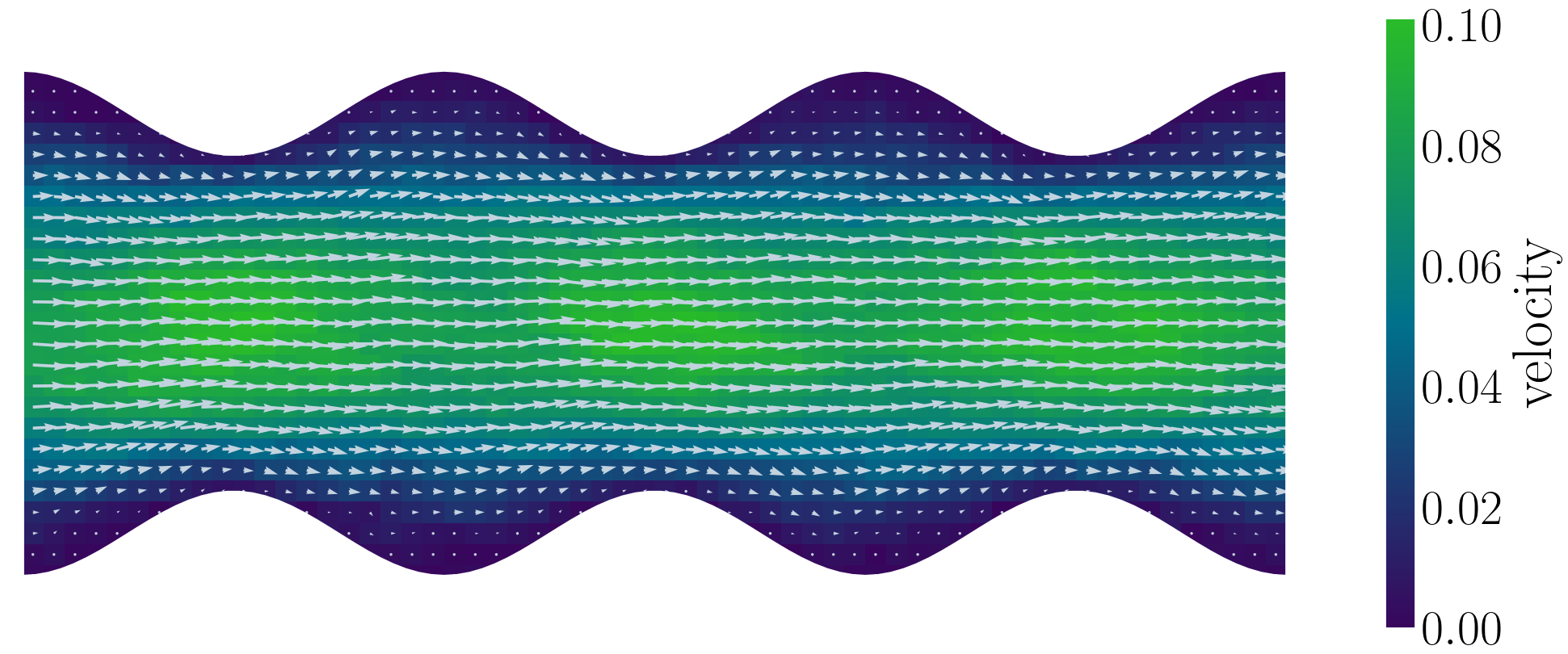}
        \caption{Flow field.}
        \label{fig:wavy_velPic}
    \end{subfigure}
    \begin{subfigure}[b]{0.495\textwidth}
        \includegraphics[width=\textwidth]{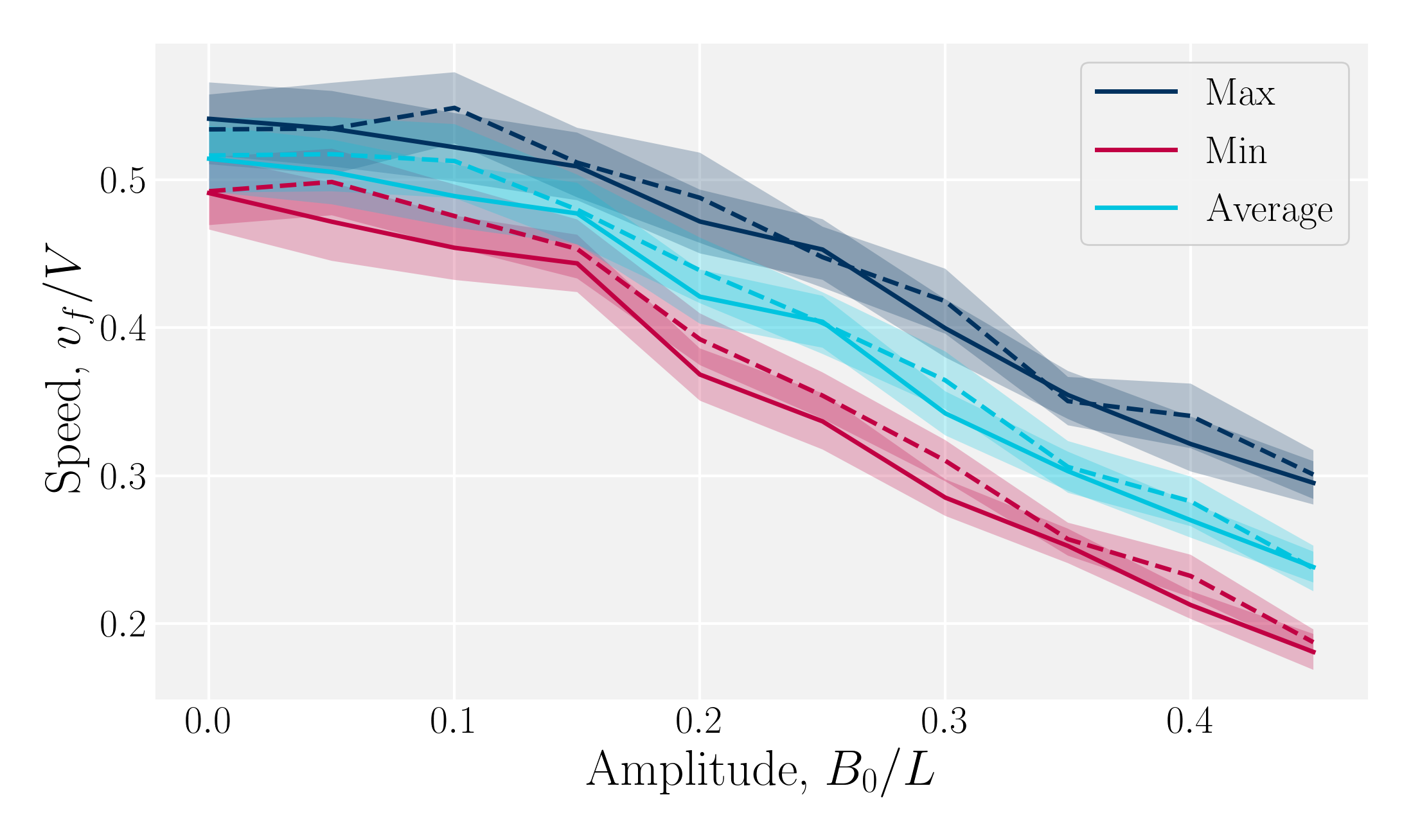}
        \caption{Characteristic velocities.}
        \label{fig:charFlow}
    \end{subfigure}
    \caption{
        Fluid velocity $\vec{v}_f\left(\vec{r}\right)$ field for a channel with wavy walls. 
        (a) Time-averaged velocity field. 
        Colour maps the \correctText{x}{$\hat{x}$}-component of the velocity in units $a/\tau$, and the arrows represent the velocity vectors. 
        Wall amplitude $B_0=2.5a$; wall wavelength $\lambda=20a$; average channel height $\av{h}_x=20a$; and pressure gradient $\vec{\nabla}P= 0.02 ma^{1-d}\tau^{-2} \hat{x}$; $U=10\kbt$; strong planar anchoring. 
        (b) The average, maximum and minimum of the velocity for flow through wavy channels of different amplitudes. 
        Solid lines for nematic phase with $U=10\kbt$ and dashed lines for isotropic phase flow. 
    }
    \label{fig:flows}
\end{figure}

Wavy channel walls complicate the hydrodynamics compared to \correctText{planar}{plane} walls (\appndx{app:planeChannel}), since the flow field $\vec{v}_\text{f}$ has both $\hat{x}$- and $\hat{y}$-components, which depend on both $y$ and $x$ due to the changing channel height $h(x)$ (\fig{fig:wavy_velPic}). 
Increasing the amplitude increases the surface area of the no-slip walls, which in turn increases hydrodynamic resistance. 
For this reason, the elution of fluid is expected to decrease compared to plane channels for constant pressure gradient. 
To quantify this, we compared the average, maximum and minimum velocities along the centerline of wavy channels for different amplitudes $B_0$. 
Consistent with \appndx{app:planeChannel}, there is nearly no difference between the flow velocities of isotropic or nematic fluids through the channels (\fig{fig:charFlow}). 
The average flow through a channel decreases \correctText{non-linearly}{monotonically} with increasing amplitude \correctText{}{for both nematic and isotropic fluids (\fig{fig:charFlow})}, suggesting that the hydrodynamic resistance increases proportionally \correctText{}{in a manner that is not strongly dependent on nematicity}. 
This is analogous to rough walls in plane channels, for which the flow through channels has been found to decrease non-linearly~\cite{Liu2019}. 

The difference between the maximum and minimum flow rates generally increases with the amplitude (\fig{fig:charFlow}). 
Because N-MPCD is a coarse-grained algorithm for fluctuating hydrodynamics, a difference exists even for $B_0=0$. 
The maximum velocity occurs between the crests where the channel is narrowest and the minimum between troughs (\fig{fig:wavy_velPic}). 
This is consistent with previous theoretical and numerical work on flow of isotropic fluids through wavy channels~\cite{Okechi2021}. 
Since the maximum velocity occurs between the crests and the minimum between troughs (\fig{fig:wavy_velPic}), an advected colloid will speed up and slow down with the fluid at these points, in the absence of nematoelastic forces, 
Since the nematic forces also tend to drive the colloid towards the center of the troughs, these dynamics can be expected to be further emphasised.

\subsection{Lock-key microfluidics}\label{sctn:colloidFlow}

\begin{figure*}[tb]
    \centering
    \includegraphics[width=\textwidth]{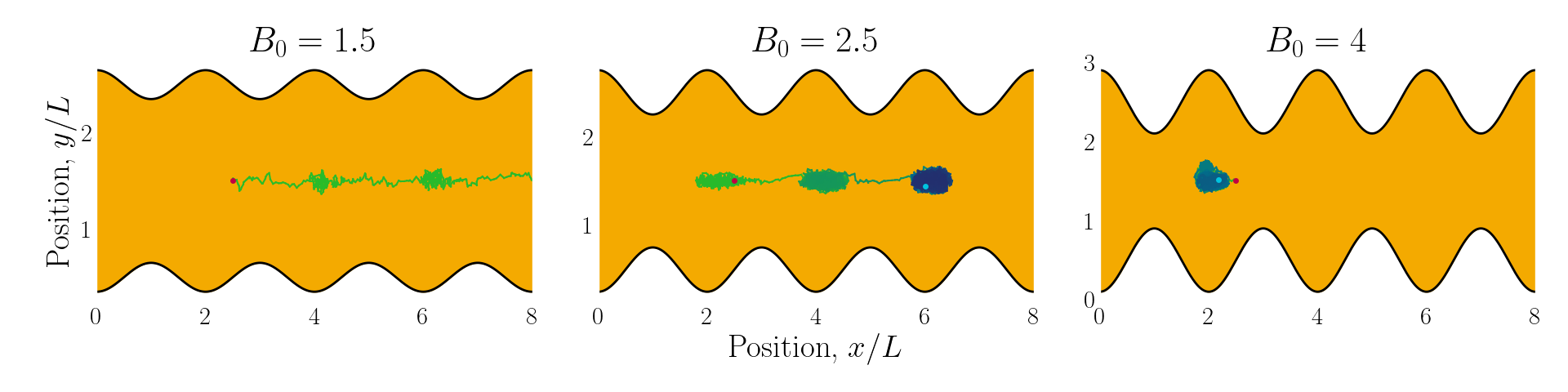}
    \caption{\correctText{}{Example trajectories for small ($B_0=1.5$; left), intermediate ($B_0=2.5$; center) and large amplitudes ($B_0=4$; right).
    Trajectories are colored by simulation time (short times in green and long in blue).}
    }
    \label{fig:xytraj}
\end{figure*}

\begin{figure}[tb]
    \centering
    \begin{subfigure}[b]{0.4\textwidth}
        \includegraphics[width=\textwidth]{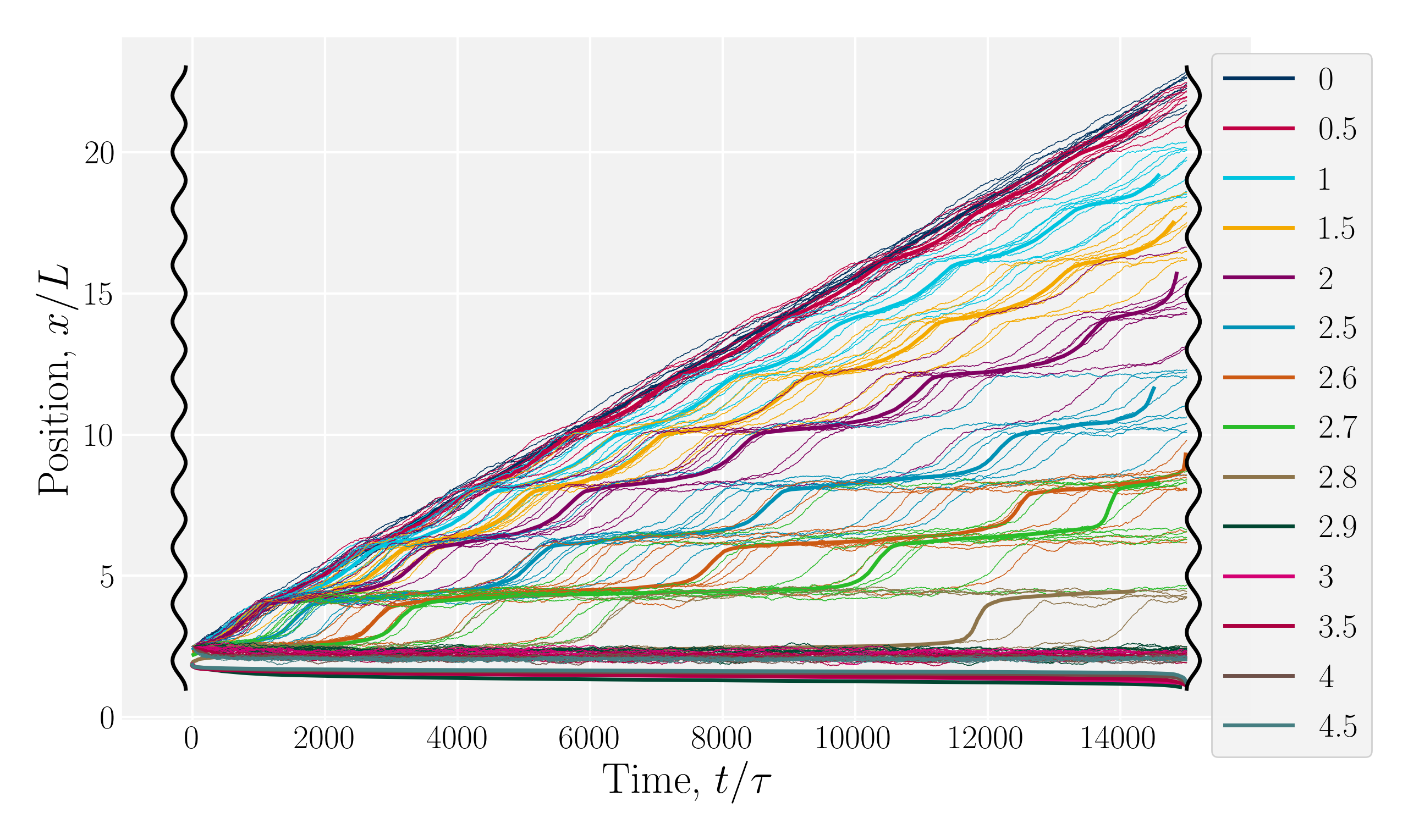}
        \caption{\correctText{Cumulative probability function of position scaled by time for each simulation (thin lines) and ensemble averages (thick lines).}{Ten ($n=10$) individual trajectories for each $B_0$ (thin lines), along with cumulative probability density function $t(x)$ (\eq{eq:cdf}; \fig{fig:CDF_pos}).}
        }
        \label{fig:cdf_runs}
    \end{subfigure}
    \begin{subfigure}[b]{0.4\textwidth}
        \includegraphics[width=\textwidth]{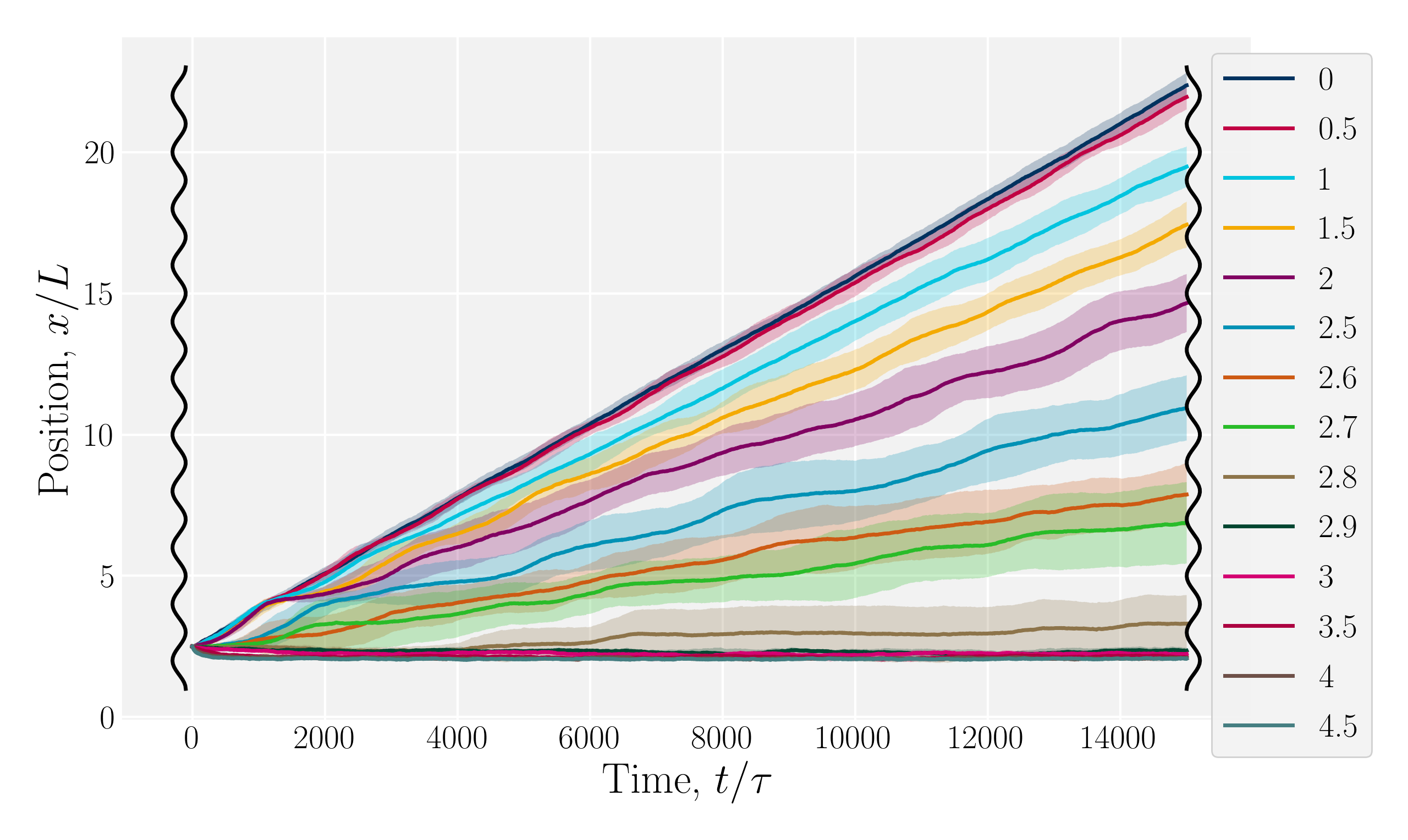}
        \caption{Ensemble averaged $x$-positions of colloids as a function of time.}
        \label{fig:av_pos}
    \end{subfigure}
    \begin{subfigure}[b]{0.4\textwidth}
        \includegraphics[width=\textwidth]{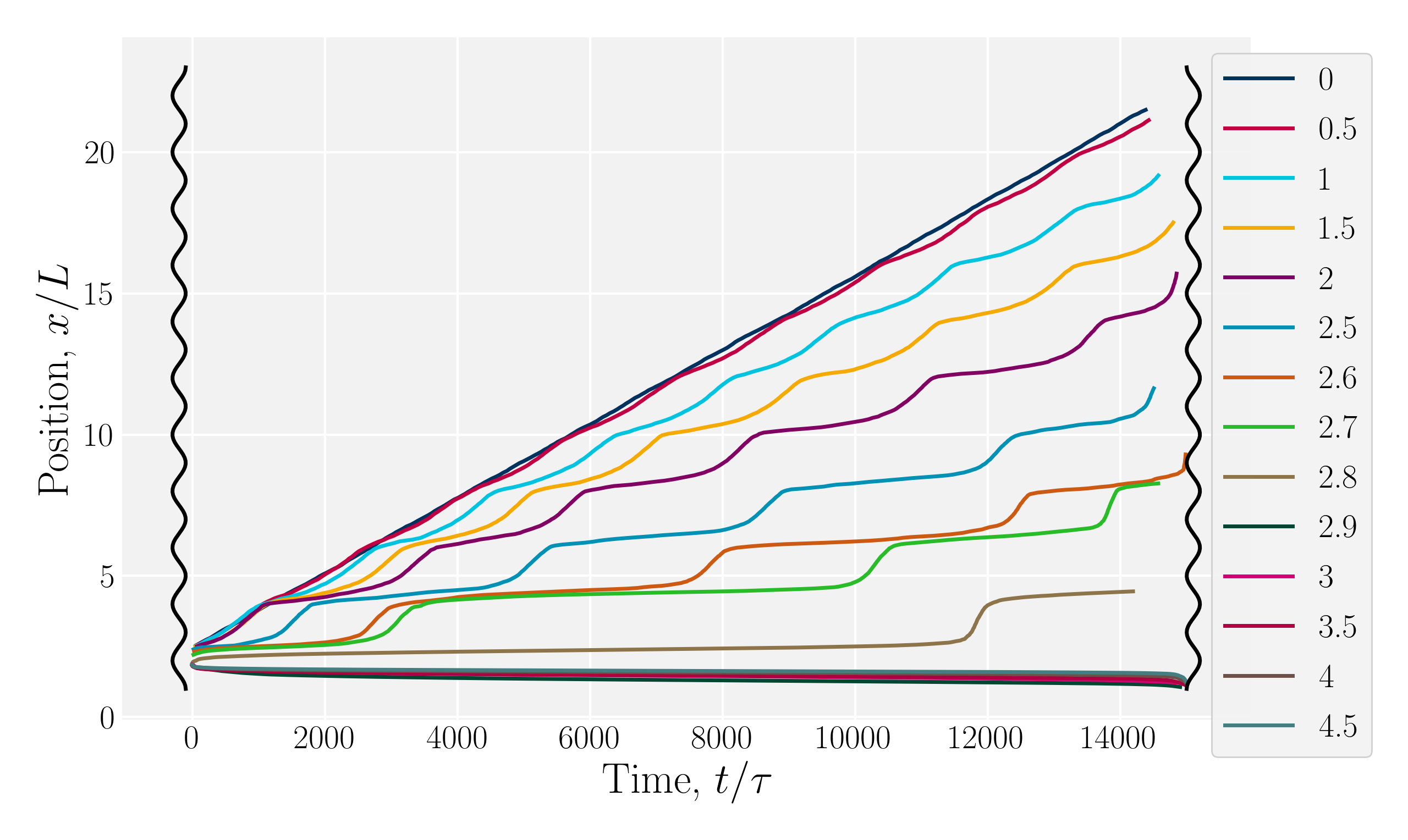}
        \caption{Ensemble averaged cumulative probability density function of position scaled by time \correctText{}{(\eq{eq:cdf})}.}
        \label{fig:CDF_pos}
    \end{subfigure}
    \caption{
        Colloid trajectories for different \correctText{}{wall} amplitudes \correctText{}{$B_0$ with homeotropic anchoring on the colloid and planar anchoring on the walls}. 
        \correctText{}{Colours specify the amplitude $B_0$.}
        Black sinusoids on the right and left present undulations of the wavy walls. 
    }
    \label{fig:pos}
\end{figure}

Having explored diffusive lock-key dynamics in wavy channels in the absence of advection (\sctn{sctn:noFlow}) and flows through wavy channels without colloids (\sctn{sctn:noColloid}), we now turn our attention to lock-key microfluidics by studying the advective dynamics of nematic colloids with homeotropic anchoring confined within wavy channels with planar anchoring. 
To investigate the colloidal transport rate through the channel, we track the position of a colloid through the channel (\fig{fig:xytraj}). 

\correctText{}{We begin by considering individual trajectories of colloids. 
Three example trajectories are shown in \fig{fig:xytraj} for small ($B_0=1.5$; left), intermediate  ($B_0=2.5$; center) and large ($B_0=4$; right) amplitudes. 
In the small amplitude case (\fig{fig:xytraj}; left), the stick slip dynamics are apparent in the $\hat{x}$-component of the trajectories (\fig{fig:cdf_runs}). 
Individual trajectories are seen to have periods within a trough (sticking events) punctuated by abrupt hops over the crest (slipping events). 
As the amplitude is increased (\fig{fig:xytraj}; center), the sticking duration increases substantially and slipping events become less frequent (\fig{fig:cdf_runs}). 
Indeed by large amplitudes (\fig{fig:xytraj}; right), the colloids lock into their initial trough and are never observed to slip forward over the crest into the next trough (\fig{fig:cdf_runs}).}

\correctText{Ensemble}{To better quantify the elution rate, we next consider the ensemble} average x-position of the colloid over time changes with the amplitude $B_0$ (\fig{fig:av_pos}). 
For $B_0=0$, the transport occurs at a constant rate of advection. 
As $B_0$ increases, the average rate decreases. 
By $B_0^*=2.9a$, the colloids are no longer advected through the channel. 
\correctText{}{By averaging over many simulation runs, the elution appears to be rather smooth for all $B_0$}
However, \correctText{}{we recognize that} the average does not fully capture the \correctText{trends}{dynamics} seen in the individual runs, which exhibit clear a stick-slip dynamics. 

\correctText{Because the colloid generally moves in a single direction, the cumulative probability scaled by the time step captures this trend. This illuminates the dynamics of each colloid.}{To better measure the stick-slip dynamics of the colloids, we consider the probability density function (PDF) $p(x)$ of finding a colloid at a given position $x$. 
The PDF can be integrated to give a cumulative probability, which gives the average time taken to arrive at a point 
\begin{align}
    t(x) &= \alpha\int_0^x p(x') dx' ,
    \label{eq:cdf}
\end{align} 
where $\alpha=\delta t/n$ is a scaling parameter converting the cumulative probability to a time for $n$ numerical realizations and time step $\delta t$. 
The time $t(x)$ better represents the stick-slip dynamics of each colloid} (\fig{fig:cdf_runs}). 
For an amplitude of $B_0=0$, the motion is completely linear and the colloids simply advect with resistance independent of $x$-position, as expected. 
For amplitudes $B_0$ smaller than $B_0^* = 2.9a$, the colloid moves along the channel in a stick-slip manner \correctText{}{ (\fig{fig:CDF_pos})}. 
\correctText{}{Stick-slip dynamics are only observed in nematic fluids and not in flowing isotropic fluids (see \appndx{app:isotropic})}
We can see that the time the colloid spends ``sticking'' to the docking site of the trough center increases with amplitude, until the repulsive forces from the crests become large enough to fully stop the colloid from advancing. 
In the case of $B_0=2.8a$, the colloid often locks in place for the entire simulation and other times hops one or two troughs over the course of the entire simulation (\fig{fig:cdf_runs}). 
The colloid permanently sticks at an amplitude of $B_0^*=2.9a$ and the colloids are never observed to advance (\fig{fig:CDF_pos}). 

\correctText{}{We have discussed colloids moving through wavy channels with planar anchoring, but colloids eluting through channels with homeotropic walls exhibit comparable stick-slip dynamics (data not shown). 
The homeotropic cases do not show qualitative differences from \fig{fig:pos}. 
Indeed, permanent locking occurs at $B_0=3.0a$---only slightly above $B_0=2.8a$, the point where it occurred for planar anchoring. 
Later, we will explicitly compare measures of colloid velocity for planar and homeotropic wavy walls and see that they are qualitatively similar.}

\begin{figure}[tb]
    \centering
    \begin{subfigure}[b]{0.35\textwidth}
        \includegraphics[width=\textwidth]{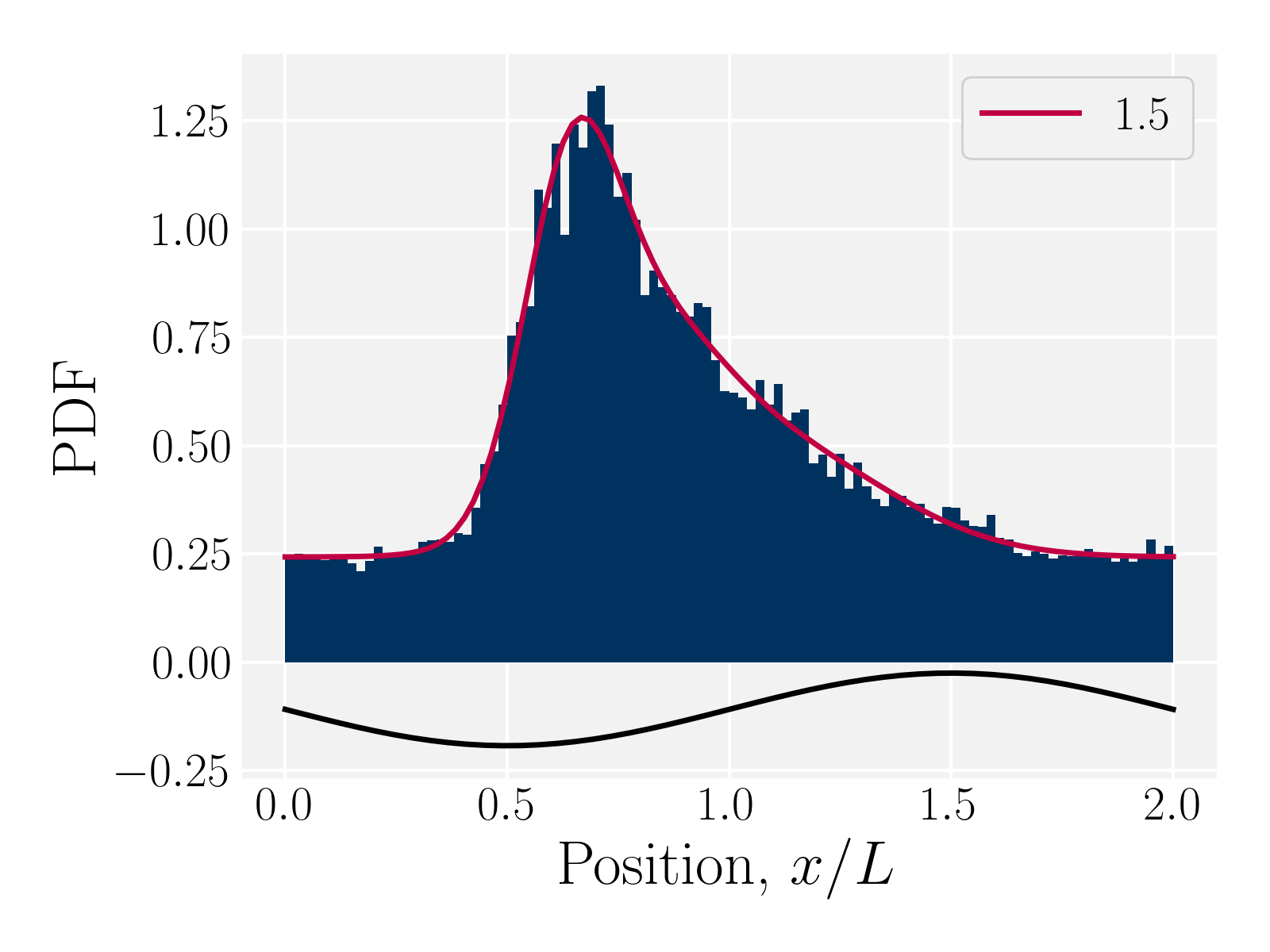}
        \caption{$B_0=1.5$.}
        \label{fig:pdf1.5}
    \end{subfigure}
    \begin{subfigure}[b]{0.35\textwidth}
        \includegraphics[width=\textwidth]{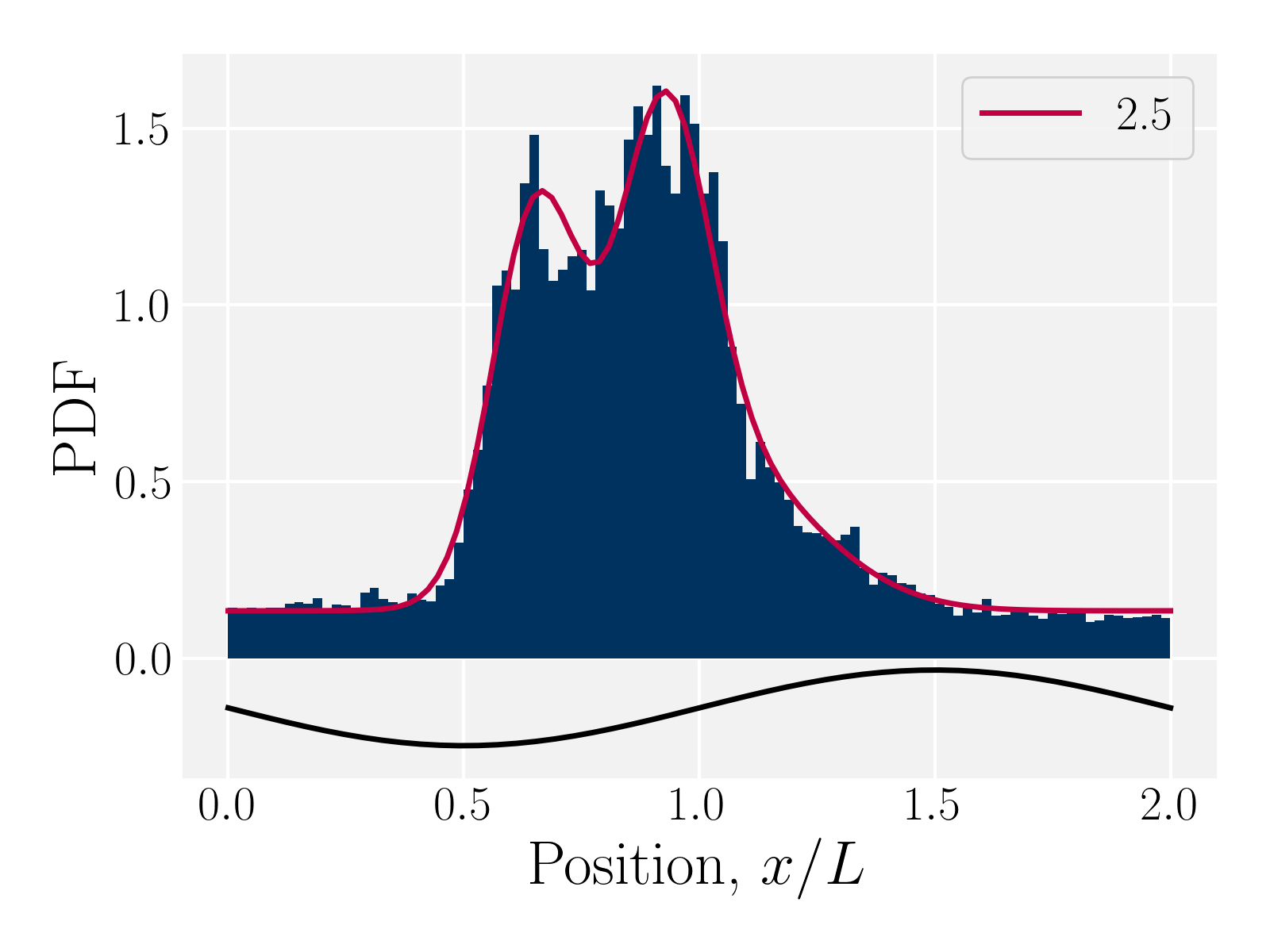}
        \caption{$B_0=2.5$.}
        \label{fig:pdf2.5}
    \end{subfigure}
    \begin{subfigure}[b]{0.35\textwidth}
        \includegraphics[width=\textwidth]{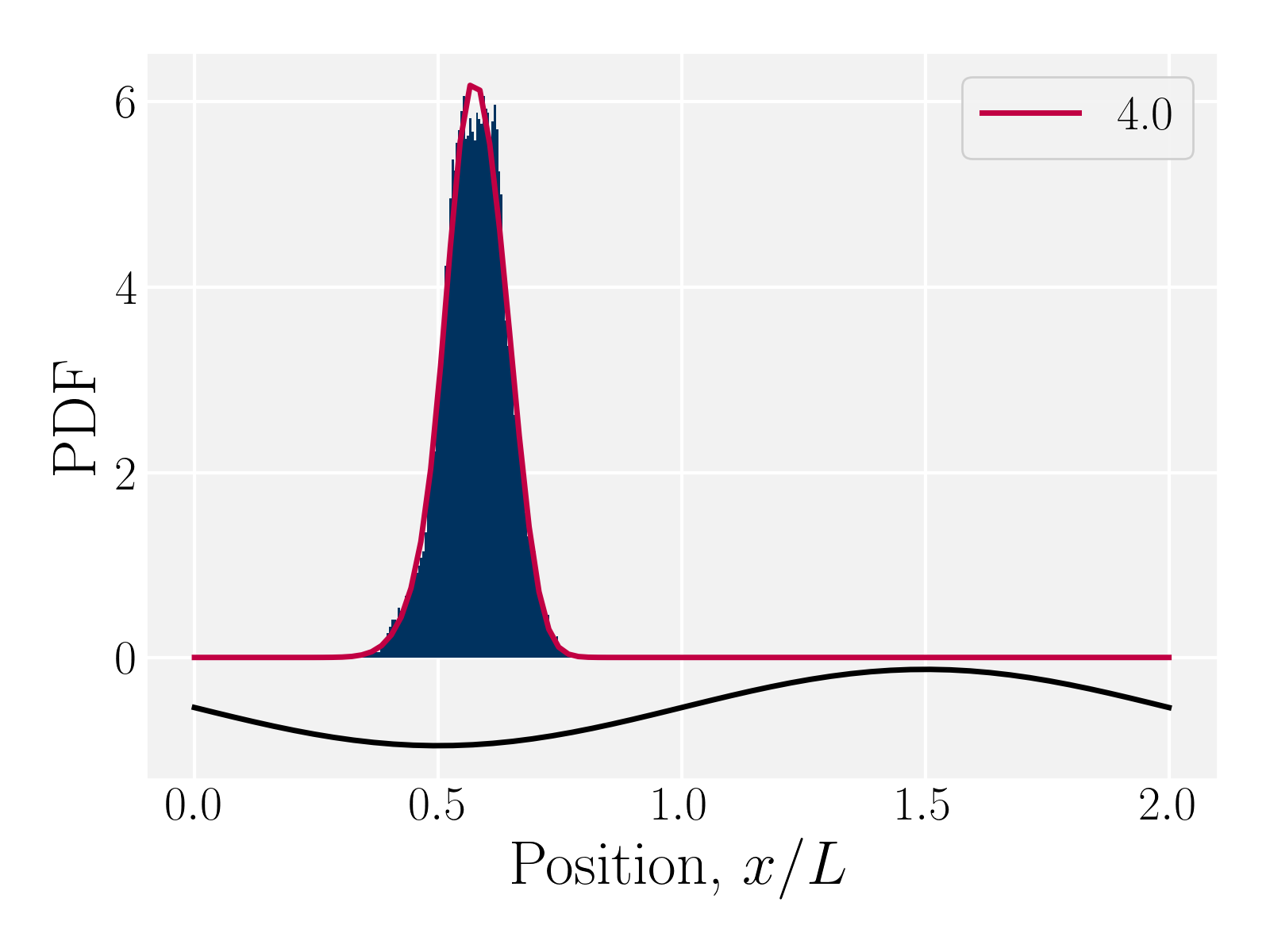}
        \caption{$B_0=4$.}
        \label{fig:pdf4}
    \end{subfigure}
    \caption{
        The probability density function (PDF) of finding the colloid \correctText{}{with homeotropic anchoring on the colloid} at a given position $x$ for three different amplitudes $B_0=\{1.5, 2.5, 4\}$ \correctText{}{with planar anchoring on the wavy walls}. 
        Magenta line shows a fit of three superimposed Gaussians.
        The black curve illustrates undulations of the wavy walls. 
    }
    \label{fig:pdf}
\end{figure}

The probability density function \correctText{(PDF)}{} of finding a colloid at a given position $x$ along the channel explains the trajectories. 
The PDFs for three different amplitudes \correctText{}{in planar anchored channels} demonstrates the three different behaviours (\fig{fig:pdf}). 
\fig{fig:pdf1.5} shows the typical shape for the low amplitudes, where the peak is slightly offset from the center of the trough at $x=\lambda/4$. 
The PDF is highly biased and skewed downstream of the trough center. 
In this case, the colloid continuously advects through the channel.

This changes as the amplitude increases. 
At intermediate amplitudes, the probability densities look close to the point where colloids mostly stop moving (stick) but still can hop over crests from time to time (slip) (\fig{fig:pdf2.5}). 
Both the height and location of the peak changes with the amplitude. 
The distribution becomes bi-modal, with both PDF peaks offset from the trough's center. 
At the largest amplitudes (\fig{fig:pdf4}), the probability density is a single peak, demonstrating that the colloid has completely ``locked'' into the deep free energy minimum at the trough location of $x=\lambda/4$. 
In this case, the probability densities are single peaks, with narrow variance but with a mean position that is slightly offset downstream of $x=\lambda/4$, due to the pressure gradient acting on the locked colloids. 
The biggest difference from smaller amplitude systems is that the probability density is zero everywhere away from the peak. 

The cumulative probabilities can be used to measure the average axial velocities of colloids $v_\text{c}\left(x\right)$ at each position $x$. 
The cumulative probability gives the time as a function of position, and so the velocity along $x$ is found by
\begin{equation}
    \label{eq:velC}
    v_\text{c}\left(x\right) = \left(\frac{d t(x)}{d x}\right)^{-1} ,
\end{equation}
where $t(x)$ is given by the cumulative distribution function (\fig{fig:pos}b). 
Since the cumulative probability density can be calculated from the probability densities $p(x)$, the velocities are found to be
\begin{equation}
    \label{eq:velC2}
    v_\text{c}\left(x\right) = \left[ \alpha p(x) \right]^{-1} ,
\end{equation}
where $\alpha$ accounts for the normalization of $p(x)$. 
The colloidal velocity data is stochastic, due to thermal diffusion. 

\begin{figure}[tb]
    \centering
    \begin{subfigure}[b]{0.35\textwidth}
        \includegraphics[width=\textwidth]{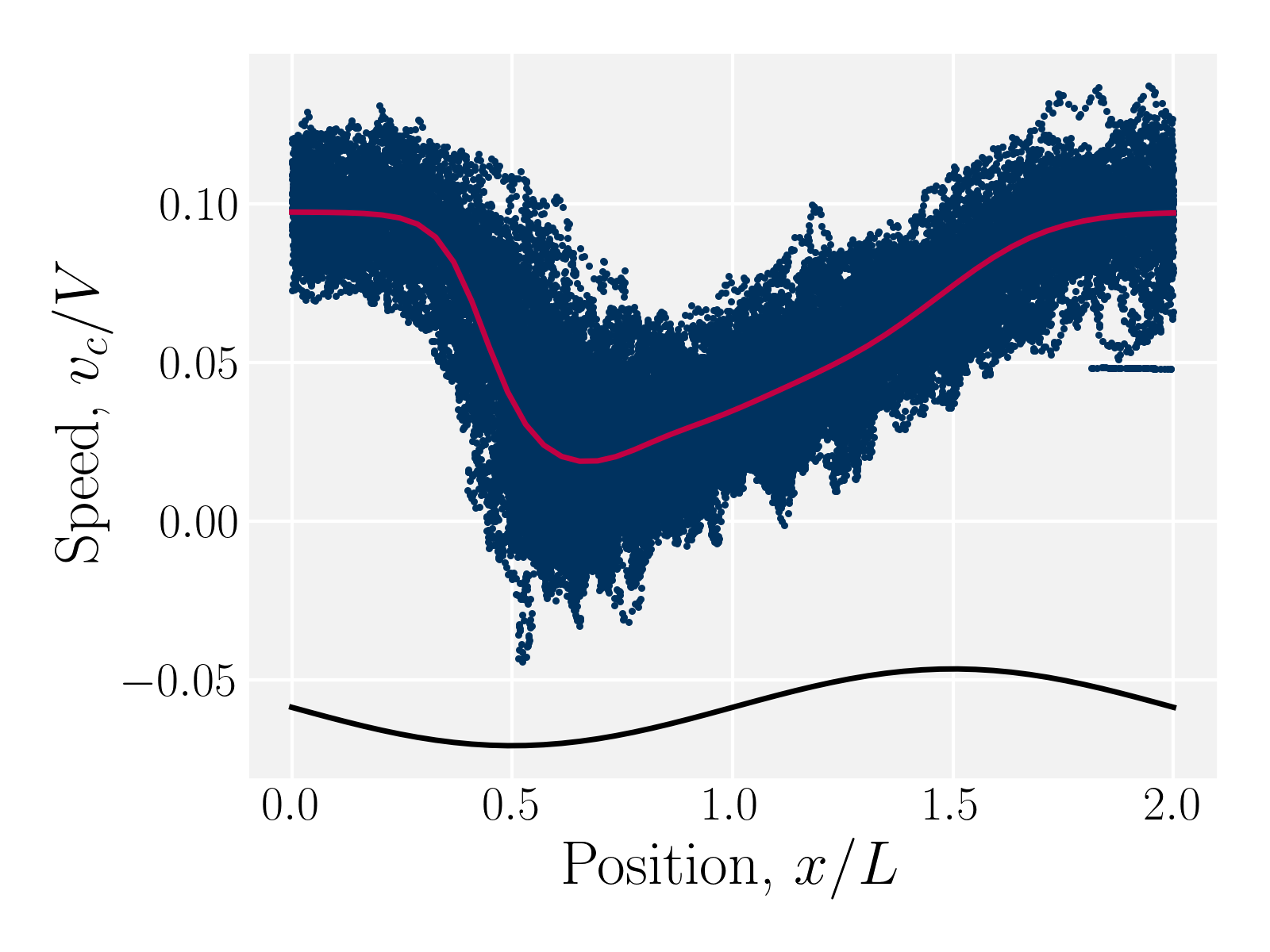}
        \caption{$B_0=1.5$.}
        \label{fig:vel1.5}
    \end{subfigure}
    \begin{subfigure}[b]{0.35\textwidth}
        \includegraphics[width=\textwidth]{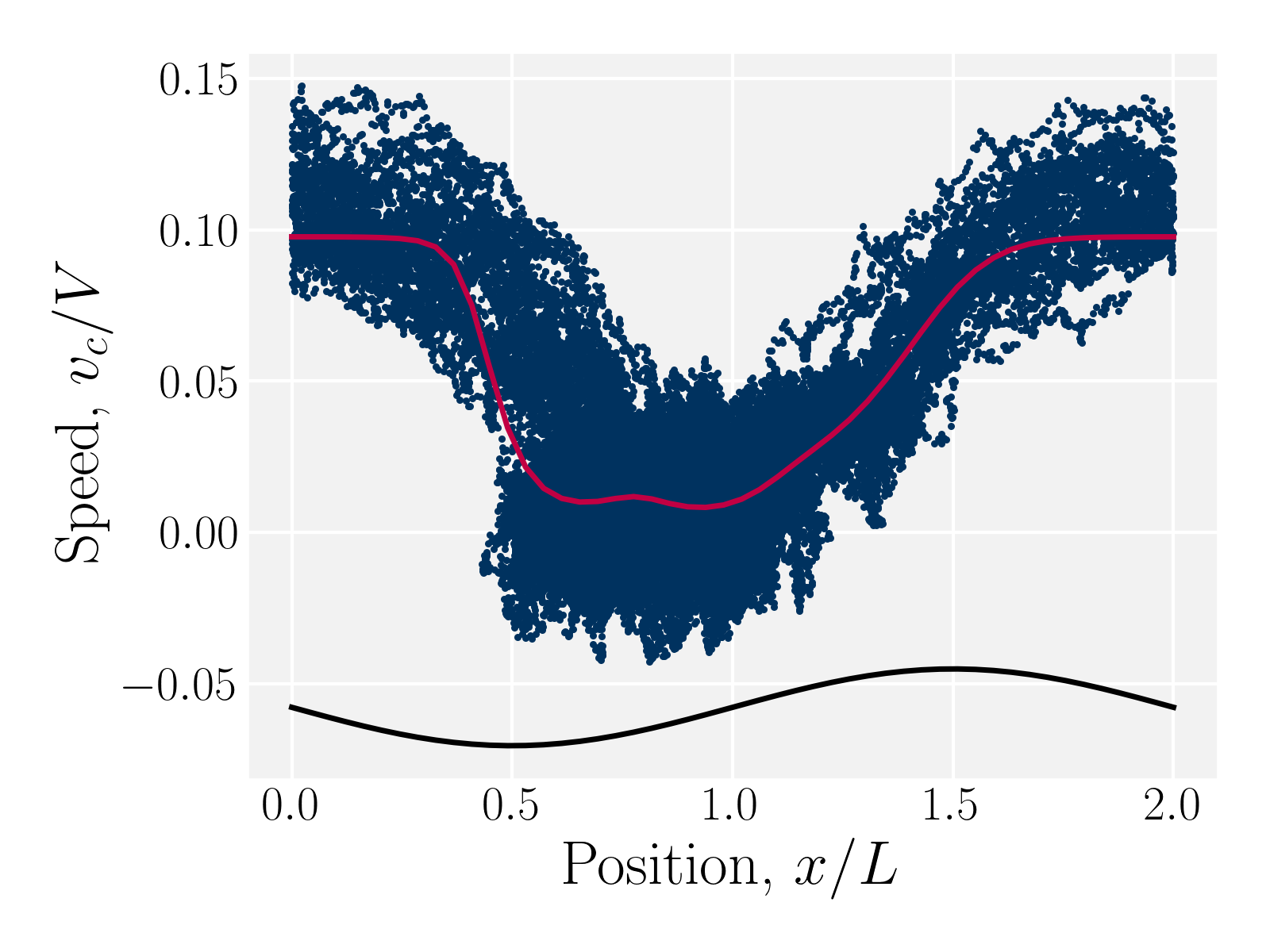}
        \caption{$B_0=2.5$.}
        \label{fig:vel2.5}
    \end{subfigure}
    \begin{subfigure}[b]{0.35\textwidth}
        \includegraphics[width=\textwidth]{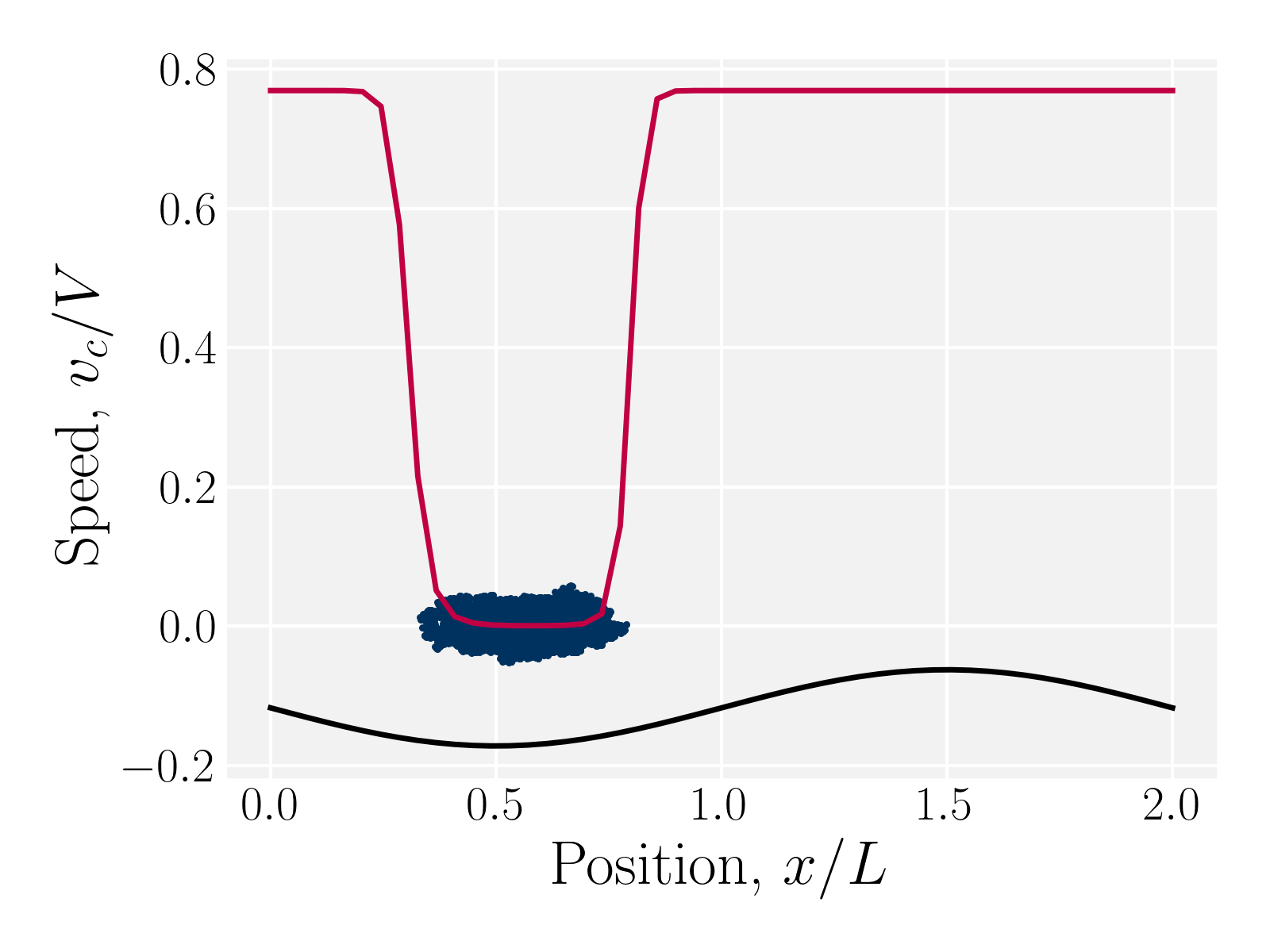}
        \caption{$B_0=4$.}
        \label{fig:vel4}
    \end{subfigure}
    \caption{
        \correctText{Colloid}{Homeotropic colloid} velocities at different positions $x$ for two different amplitudes $B_0$ \correctText{}{with planar anchoring on the wavy walls}. 
        Magenta line shows the estimated speed from the probability density.
    }
    \label{fig:vel_ind}
\end{figure}

\begin{figure*}[tb]
    \centering
    \begin{subfigure}[b]{0.32\textwidth}
        \includegraphics[width=\textwidth]{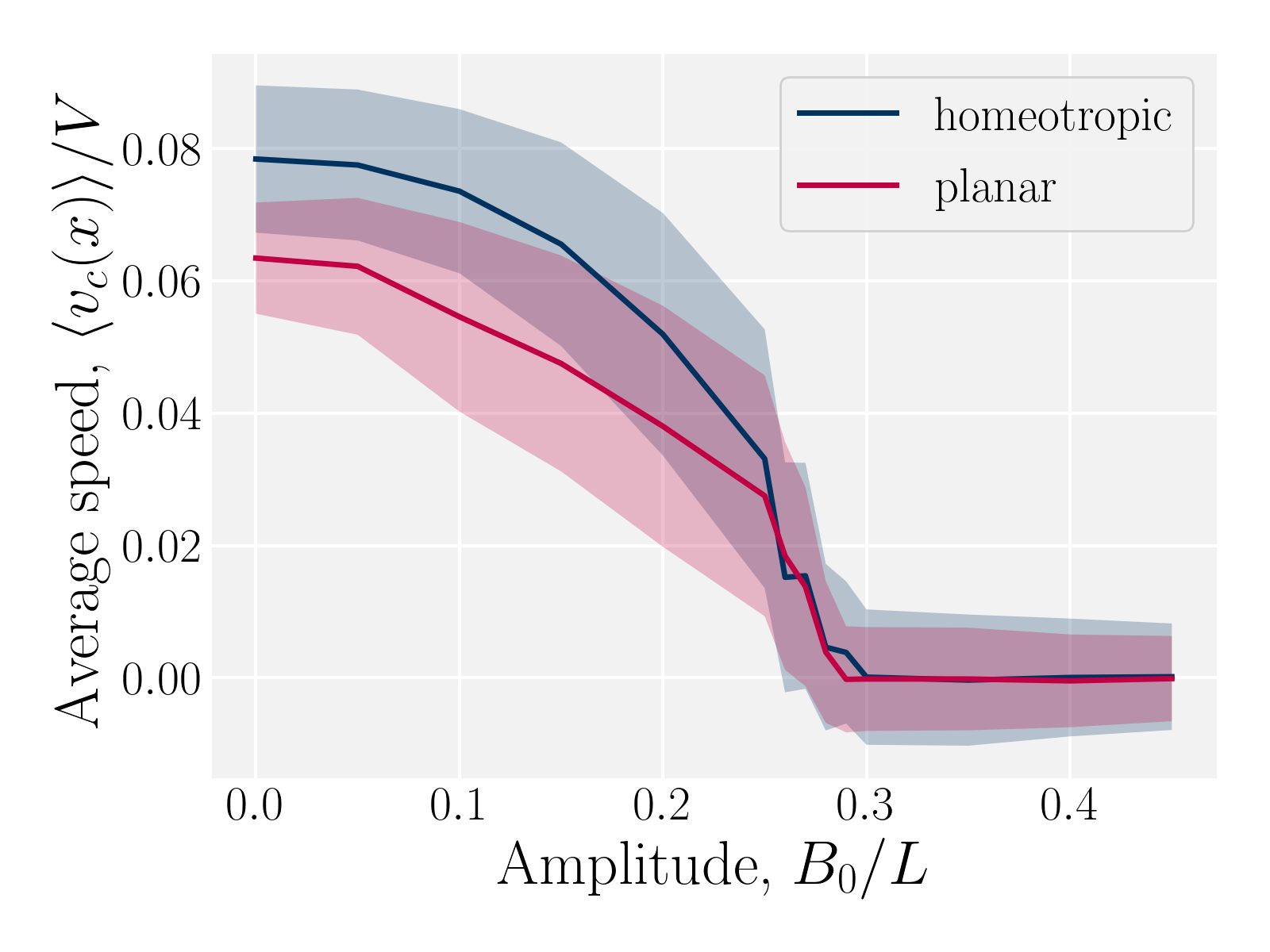}
        \caption{Average colloidal velocity $\av{v_c(x)}/V$.
        }
        \label{fig:average_vel}
    \end{subfigure}
    \begin{subfigure}[b]{0.32\textwidth}
        \includegraphics[width=\textwidth]{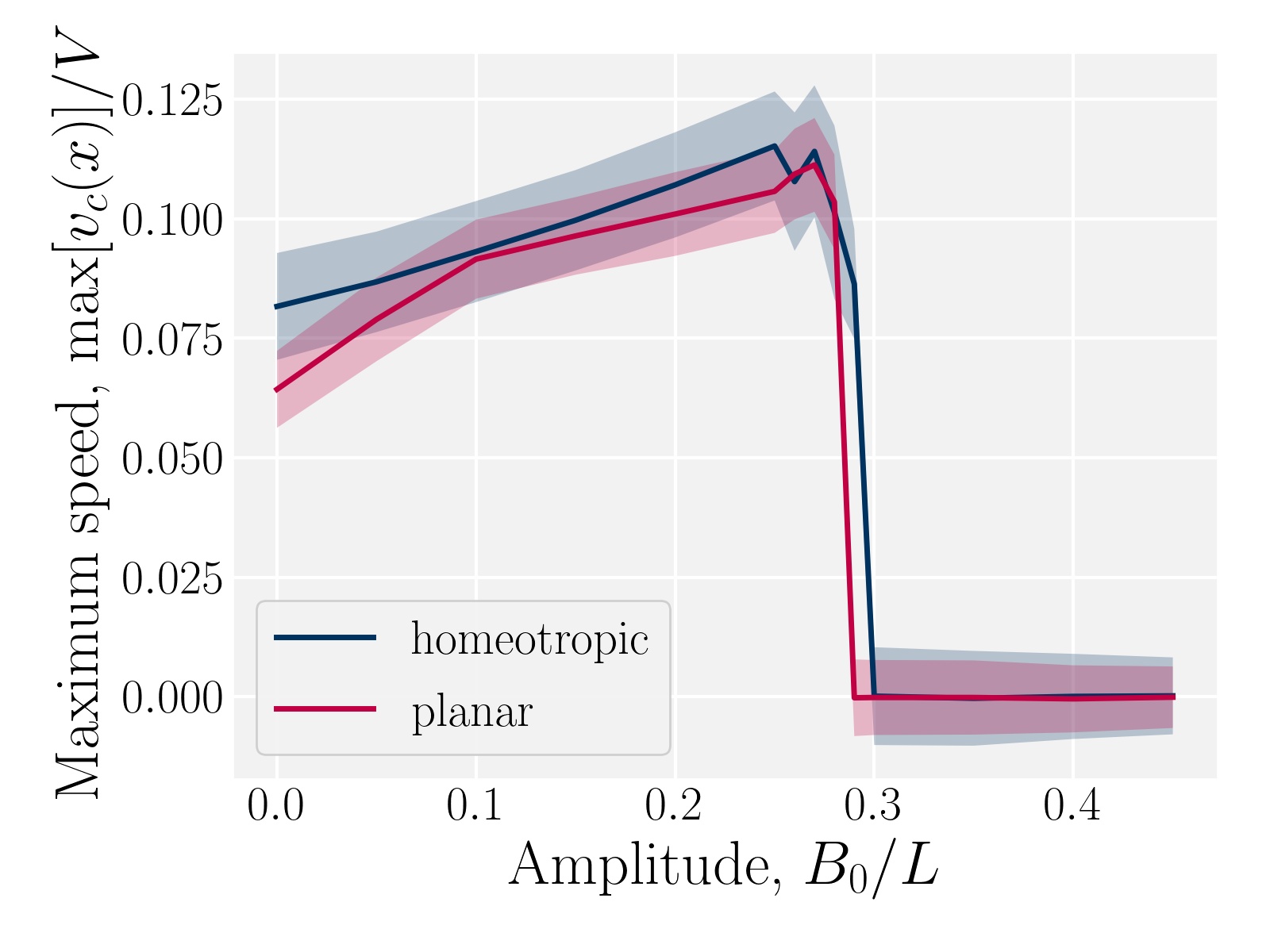}
        \caption{Maximum colloid speed $\max\left[v_\text{c}(x)\right]/V$. 
        }
        \label{fig:max_vel}
    \end{subfigure}
    \begin{subfigure}[b]{0.32\textwidth}
        \includegraphics[width=\textwidth]{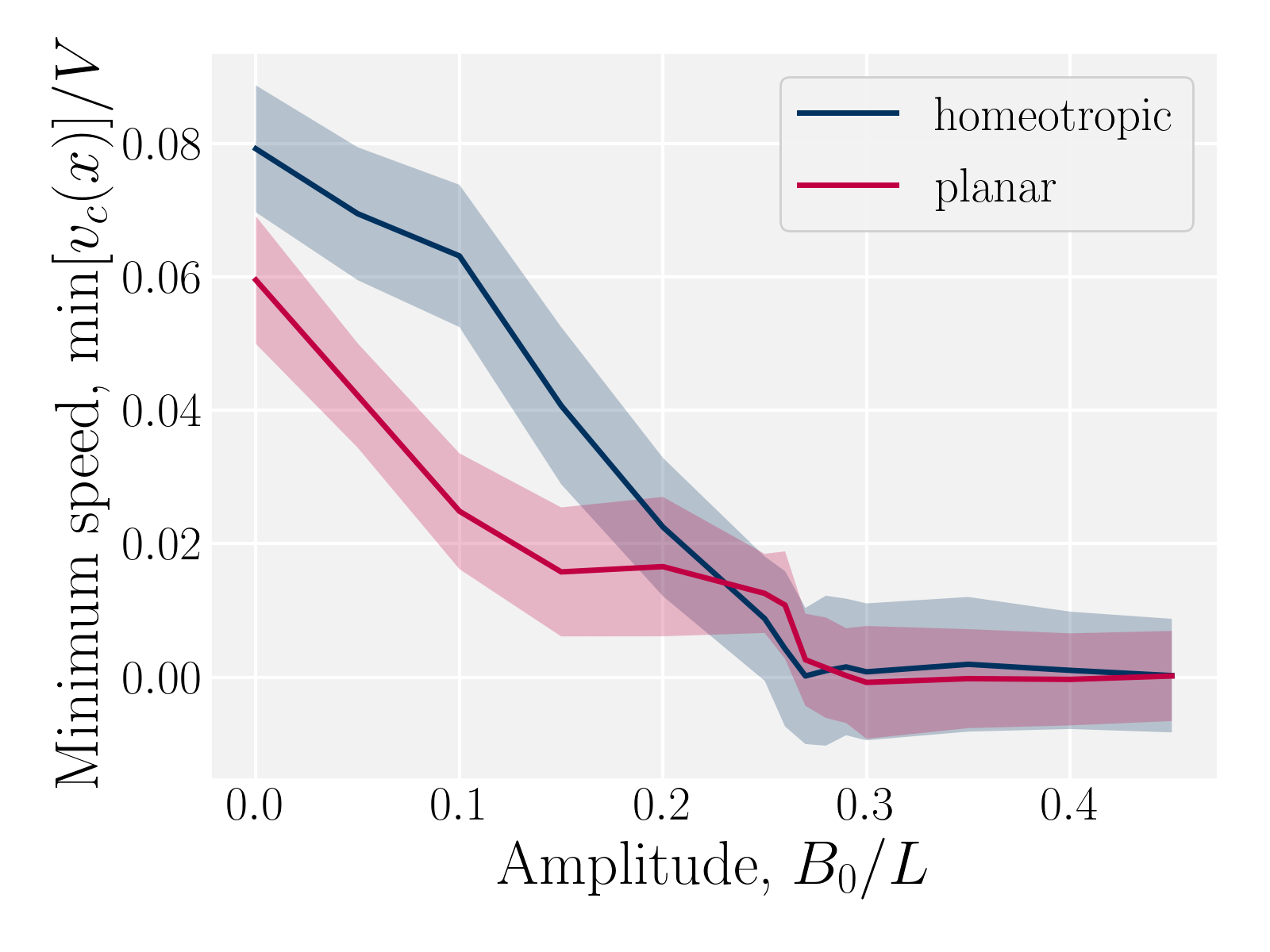}
        \caption{Minimum colloid speed $\min\left[v_\text{c}(x)\right]/V$.
        }
        \label{fig:min_vel}
    \end{subfigure}
    \caption{
        \correctText{Colloid}{Homeotropic colloid} velocity measures with varying amplitude $B_0$ of the wavy boundaries with planar and homeotropic anchoring. 
    }
    \label{fig:avmaxmin_vel}
\end{figure*}

The stick-slip dynamics exhibit three different scenarios (\fig{fig:vel_ind}). 
Either the colloid advects smoothly \correctText{with regions of moves along}{} with the flow (\fig{fig:vel1.5}) or the dynamics are stick-slip (\fig{fig:vel2.5}) or the colloids locks into a trough (\fig{fig:vel4}). 
For the continuous advection case (\fig{fig:vel1.5}), the maximum velocity occurs as the channel widens. 
This coincides with the maximum of the flow, and is in the region where the nematoelastic forces act in the same direction as the flow. 
The minimum occurs slightly offset from the bottom of the trough (\fig{fig:pdf}). 
For the locked scenario (\fig{fig:vel4}), the colloid stays in the trough and the velocity is essentially zero. 
\correctText{}{In the locked case, the measured probability $p(x)$ is zero away from the trough, with the sticking time appearing to have diverged; however, longer simulations may eventually find rare crossing events. 
Any such rare crossing events would necessarily be accompanied by an unlikely thermal fluctuation in colloid speed to overcome the crest (\eq{eq:velC2}).}

The speed profiles lead to interesting average colloid dynamics, which \correctText{will}{} demonstrate that homeotropic or planar \correctText{}{wall} anchoring only change quantitative details and not the qualitative behaviour. 
To quantify the stick-slip translation of the colloid, we track the average speed of the colloid $\av{v_\text{c}(x)}$, the maximum $\max\left[v_\text{c}(x)\right]$ and the minimum $\min\left[v_\text{c}(x)\right]$ (\fig{fig:avmaxmin_vel}). 
Colloids \correctText{with homeotropic anchoring}{in channels with homeotropic anchoring} elute more quickly than \correctText{planar colloids on average}{colloids in channels with planar walls} (\fig{fig:average_vel}). 
The average speed $\av{v_\text{c}(x)}$ of both decreases as amplitude increases until hitting zero at $B_0^*=2.9a$. 
In general, the average speed of the colloid decreases with an increasing amplitude until it reaches zero (\fig{fig:average_vel}). 
\correctText{This is primarily due to the slower flow of the nematic through the channel for higher amplitudes (Fig.6b). 
Since the colloid is advected by the fluid, it lags behind the average fluid spped. 
However, the average difference becomes smaller with increasing amplitude $B_0$.}{This occurs for two reasons. 
Firstly, as the amplitude increases the flow through the channel decreases even in the absence of the colloid (\fig{fig:charFlow}b).
Secondly, the strength of the elastic forces increases with amplitude and so the time spent crossing the free energy barrier increases.}

The average elution rate is different for planar and homeotropic anchoring on the walls. 
While the general shapes are the same, suggesting similar behaviours, the elastic forces from the wavy boundaries appear to be weaker for homeotropic anchoring than for planar. 
This is likely due to the different positioning of the defects. 
In the planar case, when the colloid is situated directly above the troughs, there is more space for the defects, compared to when it is situated above the crest where they are shifted closer to the colloid. 
However, for the homeotropic case, the position of the colloid has no effect on the defect positions --- they always reside directly up- and downstream (\fig{fig:def_homeo}). 
The shape of the colloid matches much better with the walls when situated above the trough, meaning it is a more stable point. 
Thus, the mechanisms responsible for the motions of the colloid are different in the different anchoring cases.

The maximum speed of the colloids behaves qualitatively differently from the average. 
It suddenly decreases to $\max\left[v_\text{c}(x)\right]=0$ (\fig{fig:max_vel}). 
This sudden drop occurs at the same $B_0^*$ as the average going to zero and signals the transition to a locked state that no longer moves along the channel. 
This once again illustrates the two distinct states of stick-slip and locked, with a discontinuous transition between them at $B_0^*$. 
Prior to the sudden stop, $\max\left[v_\text{c}(x)\right]$ increases with the amplitude, which is a consequence of Bernoulli's principle. 
The situation is nearly identical for homeotropic and planar anchoring (\fig{fig:max_vel}). 
Similarly, we can consider the minimum colloidal speed (\fig{fig:min_vel})\correctText{}{, which likewise shows similar behaviour for homeotropic and planar anchoring on the wall. 
These results indicate that homeotropic and planar anchoring result in very similar average elution rates and as well as similar maximum and minimum colloid velocities.}

\correctText{The}{However, the} minimum speed decreases with the amplitude. 
The minimum speed occurs at different points along the channel for different anchoring conditions and for the flow. 
For homeotropic anchoring, the colloid has its minimum-velocity position much higher up on the side of the trough compared to both the planar anchoring and the flow. 
Thus, it is nematic interactions with the wavy wall that set the position of the minimum-speed.
\correctText{}{Through its no-slip boundary conditions, the colloid also impacts the average speed of the fluid. 
As the nematic forces push the colloid towards a free energy minimum, it applies a drag force to the fluid. 
When the colloid sticks or slows between the trough and the crest, the nematic interactions work against the pressure gradient and the fluid slows (\appndx{app:feedback}). 
On the other hand, the colloid speeds up once it crosses the barrier due to the nematic force pushing it towards the next free energy minimum. 
During this slip phase of motion, the colloid drags the fluid forward and increases the fluid flow (\appndx{app:feedback}).}

In addition to comparing planar and homeotropic anchoring on the walls, we also compared different mean field potentials $U$ for planar anchoring on the walls. 
The maximum colloid speed has the same general shape as we previously found (\fig{fig:max_velU}). 
As the flow through the channel is independent of the mean field potential, this is expected. 
However, stronger mean field potentials increase the nematoelastic interactions and so the velocities must be scaled by $U$, which does not collapse the curves for small amplitudes because the fluid flow is independent of $U$. 
Most clearly, the transition amplitude $B_0^*$ from stick-slip to locked moves to smaller values as the mean field potential $U$ increases (\fig{fig:max_velU}). 
This suggests a mechanism by which fine control of the stick-slip transition to locking can be controlled.

\section{Conclusion}\label{sctn:conclusion}
In this work, we have considered a simple lock-key microfluidic system. 
We chose a colloid size commensurate with the walls' wavelength, creating periodically repeating barriers and docking sites, which exhibit interesting dynamics arising from the competition between nematoelastic free energy minimization and nematohydrodynamic advection. 
To tackle these complex dynamics, we first neglected flow (\sctn{sctn:noFlow}). 
In wavy channels, colloids with homeotropic anchoring migrate to the widest point in the channel (the troughs), \correctText{irregardless}{regardless} of whether the anchoring conditions on the wall are homeotropic or planar (\fig{fig:mean_trajs}). 
However, the pair of defects associated with the colloids take a different configurations for the different anchoring (\fig{fig:defects_res}). 
\correctText{}{Our results are not comprehensive and indicate that future work is required to fully understand how wall structure and defect configuration lead to equilibrium colloid positions.}
We then considered flows of nematics within wavy channels in the absence of the colloids (\sctn{sctn:noColloid}). 
Since the transport coefficient is constant, the flow rate through the channel is observed to decrease as the the surface area of the boundary surface increased with increasing amplitude of the wavy channel (\fig{fig:flows}). 

\begin{figure}[tb]
    \centering
    \includegraphics[width=0.475\textwidth]{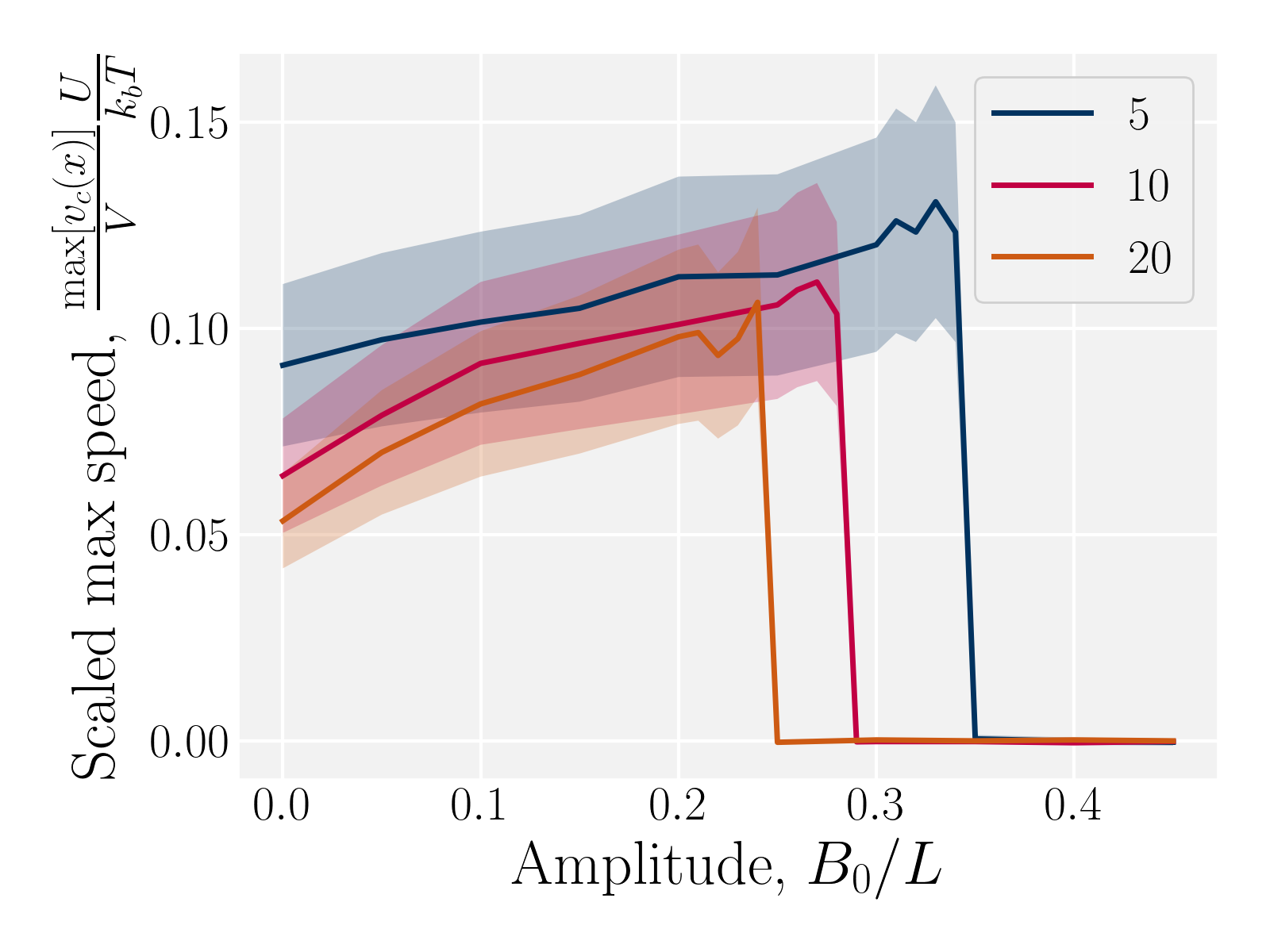}
    \caption{
        Maximum \correctText{}{homeotropic} colloid speed $\max\left[v_\text{c}(x)\right]/V$ \correctText{}{in wavy channels with planar anchoring}. 
        Sudden transition from stick-slip dynamics to locking. 
        Same as \fig{fig:max_vel} but for various mean field potentials $U$. 
        The velocity has been multiplied by the mean field potential, $U$, to collapse the curves. 
    }
    \label{fig:max_velU}
\end{figure}

Both these results suggest the qualitative dynamics of colloid transport slows as the amplitude to the wavy walls increases, which is indeed demonstrated in simulations (\fig{fig:pos}). 
We find \correctText{this decreases arises specifically from}{the elution slows because of} stick-slip dynamics, with colloids ``sticking'' to the docking sites for longer durations as a function of boundary amplitude (\fig{fig:pos}-\ref{fig:pdf}). 
Despite this net decrease in the transport rate, the maximum velocity, of when the colloid hops over the crests from docking site to docking site, increases with amplitude until the colloid suddenly locks into the troughs and no longer moves (\fig{fig:vel_ind}). 
Modifying the material parameters of the nematic fluid can shift this lock transition point (\fig{fig:max_velU}), with nematoelastic forces increasing with the mean field potential to nonlinearly shift the locking transition to smaller amplitudes. 
\correctText{}{The transition between locking and flowing is essentially a type of elastic ‘snap-through’ event~\cite{gomez2017,Radisson2023}, which would suggest a critical slowing and may allow simple-but-accurate models of the stick-slip dynamics.} 
\correctText{}{Such considerations might lead to the development of quantitative methods to classify different trajectories, as well as techniques to accurately measure the fraction of time a colloid spends sticking or slipping.} 
\correctText{}{We hope these results will motivate analytical work on the nematic interactions between mobile colloids and complex confining walls that go beyond interactions between colloids and plane walls~\cite{Chernyshuk2011}.} 

\correctText{}{This study has considered the effect of wavy wall amplitude on the stick-slip dynamics of eluting colloids in 2D. 
In addition to amplitude there are many more parameters that are expected to impact elution rate. 
The colloid size, channel height and wall wavelength are all length scales whose combinations will modify elution rate. 
The geometry could be further complicated by allowing the phase of the top and bottom wall to differ. 
Additionally, the strength of the wall/colloid interactions could also be tuned by modifying the Frank elastic constant of the nematic or the anchoring strength on the walls or colloids, just as the driving force could be changed by varying the pressure gradient. 
State diagrams for stick-slip or locking behaviors could be constructed for each pairing of such parameters. 
Future studies can exploit \appndx{app:bc} to simulate three dimensional systems confined by wavy walls or study the role of colloid surface structure (\appndx{app:wavyspheres}).}

Our results show that wavy walls not only have potential to direct colloids to specific docking sites but also to control site-specific resting duration and intermittent elution, allowing autonomic temporal control. 
They suggest how experimental systems can be engineered to hold particles in place for a given time by combining surface structure and advective flow. 
\correctText{}{By coupling site-specific resting duration with position-dependent pressure gradients, future studies could explore feedback loops to reinforce or diminish stick-and-slip motion.}
More complex surface structures could be designed such that particles stick for longer or shorter periods at different docking sites. 
Such site-specific resting times could be used to illuminate or irradiate particles for a given time without switching the light source. 
By combining these lock-key microfluidics with different microreactors at each docking site, efficient series chemical reactions could be designed to occur for autonomous-but-responsive durations. 

\section*{Conflicts of interest}
There are no conflicts to declare.

\section*{Acknowledgements}
This work was supported by the European Research Council (ERC) under the European Union’s Horizon 2020 research and innovation program (grant agreement no. 851196).
For the purpose of open access, the authors have applied a Creative Commons Attribution (CC BY) license to any Author Accepted Manuscript version arising from this submission.

\bibliographystyle{unsrt}
\bibliography{ref}

\begin{thebibliography}{100}

\bibitem{Liu2022}
Yun Liu, Guangze Yang, Yue Hui, Supun Ranaweera, and Chun-Xia Zhao.
\newblock Microfluidic nanoparticles for drug delivery.
\newblock {\em Small}, 18(36):2106580, 2022.

\bibitem{Sivaramakrishnan2020}
Muthusaravanan Sivaramakrishnan, Ram Kothandan, Deenadayalan~Karaiyagowder
  Govindarajan, Yogesan Meganathan, and Kumaravel Kandaswamy.
\newblock Active microfluidic systems for cell sorting and separation.
\newblock {\em Current Opinion in Biomedical Engineering}, 13:60--68, 2020.

\bibitem{Burklund2020}
Alison Burklund, Amogha Tadimety, Yuan Nie, Nanjing Hao, and John~X.J. Zhang.
\newblock Advances in diagnostic microfluidics.
\newblock volume~95 of {\em Advances in Clinical Chemistry}, pages 1--72.
  Elsevier, 2020.

\bibitem{Shendruk2013}
Tyler~N. Shendruk, Radin Tahvildari, Nicolas~M. Catafard, Lukasz Andrzejewski,
  Christian Gigault, Andrew Todd, Laurent Gagne-Dumais, Gary~W. Slater, and
  Michel Godin.
\newblock Field-flow fractionation and hydrodynamic chromatography on a
  microfluidic chip.
\newblock {\em Analytical Chemistry}, 85(12):5981--5988, 2013.

\bibitem{Zhang2016}
Jun Zhang, Sheng Yan, Dan Yuan, Gursel Alici, Nam-Trung Nguyen, Majid
  Ebrahimi~Warkiani, and Weihua Li.
\newblock Fundamentals and applications of inertial microfluidics: a review.
\newblock {\em Lab Chip}, 16:10--34, 2016.

\bibitem{Zhang2018}
Haiying Zhang, Daniela Freitas, Han~Sang Kim, Kristina Fabijanic, Zhong Li,
  Haiyan Chen, Milica~Tesic Mark, Henrik Molina, Alberto~Benito Martin, Linda
  Bojmar, et~al.
\newblock Identification of distinct nanoparticles and subsets of extracellular
  vesicles by asymmetric flow field-flow fractionation.
\newblock {\em Nature Cell Biology}, 20(3):332--343, 2018.

\bibitem{Zhang2020}
Peiran Zhang, Hunter Bachman, Adem Ozcelik, and Tony~Jun Huang.
\newblock Acoustic microfluidics.
\newblock {\em Annual Review of Analytical Chemistry}, 13(1):17--43, 2020.

\bibitem{Sengupta2013b}
Anupam Sengupta, Christian Bahr, and Stephan Herminghaus.
\newblock Topological microfluidics for flexible micro-cargo concepts.
\newblock {\em Soft Matter}, 9:7251--7260, 2013.

\bibitem{Lavrentovich2020}
Oleg~D. Lavrentovich.
\newblock Design of nematic liquid crystals to control microscale dynamics.
\newblock {\em Liquid Crystals Reviews}, 8(2):59--129, 2020.

\bibitem{fedorowicz2023}
Kamil Fedorowicz, Robert Prosser, and Anupam Sengupta.
\newblock Curvature-mediated programming of liquid crystal microflows.
\newblock {\em arXiv preprint arXiv:2304.02759}, 2023.

\bibitem{dalby2022}
James Dalby, Yucen Han, Apala Majumdar, and Lidia Mrad.
\newblock A multi-faceted study of nematic order reconstruction in microfluidic
  channels.
\newblock {\em arXiv preprint arXiv:2204.07808}, 2022.

\bibitem{Sengupta2020}
Anupam Sengupta and Marco~G. Mazza.
\newblock {\em Liquid Crystals at Interfaces and Under Flow: Recent Advances
  and Trends}, chapter~6, pages 183--226.
\newblock 2020.

\bibitem{Smalyukh2018}
Ivan~I. Smalyukh.
\newblock Liquid crystal colloids.
\newblock {\em Annual Review of Condensed Matter Physics}, 9(1):207--226, 2018.

\bibitem{stark2001}
Holger Stark.
\newblock Physics of colloidal dispersions in nematic liquid crystals.
\newblock {\em Physics Reports}, 351(6):387--474, 2001.

\bibitem{han2023}
Yucen Han and Apala Majumdar.
\newblock Multistability for a reduced nematic liquid crystal model in the
  exterior of 2{D} polygons.
\newblock {\em Journal of Nonlinear Science}, 33(2):24, 2023.

\bibitem{han2022}
Yucen Han, Baoming Shi, Lei Zhang, and Apala Majumdar.
\newblock A reduced {L}andau-de {G}ennes study for nematic equilibria in
  three-dimensional prisms.
\newblock {\em arXiv preprint arXiv:2211.07536}, 2022.

\bibitem{chen2021}
Han-Qing Chen, Xi-Yuan Wang, Hari~Krishna Bisoyi, Lu-Jian Chen, and Quan Li.
\newblock Liquid crystals in curved confined geometries: Microfluidics bring
  new capabilities for photonic applications and beyond.
\newblock {\em Langmuir}, 37(13):3789--3807, 2021.

\bibitem{Dogic2014}
Zvonimir Dogic, Prerna Sharma, and Mark~J. Zakhary.
\newblock Hypercomplex liquid crystals.
\newblock {\em Annual Review of Condensed Matter Physics}, 5(1):137--157, 2014.

\bibitem{Smalyukh2020}
Ivan~I Smalyukh.
\newblock Review: {K}nots and other new topological effects in liquid crystals
  and colloids.
\newblock {\em Reports on Progress in Physics}, 83(10):106601, sep 2020.

\bibitem{Mundoor2018}
Haridas Mundoor, Sungoh Park, Bohdan Senyuk, Henricus~H. Wensink, and Ivan~I.
  Smalyukh.
\newblock Hybrid molecular-colloidal liquid crystals.
\newblock {\em Science}, 360(6390):768--771, 2018.

\bibitem{paget2023}
Jack Paget, Marco~G Mazza, Andrew~J Archer, and Tyler~N Shendruk.
\newblock Complex-tensor theory of simple smectics.
\newblock {\em Nature Communications}, 14(1):1048, 2023.

\bibitem{yuan2019}
Ye~Yuan, Qingkun Liu, Bohdan Senyuk, and Ivan~I Smalyukh.
\newblock Elastic colloidal monopoles and reconfigurable self-assembly in
  liquid crystals.
\newblock {\em Nature}, 570(7760):214--218, 2019.

\bibitem{Senyuk2017}
B.~Senyuk, Q.~Liu, P.~D. Nystrom, and I.~I. Smalyukh.
\newblock Repulsion–attraction switching of nematic colloids formed by liquid
  crystal dispersions of polygonal prisms.
\newblock {\em Soft Matter}, 13:7398--7405, 2017.

\bibitem{muvsevivc2019}
Igor Mu{\v{s}}evi{\v{c}}.
\newblock Interactions, topology and photonic properties of liquid crystal
  colloids and dispersions.
\newblock {\em The European Physical Journal Special Topics}, 227:2455--2485,
  2019.

\bibitem{yoshida2015}
Hiroyuki Yoshida, K~Asakura, J-i Fukuda, and M~Ozaki.
\newblock Three-dimensional positioning and control of colloidal objects
  utilizing engineered liquid crystalline defect networks.
\newblock {\em Nature Communications}, 6(1):7180, 2015.

\bibitem{nikkhou2015}
Maryam Nikkhou, Miha {\v{S}}karabot, Simon {\v{C}}opar, Miha Ravnik, Slobodan
  {\v{Z}}umer, and Igor Mu{\v{s}}evi{\v{c}}.
\newblock Light-controlled topological charge in a nematic liquid crystal.
\newblock {\em Nature Physics}, 11(2):183--187, 2015.

\bibitem{Hung2007}
Francisco~R. Hung, Brian~T. Gettelfinger, Jr. Koenig, Gary~M., Nicholas~L.
  Abbott, and Juan~J. de~Pablo.
\newblock {Nanoparticles in nematic liquid crystals: Interactions with
  nanochannels}.
\newblock {\em The Journal of Chemical Physics}, 127(12), 09 2007.

\bibitem{Lapointe2008}
Clayton~P. Lapointe, Daniel~H. Reich, and Robert~L. Leheny.
\newblock Manipulation and organization of ferromagnetic nanowires by patterned
  nematic liquid crystals.
\newblock {\em Langmuir}, 24(19):11175--11181, 2008.

\bibitem{Silvestre2004}
N.~M. Silvestre, P.~Patr{\'{i}}cio, and M.~M. Telo~da Gama.
\newblock {Key-lock mechanism in nematic colloidal dispersions}.
\newblock {\em Physical Review E}, 69(6):6, 2004.

\bibitem{Cheung2008}
David~L. Cheung and Michael~P. Allen.
\newblock {Effect of substrate geometry on liquid-crystal-mediated
  nanocylinder-substrate interactions}.
\newblock {\em The Journal of Chemical Physics}, 129(11), 09 2008.

\bibitem{Eskandari2014}
Z.~Eskandari, N.~M. Silvestre, M.~M. Telo~da Gama, and M.~R. Ejtehadi.
\newblock Particle selection through topographic templates in nematic colloids.
\newblock {\em Soft Matter}, 10:9681--9687, 2014.

\bibitem{Luo2016}
Y.~Luo, F.~Serra, and K.~J. Stebe.
\newblock {Experimental realization of the ``lock-and-key'' mechanism in liquid
  crystals}.
\newblock {\em Soft Matter}, 12(28):6027--6032, 2016.

\bibitem{Boniello2019}
G.~Boniello, Y.~Luo, D.~A. Beller, F.~Serra, and K.~J. Stebe.
\newblock {Colloids in confined liquid crystals: A plot twist in the
  lock-and-key mechanism}.
\newblock {\em Soft Matter}, 15(26):5220--5226, 2019.

\bibitem{RojasGomez2017}
Óscar A~Rojas-Gómez, José~M Romero-Enrique, Nuno~M Silvestre, and Margarida
  M~Telo da~Gama.
\newblock Pattern-induced anchoring transitions in nematic liquid crystals.
\newblock {\em Journal of Physics: Condensed Matter}, 29(6):064002, dec 2016.

\bibitem{Luo2019}
Y.~Luo, T.~Yao, D.~Beller, F.~Serra, and K.~Stebe.
\newblock Deck the walls with anisotropic colloids in nematic liquid crystals.
\newblock {\em Langmuir}, 35:9274--9285, 2019.

\bibitem{Yao2022}
Tianyi Yao, Žiga Kos, Qi~Xing Zhang, Yimin Luo, Francesca Serra, Edward~B.
  Steager, Miha Ravnik, and Kathleen~J. Stebe.
\newblock Nematic colloidal micro-robots as physically intelligent systems.
\newblock {\em Advanced Functional Materials}, 32(44):2205546, 2022.

\bibitem{Luo2018}
Y.~Luo, D.~A. Beller, G.~Boniello, F.~Serra, and K.~J. Stebe.
\newblock {Tunable colloid trajectories in nematic liquid crystals near wavy
  walls}.
\newblock {\em Nature Communications}, 9(1):1--11, 2018.

\bibitem{Chen2018}
Kui Chen, Olivia~J. Gebhardt, Raghavendra Devendra, German Drazer, Randall~D.
  Kamien, Daniel~H. Reich, and Robert~L. Leheny.
\newblock Colloidal transport within nematic liquid crystals with arrays of
  obstacles.
\newblock {\em Soft Matter}, 14:83--91, 2018.

\bibitem{kapral2008}
Raymond Kapral.
\newblock Multiparticle collision dynamics: Simulation of complex systems on
  mesoscales.
\newblock {\em Advances in Chemical Physics}, 140:89, 2008.

\bibitem{Gompper2009}
G.~Gompper, T.~Ihle, D.~M. Kroll, and R.~G. Winkler.
\newblock {\em Multi-Particle Collision Dynamics: A Particle-Based Mesoscale
  Simulation Approach to the Hydrodynamics of Complex Fluids}, pages 1--87.
\newblock Springer Berlin Heidelberg, Berlin, Heidelberg, 2009.

\bibitem{Slater2009}
Gary~W. Slater, Christian Holm, Mykyta~V. Chubynsky, Hendrick~W. de~Haan,
  Antoine Dubé, Kai Grass, Owen~A. Hickey, Christine Kingsburry, David Sean,
  Tyler~N. Shendruk, and Lixin Zhan.
\newblock Modeling the separation of macromolecules: A review of current
  computer simulation methods.
\newblock {\em Electrophoresis}, 30(5):792--818, 2009.

\bibitem{Howard2019}
Michael~P Howard, Arash Nikoubashman, and Jeremy~C Palmer.
\newblock Modeling hydrodynamic interactions in soft materials with
  multiparticle collision dynamics.
\newblock {\em Current Opinion in Chemical Engineering}, 23:34--43, 2019.
\newblock Frontiers of Chemical Engineering: Molecular Modeling.

\bibitem{Sayyidmousavi2018}
Alireza Sayyidmousavi and Katrin Rohlf.
\newblock Reactive multi-particle collision dynamics with reactive boundary
  conditions.
\newblock {\em Physical Biology}, 15(4):046007, may 2018.

\bibitem{Reigh2020}
Shang~Yik Reigh, Mu-Jie Huang, Hartmut Löwen, Eric Lauga, and Raymond Kapral.
\newblock Active rotational dynamics of a self-diffusiophoretic colloidal
  motor.
\newblock {\em Soft Matter}, 16:1236--1245, 2020.

\bibitem{Hickey2012}
Owen~A. Hickey, Tyler~N. Shendruk, James~L. Harden, and Gary~W. Slater.
\newblock Simulations of free-solution electrophoresis of polyelectrolytes with
  a finite debye length using the debye-h\"uckel approximation.
\newblock {\em Physical Review Letters}, 109:098302, Aug 2012.

\bibitem{Shendruk2015b}
Tyler~N. Shendruk, Martin Bertrand, and Gary~W. Slater.
\newblock Electrophoretic mobility of polyelectrolytes within a confining well.
\newblock {\em ACS Macro Letters}, 4(4):472--476, 2015.

\bibitem{Burelbach2018}
Jérôme Burelbach, David~B. Brückner, Daan Frenkel, and Erika Eiser.
\newblock Thermophoretic forces on a mesoscopic scale.
\newblock {\em Soft Matter}, 14:7446--7454, 2018.

\bibitem{zottl2018}
Andreas Z{\"o}ttl and Holger Stark.
\newblock Simulating squirmers with multiparticle collision dynamics.
\newblock {\em The European Physical Journal E}, 41:1--11, 2018.

\bibitem{Kuhr2019}
Jan-Timm Kuhr, Felix Rühle, and Holger Stark.
\newblock Collective dynamics in a monolayer of squirmers confined to a
  boundary by gravity.
\newblock {\em Soft Matter}, 15:5685--5694, 2019.

\bibitem{Clopes2020}
Judit Clopés, Gerhard Gompper, and Roland~G. Winkler.
\newblock Hydrodynamic interactions in squirmer dumbbells: active
  stress-induced alignment and locomotion.
\newblock {\em Soft Matter}, 16:10676--10687, 2020.

\bibitem{Lamura2021}
Antonio Lamura, Roland~G. Winkler, and Gerhard Gompper.
\newblock {Wall-anchored semiflexible polymer under large amplitude oscillatory
  shear flow}.
\newblock {\em The Journal of Chemical Physics}, 154(22), 06 2021.
\newblock 224901.

\bibitem{Clopes2022}
Judit Clop\'es~Llah\'{\i}, Aitor Mart\'{\i}n-G\'omez, Gerhard Gompper, and
  Roland~G. Winkler.
\newblock Simulating wet active polymers by multiparticle collision dynamics.
\newblock {\em Physical Review E}, 105:015310, Jan 2022.

\bibitem{Choi2023}
Kisuk Choi, Jaebin Lee, Woo~Jin Choi, and Jin~Suk Myung.
\newblock Dynamics of semi-flexible and breakable fibers under poiseuille flow.
\newblock {\em Polymer Engineering \& Science}, 63(3):1032--1040, 2023.

\bibitem{wang2023}
Zhenhua Wang, Zhen-Gang Wang, An-Chang Shi, Yuyuan Lu, and Lijia An.
\newblock Behaviors of a polymer chain in channels: From zimm to rouse
  dynamics.
\newblock {\em Macromolecules}, 56(6):2447--2453, 2023.

\bibitem{Wani2022}
Yashraj~M. Wani, Penelope~Grace Kovakas, Arash Nikoubashman, and Michael~P.
  Howard.
\newblock {Diffusion and sedimentation in colloidal suspensions using
  multiparticle collision dynamics with a discrete particle model}.
\newblock {\em The Journal of Chemical Physics}, 156(2), 01 2022.

\bibitem{Chen2019}
Renjie Chen, Ryan Poling-Skutvik, Michael~P. Howard, Arash Nikoubashman,
  Sergei~A. Egorov, Jacinta~C. Conrad, and Jeremy~C. Palmer.
\newblock Influence of polymer flexibility on nanoparticle dynamics in
  semidilute solutions.
\newblock {\em Soft Matter}, 15:1260--1268, 2019.

\bibitem{Eisenstecken2018}
Thomas Eisenstecken, Raphael Hornung, Roland~G. Winkler, and Gerhard Gompper.
\newblock Hydrodynamics of binary-fluid mixtures —an augmented multiparticle
  collison dynamics approach.
\newblock {\em Europhysics Letters}, 121(2):24003, mar 2018.

\bibitem{Toneian2019}
David Toneian, Gerhard Kahl, Gerhard Gompper, and Roland~G. Winkler.
\newblock {Hydrodynamic correlations of viscoelastic fluids by multiparticle
  collision dynamics simulations}.
\newblock {\em The Journal of Chemical Physics}, 151(19), 11 2019.

\bibitem{Ilg2022}
Patrick Ilg.
\newblock {Multiparticle collision dynamics for ferrofluids}.
\newblock {\em The Journal of Chemical Physics}, 156(14), 04 2022.
\newblock 144905.

\bibitem{Ilg2022b}
Patrick Ilg.
\newblock Simulating the flow of interacting ferrofluids with multiparticle
  collision dynamics.
\newblock {\em Physical Review E}, 106:064605, Dec 2022.

\bibitem{dicintio2020}
P.~Di~Cintio, M.~Pasquato, L.~Barbieri, H.~Bufferand, L.~Casetti, G.~Ciraolo,
  U.~N. di~Carlo, P.~Ghendrih, J.~P. Gunn, S.~Gupta, and et~al.
\newblock Multiparticle collision simulations of dense stellar systems and
  plasmas.
\newblock {\em Proceedings of the International Astronomical Union},
  16(S362):134–140, 2020.

\bibitem{dicintio2022a}
Pierfrancesco Di~Cintio, Mario Pasquato, Hyunwoo Kim, and Suk-Jin Yoon.
\newblock Introducing a new multi-particle collision method for the evolution
  of dense stellar systems-crash-test {N}-body simulations.
\newblock {\em Astronomy \& Astrophysics}, 649:A24, 2021.

\bibitem{dicintio2022b}
Pierfrancesco Di~Cintio, Mario Pasquato, Alicia Simon-Petit, and Suk-Jin Yoon.
\newblock Introducing a new multi-particle collision method for the evolution
  of dense stellar systems-ii. core collapse.
\newblock {\em Astronomy \& Astrophysics}, 659:A19, 2022.

\bibitem{shendruk2015}
T.~N. Shendruk and J.~M. Yeomans.
\newblock {Multi-particle collision dynamics algorithm for nematic fluids}.
\newblock {\em Soft Matter}, 11(25), 2015.

\bibitem{Mandal2019}
Shubhadeep Mandal and Marco~G. Mazza.
\newblock Multiparticle collision dynamics for tensorial nematodynamics.
\newblock {\em Physical Review E}, 99:063319, Jun 2019.

\bibitem{ReyesArango2020}
Denisse Reyes-Arango, Jacqueline Quintana-H., Julio~C. Armas-Pérez, and
  Humberto Híjar.
\newblock Defects around nanocolloids in nematic solvents simulated by
  multi-particle collision dynamics.
\newblock {\em Physica A}, 547:123862, 2020.

\bibitem{Hijar2020}
Humberto H\'{\i}jar.
\newblock Dynamics of defects around anisotropic particles in nematic liquid
  crystals under shear.
\newblock {\em Physical Review E}, 102:062705, Dec 2020.

\bibitem{Armendariz2021b}
Jos\'{e} Armend\'{a}riz and Humberto H\'{\i}jar.
\newblock Magnetic anisotropic colloids in nematic liquid crystals: Fluctuating
  dynamics simulated by multi-particle collision dynamics.
\newblock {\em International Journal of Modern Physics B}, 35(08):2150124,
  2021.

\bibitem{Armendariz2021}
José Armendáriz and Humberto Híjar.
\newblock Nonequilibrium dynamics of a magnetic nanocapsule in a nematic liquid
  crystal.
\newblock {\em Materials}, 14(11), 2021.

\bibitem{Mandal2021}
Shubhadeep Mandal and Marco~G Mazza.
\newblock Multiparticle collision dynamics simulations of a squirmer in a
  nematic fluid.
\newblock {\em The European Physical Journal E}, 44(5):64, 2021.

\bibitem{kozhukhov2022}
Timofey Kozhukhov and Tyler~N. Shendruk.
\newblock Mesoscopic simulations of active nematics.
\newblock {\em Science Advances}, 8(34):eabo5788, 2022.

\bibitem{kozhukhov2023}
Timofey Kozhukhov, Benjamin Loewe, and Tyler~N. Shendruk.
\newblock Active modulation for mitigation of density fluctuations in active
  simulations.
\newblock {\em In Preparation}, 2023.

\bibitem{Keogh2023}
Ryan~R Keogh, Timofey Kozhukhov, Kristian Thijssen, and Tyler~N Shendruk.
\newblock Active darcy's law.
\newblock {\em arXiv}, arXiv:2308.05462, 2023.

\bibitem{Malevanets1999}
Anatoly Malevanets and Raymond Kapral.
\newblock {Mesoscopic model for solvent dynamics}.
\newblock {\em The Journal of Chemical Physics}, 110(17):8605--8613, 05 1999.

\bibitem{Malevanets2000}
Anatoly Malevanets and Raymond Kapral.
\newblock {Solute molecular dynamics in a mesoscale solvent}.
\newblock {\em The Journal of Chemical Physics}, 112(16):7260--7269, 04 2000.

\bibitem{Ihle2001}
T.~Ihle and D.~M. Kroll.
\newblock Stochastic rotation dynamics: A {G}alilean-invariant mesoscopic model
  for fluid flow.
\newblock {\em Physical Review E}, 63:020201, Jan 2001.

\bibitem{Noguchi2007}
H.~Noguchi, N.~Kikuchi, and G.~Gompper.
\newblock Particle-based mesoscale hydrodynamic techniques.
\newblock {\em Europhysics Letters}, 78(1):10005, mar 2007.

\bibitem{Noguchi2008}
Hiroshi Noguchi and Gerhard Gompper.
\newblock Transport coefficients of off-lattice mesoscale-hydrodynamics
  simulation techniques.
\newblock {\em Physical Review E}, 78:016706, Jul 2008.

\bibitem{Hijar2019}
Humberto H\'{\i}jar.
\newblock Hydrodynamic correlations in isotropic fluids and liquid crystals
  simulated by multi-particle collision dynamics.
\newblock {\em Condensed Matter Physics}, 22(1):13601, 2019.

\bibitem{Lamura2001}
A.~Lamura, G.~Gompper, T.~Ihle, and D.~M. Kroll.
\newblock Multi-particle collision dynamics: Flow around a circular and a
  square cylinder.
\newblock {\em Europhysics Letters}, 56(3):319, nov 2001.

\bibitem{Whitmer2010}
Jonathan~K Whitmer and Erik Luijten.
\newblock Fluid–solid boundary conditions for multiparticle collision
  dynamics.
\newblock {\em Journal of Physics: Condensed Matter}, 22(10):104106, feb 2010.

\bibitem{Bolintineanu2012}
Dan~S. Bolintineanu, Jeremy~B. Lechman, Steven~J. Plimpton, and Gary~S. Grest.
\newblock No-slip boundary conditions and forced flow in multiparticle
  collision dynamics.
\newblock {\em Physical Review E}, 86:066703, Dec 2012.

\bibitem{Louise}
Louise Head, Davide Marenduzzo, and Tyler~N. Shendruk.
\newblock Entangled nematic disclinations using multi-particle collision
  dynamics.
\newblock Technical report, 2023.

\bibitem{Grollau2003}
S.~Grollau, N.~L. Abbott, and J.~J. de~Pablo.
\newblock Spherical particle immersed in a nematic liquid crystal: Effects of
  confinement on the director field configurations.
\newblock {\em Phys. Rev. E}, 67:011702, Jan 2003.

\bibitem{Andrienko2001}
Denis Andrienko, Guido Germano, and Michael~P. Allen.
\newblock Computer simulation of topological defects around a colloidal
  particle or droplet dispersed in a nematic host.
\newblock {\em Phys. Rev. E}, 63:041701, Mar 2001.

\bibitem{Sussman2019}
Daniel~M. Sussman and Daniel~A. Beller.
\newblock Fast, scalable, and interactive software for landau-de gennes
  numerical modeling of nematic topological defects.
\newblock {\em Frontiers in Physics}, 7, 2019.

\bibitem{fukuda2001}
Junichi Fukuda and H~Yokoyama.
\newblock Director configuration and dynamics of a nematic liquid crystal
  around a two-dimensional spherical particle: Numerical analysis using
  adaptive grids.
\newblock {\em The European Physical Journal E}, 4:389--396, 2001.

\bibitem{tasinkevych2002}
M~Tasinkevych, NM~Silvestre, Pedro Patricio, and MM~Telo~da Gama.
\newblock Colloidal interactions in two-dimensional nematics.
\newblock {\em The European Physical Journal E}, 9:341--347, 2002.

\bibitem{Silvestre2014}
Nuno~M. Silvestre, Qingkun Liu, Bohdan Senyuk, Ivan~I. Smalyukh, and Mykola
  Tasinkevych.
\newblock Towards template-assisted assembly of nematic colloids.
\newblock {\em Phys. Rev. Lett.}, 112:225501, Jun 2014.

\bibitem{Liu2019}
Y.~Liu, J.~Li, and A.~J. Smits.
\newblock {Roughness effects in laminar channel flow}.
\newblock {\em Journal of Fluid Mechanics}, 876:1129--1145, 2019.

\bibitem{Okechi2021}
N.~F. Okechi and S.~Asghar.
\newblock {Viscoelastic flow through a wavy curved channel}.
\newblock {\em Chinese Journal of Physics}, 74:144--156, 2021.

\bibitem{gomez2017}
Michael Gomez, Derek~E Moulton, and Dominic Vella.
\newblock Critical slowing down in purely elastic
  ‘snap-through’instabilities.
\newblock {\em Nature Physics}, 13(2):142--145, 2017.

\bibitem{Radisson2023}
Basile Radisson and Eva Kanso.
\newblock Elastic snap-through instabilities are governed by geometric
  symmetries.
\newblock {\em Phys. Rev. Lett.}, 130:236102, Jun 2023.

\bibitem{Chernyshuk2011}
S.~B. Chernyshuk and B.~I. Lev.
\newblock Theory of elastic interaction of colloidal particles in nematic
  liquid crystals near one wall and in the nematic cell.
\newblock {\em Phys. Rev. E}, 84:011707, Jul 2011.

\bibitem{Allahyarov2002}
E.~Allahyarov and G.~Gompper.
\newblock Mesoscopic solvent simulations: Multiparticle-collision dynamics of
  three-dimensional flows.
\newblock {\em Physical Review E}, 66:036702, Sep 2002.

\bibitem{lamura2002}
A~Lamura and G~Gompper.
\newblock Numerical study of the flow around a cylinder using multi-particle
  collision dynamics.
\newblock {\em The European Physical Journal E}, 9:477--485, 2002.

\bibitem{prohm2012}
C~Prohm, M~Gierlak, and Holger Stark.
\newblock Inertial microfluidics with multi-particle collision dynamics.
\newblock {\em The European Physical Journal E}, 35:1--10, 2012.

\bibitem{liu2020}
Aiqing Liu, Zhenyue Yang, Lijun Liu, Jizhong Chen, and Lijia An.
\newblock Role of functionality in cross-stream migration, structures, and
  dynamics of star polymers in {P}oiseuille flow.
\newblock {\em Macromolecules}, 53(22):9993--10004, 2020.

\bibitem{Qi2020}
Kai Qi, Hemalatha Annepu, Gerhard Gompper, and Roland~G. Winkler.
\newblock Rheotaxis of spheroidal squirmers in microchannel flow: Interplay of
  shape, hydrodynamics, active stress, and thermal fluctuations.
\newblock {\em Physical Review Research}, 2:033275, Aug 2020.

\bibitem{bruus2007}
Henrik Bruus.
\newblock {\em Theoretical microfluidics}, volume~18.
\newblock Oxford University Press, 2007.

\bibitem{Brochard1973}
F.~Brochard.
\newblock {Backflow Effects in Nematic Liquid Crystals}.
\newblock {\em Molecular Crystals and Liquid Crystals}, 23(1-2):51--58, 1973.

\bibitem{Copar2020}
S.~{\v{C}}opar, Ž. Kos, T.~Emer{\v{s}}i{\v{c}}, and U.~Tkalec.
\newblock {Microfluidic control over topological states in channel-confined
  nematic flows}.
\newblock {\em Nature Communications}, 11(1), 2020.

\bibitem{Anderson2015}
T.~G. Anderson, E.~Mema, L.~Kondic, and L.~J. Cummings.
\newblock {Transitions in Poiseuille flow of nematic liquid crystal}.
\newblock {\em International Journal of Non-Linear Mechanics}, 75:15--21, 2015.

\bibitem{Batista2015}
V.~M.~O. Batista, M.~L. Blow, and M.~M. Telo Da~Gama.
\newblock {The effect of anchoring on the nematic flow in channels}.
\newblock {\em Soft Matter}, 11(23):4674--4685, 2015.

\end{thebibliography}

\appendix

\section{Numerical Methods}
\label{app:methods}
This study utilizes a nematic multi-particle collision dynamics approach, which is chosen for its ability to simulate moderate P\'{e}clet numbers, mobile colloids, complex boundaries and fluctuating nematohydrodynamics. 

\subsection{Multi-particle Collision Dynamics Algorithm}\label{app:mpcd}

The MPCD algorithm essentially consists of two steps~\cite{Malevanets1999,Malevanets2000}: 
    \textit{(i)} streaming and 
    \textit{(ii)} collision. 

\subsubsection{Streaming Step}\label{app:streaming}
The streaming step updates each particle position assuming ballistic motion
\begin{equation}
    \label{eq:posMPCD}
    \vec{r}_i(t+\delta t) = \vec{r}_i(t) + \vec{v}_i(t) \delta t ,
\end{equation}
where $\vec{r}_i$ is the particle position, $t$ is the current time and $\delta t$ is the time step. 

\subsubsection{Collision Step}\label{app:collision}
\paragraph{Momentum exchange:}
The particle velocities are updated by the collision operator $\Xi_{i,c}(t)$ for particle $i$ in cell $c$ at time $t$, which is used to update the velocities
\begin{equation}
    \vec{v}_i(t+\delta t) = \vec{v}_{c}(t) + \vec{\Xi}_{i,c}(t), 
    \label{eq:velMPCD}
\end{equation}
where $\vec{v}_{c} = \av{\vec{v}_i}_c$ is the center of mass velocity of the cell and $\av{\cdot}_c$ is the average within cell $c$. 
A modified Andersen-thermostatted collision operator~\cite{Noguchi2007,Noguchi2008} is employed
\begin{equation}
    \vec{\Xi}_{i,c} = \vec{\xi}_i - \av{ \vec{\xi}_j }_{c} + \left(\tens{I}^{-1}_c \cdot \left( \delta \vec{L}_c + \delta \vec{\mathcal{L}}_c\right) \right) \times \vec{r}^\prime_i ,
    \label{eq:collisionOperator}
\end{equation}
where each $\vec{\xi}_i$ is a randomly generated velocity with magnitude drawn from a Boltzmann distribution with energy $\kbt$. 
To conserve momentum and keep the average velocity of the cell constant, the average of the random velocities $\av{ \vec{\xi}_j }_{c}$ is subtracted. 
The last term ensures that angular momentum is conserved by removing any spurious angular velocity generated by the collision operation. 
The moment of inertia of the cell is 
$\tens{I}_c(t) = \av{ \left(\vec{r}_{j}^{\prime}\cdot\vec{r}_{j}^{\prime}\right) \tens{1} - \vec{r}^\prime_j \otimes \vec{r}^\prime_j}_c N_c$, 
where $\tens{1}$ is the identity matrix, $\vec{r}_c=\av{\vec{r}_j}_c$ is the center of mass position and $\vec{r}^\prime_{k}=\vec{r}_k-\vec{r}_c$. 
The spurious angular momentum can arise from the randomly generated velocities $\delta \vec{L}_c(t) = \av{m\left[\vec{r}^\prime_{j}\times\left(\vec{v}_j-\vec{\xi}_j\right)\right]}_c N_c$ or changes in orientation $\delta \vec{\mathcal{L}}_c(t)$, which will be discussed below. 

\paragraph{Acceleration:} Pressure gradients can be written as an effective external acceleration $\grad P = \rho \vec{g}$ and the external acceleration $\vec{g}$ can be included in the collision operator~\cite{Allahyarov2002} as an impulse $\vec{v}_i(t+\delta t)  = \vec{v}_{c}(t) + \vec{\Xi}_{i,c}(t) + \vec{g} \delta t$. 
This approach has been employed to simulate flows around obstacles~\cite{Lamura2001,lamura2002}, colloid/polymer transport~\cite{prohm2012,Shendruk2013,liu2020}, and rheotaxis of microswimmers~\cite{Qi2020}. 

\paragraph{Orientation exchange:}\label{app:ori}
Particle orientations are updated according to a nematic collision operator \correctText{$\vec{\Psi}_{i,c}(t)$}{$\Psi_{i,c}(t)$} for particle $i$ in cell $c$~\cite{shendruk2015}. 
 After the collision\correctText{}{,} the updated orientation is 
 \correctText{$\vec{u}_i(t+\delta t) = \vec{n}_{c}(t) + \vec{\Psi}_{i,c}(t)$}{\begin{equation}
    \vec{u}_i(t+\delta t) = \Psi_{i,c}\vec{u}_i(t) .
    \label{eq:ori}
\end{equation}
The collision operator is a rotation operator causes the orientation of each N-MPCD particle to change as $\dot{\vec{u}}_i$ over the time step $\delta t$. 
The collision operator and resulting re-rotation can be decomposed into two contributions $\dot{\vec{u}}_i = \delta \vec{u}^\text{MS}_i / \delta t + \delta \vec{u}_i^\text{J} / \delta t$ from the thermostat ($\delta \vec{u}^\text{MS}_i / \delta t$) and from the flows ($\delta \vec{u}_i^\text{J} / \delta t$).}
\begin{enumerate}
    \item \correctText{}{\underline{A local thermostat:}} The local thermostat draws random orientations from the Maier-Saupe distribution centered about $\vec{n}_c$
    \begin{equation}
        p_c(\vec{u}_{\correctText{}{\text{MS}}}) \propto \exp{\left(U S_c (\vec{n}_c \cdot \vec{u}_{\correctText{}{\text{MS}}})^2/\kbt \right)} ,
        \label{eq:maierSaupe}
    \end{equation}
    where $U$ is the mean field potential, which determines how strongly the orientations align. 
    This operation updates all particles orientations without changing the overall cell director. 
    \correctText{}{To determine each cell's director $\vec{n}_c$, the tensor
    \begin{equation}
        \tens{Q}_c = \av{d\vec{u}_i\otimes \vec{u}_i - \tens{1} }_{c} / \left(d-1\right) 
        \label{eq:Q}
    \end{equation}
    is calculated. 
    The largest eigenvalue of $\tens{Q}_c$ is the scalar order parameter $S_c$ and the corresponding eigenvector is parallel to the local director $\vec{n}_c$ for the cell. 
    The change in orientation is $\delta\vec{u}_i^\text{MS} = \vec{u}_i(t)-\vec{u}_i^\text{MS}$.}
    \item \correctText{}{\underline{Rotations due to flows:}} The nematodynamic response models the nematogens' response to velocity gradients through Jeffery's equation
    \begin{align}
        \delta \vec{u}_i^\text{J} &= \delta t \chi \left[ \vec{u}_i\cdot\tens{\Omega}_c + \xi\left( \vec{u}_i\cdot\tens{E}_c - \vec{u}_i\vec{u}_i\vec{u}_i:\tens{E}_c \right) \right],
        \label{eq:jeffery}
    \end{align}
    for a bare tumbling parameter $\xi$ and heuristic shear coupling coefficient $\chi$ in a flow with shear rate tensor $2\tens{E}_c(t)=\grad \ \vec{v}_c+\left(\grad \ \vec{v}_c\right)^T$ and vorticity tensor $2\tens{\Omega}_c(t)=\grad \ \vec{v}_c-\left(\grad \ \vec{v}_c\right)^T$. 
    The director dynamics are coupled back to the velocity field through the change in angular momentum \correctText{$\delta \vec{\mathcal{L}}_c(t) = - \gamma_\text{R} \sum_i^{N_c} \vec{u}_i(t) \times \left[ \dot{\vec{u}}_i + \dot{\vec{u}}_i^\text{J} \right]$}{$\delta \vec{\mathcal{L}}_c(t) = - \gamma_\text{R} \sum_i^{N_c} \vec{u}_i(t) \times \dot{\vec{u}}_i$}, where $\gamma_\text{R}$ is a rotational \correctText{mobility}{friction} coefficient~\cite{shendruk2015,Hijar2020,Armendariz2021}. \correctText{}{By choosing a sufficiently small value of $\gamma_\text{R}$ backflow effects can be kept small, even in the limit of small Ericksen numbers.}
\end{enumerate}
\correctText{To determine each cell's director $\vec{n}_c$, the tensor $\tens{Q}_c = \av{d\vec{u}_i\otimes \vec{u}_i - \tens{1} }_{c} / \left(d-1\right)$ is calculated. 
The largest eigenvalue of $\tens{Q}_c$ is the scalar order parameter $S_c$ and the corresponding eigenvector is parallel to the local director $\vec{n}_c$ for the cell.}{}
\correctText{The collision operator $\vec{\Psi}_{i,c}(t)$ can be decomposed into two contributions:
    \textit{(a)} a local thermostat,
    \textit{(b)} rotations due to flows.}{}

\section{Boundary Conditions}\label{app:bc}
\correctText{}{The main text considers simulations performed in 2D. 
In this section, we present the boundary conditions generalized to 3D and state the simplifying condition for the 2D surfaces used in the simulation results.}
To create the wavy channel, particles are subjected to boundary conditions, each of which is composed of two parts:
the boundary surface and boundary rules. 

\subsection{Boundary Surfaces}\label{app:bs}
Each boundary surface labelled by index $b$ is represented as a mathematical surface where $\mathcal{S}_b(\vec{r})=0$. 
The baseline form employed for $\mathcal{S}_b(\vec{r})$ can represent planes, ellipsoids and squircles
\begin{equation}
    \label{eq:surface_original}
    \mathcal{S}_{b,0}(\vec{r}) = \sum_{j=0}^d \left[A_{b,j}\left(r_j-q_{b,j})\right)\right]^{p_x} - R_b^{p_{b,R}} ,
\end{equation}
where $\vec{r}=(x,y,z)$ is the position on the surface, $\vec{q}_b=(q_{b,x}, q_{b,y}, q_{b,z})$ defines the position of the $b^\textrm{th}$ boundary, and $\vec{A}_b=(A_{b,x}, A_{b,y}, 
A_{b,z})$, $\vec{p}_b=(p_{b,x}, p_{b,y}, p_{b,z}, p_{b,R})$ and $R_b$ define the shape. 

In this study, the surface has been extended to allow for wavy boundaries by adding a second term, which allows for waves along two perpendicular directions with amplitude $B_{b,0}$ and frequencies $B_{b,1}$ and $B_{b,2}$. 
We represent these surfaces as
\begin{equation}
    \label{eq:surface}
    \mathcal{S}_b = \mathcal{S}_{b,0} + B_{b,0}\cos{\left(B_{b,1}\mathcal{S}_{b,1}\right)}\cos{\left(B_{b,2}\mathcal{S}_{b,2}\right)} ,
\end{equation}
where $\mathcal{S}_{b,0}(\vec{r})$ defines the basic surface given by \eq{eq:surface_original}, while $\mathcal{S}_{b,1}(\vec{r})$ and $\mathcal{S}_{b,2}(\vec{r})$ define the orientations of the waves, and thus depend on $\mathcal{S}_{b,0}(\vec{r})$. 
\correctText{}{For 2D simulations, $B_{b,2}=0$.}

\subsubsection{Wavy Planes}

\begin{figure}[tb]
    \centering
    \begin{subfigure}[b]{0.235\textwidth}
        \centering
        \includegraphics[width=\textwidth]{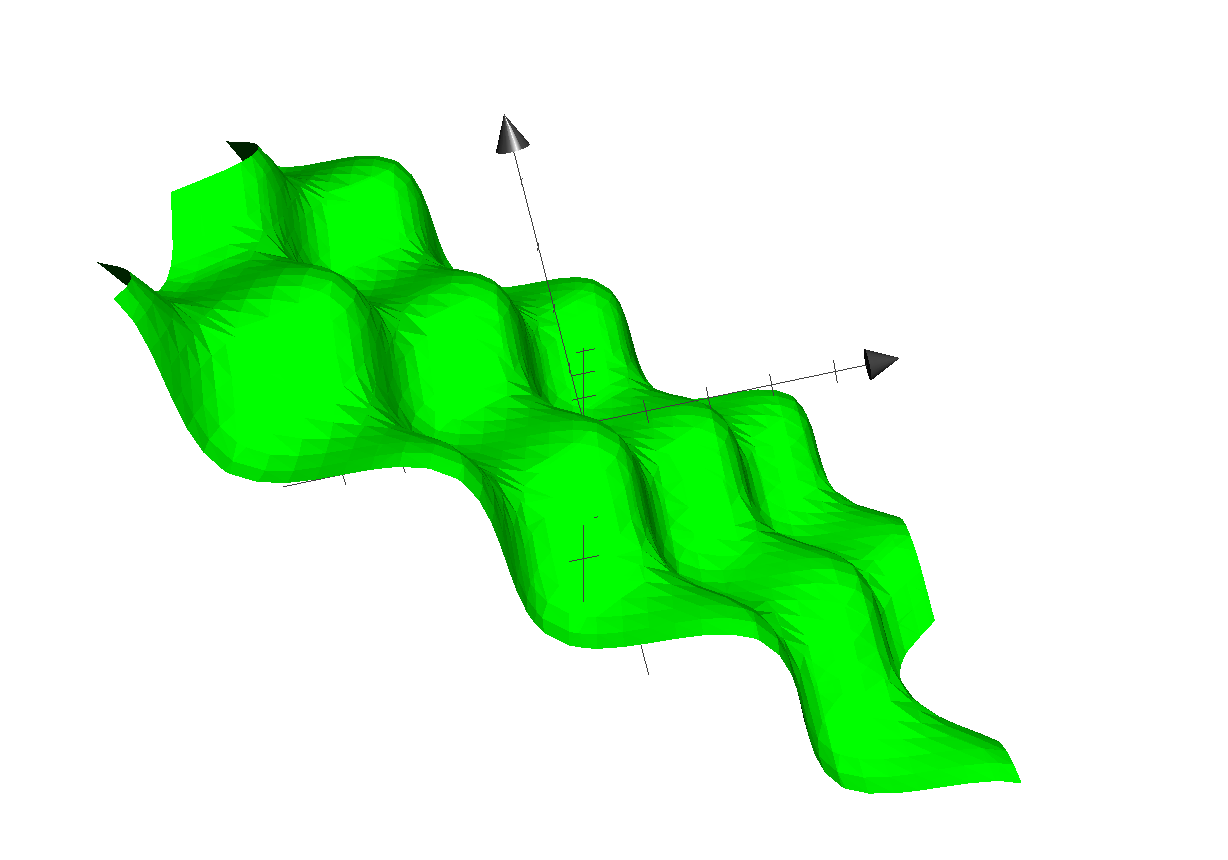}
        \caption{Egg carton plane \correctText{}{wall}.}
        \label{fig:eggCarton}
    \end{subfigure}
    \hfill
    \begin{subfigure}[b]{0.235\textwidth}
        \centering
        \includegraphics[width=\textwidth]{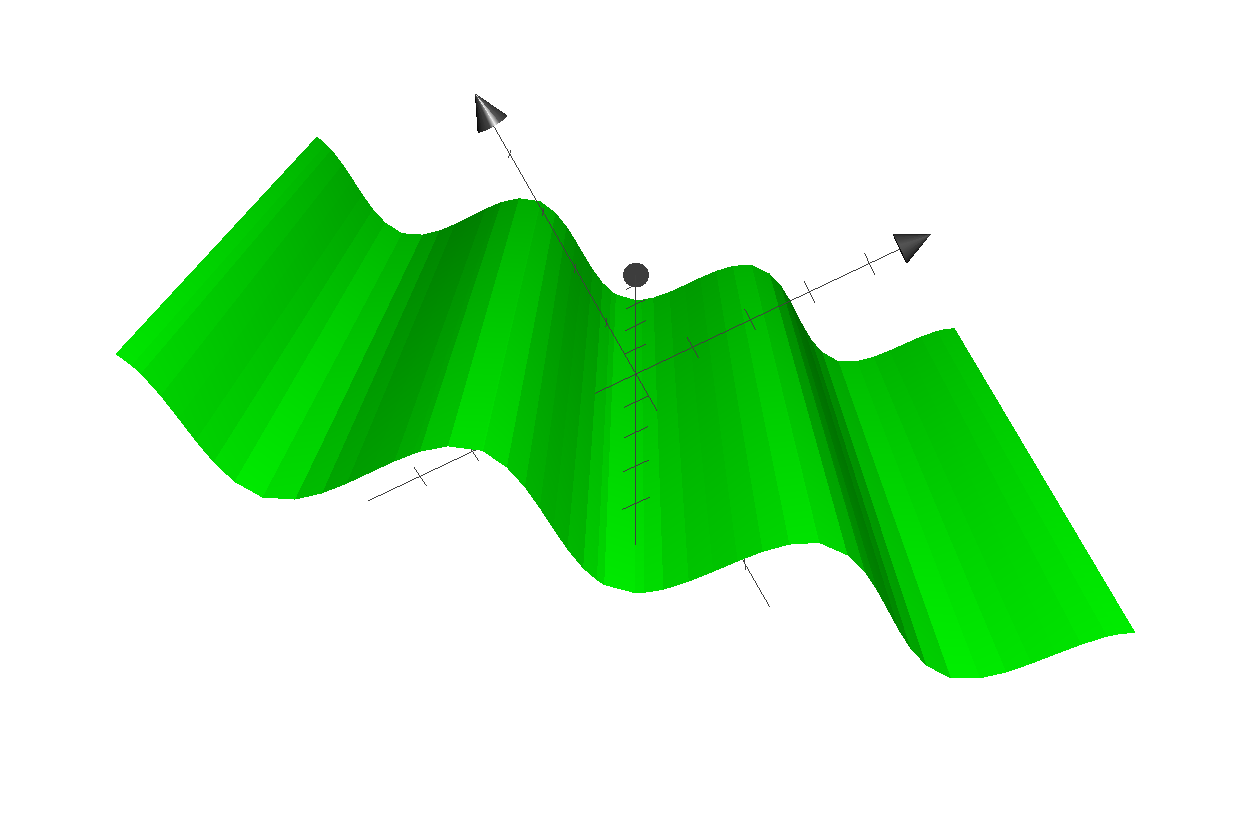}
        \caption{Corrugated plane \correctText{}{wall}.}
        \label{fig:corrPlane}
    \end{subfigure}
    \caption{The types of wavy \correctText{}{plane} walls implemented in this study. }
    \label{fig:wavyPlanes}
\end{figure}

For planes, $\vec{p}_b=\vec{1}$, $\vec{A}_b$ is the normal vector and $R=0$. 
For wavy planes, the functions $\mathcal{S}_{b,1}(\vec{r})$ and $\mathcal{S}_{b,2}(\vec{r})$ can define
    \textit{(i)} an egg carton surface (\fig{fig:eggCarton}) or
    \textit{(ii)} a corrugation (\fig{fig:corrPlane}).
To get the egg carton-like surface, $\mathcal{S}_{b,0}(\vec{r})$, $\mathcal{S}_{b,1}(\vec{r})$ and $\mathcal{S}_{b,2}(\vec{r})$ must be mutually perpendicular. 
A decision has to be made to make $\mathcal{S}_{b,1}(\vec{r})$ independent of $z$ and take the form
\begin{equation}
    \label{eq:f1}
    \mathcal{S}_{b,1}(\vec{r}) = \frac{A_{\correctText{}{b},y} x-A_{\correctText{}{b},x} y}{\sqrt{A_{\correctText{}{b},x}^2+A_{\correctText{}{b},y}^2}} .
\end{equation}
The plane is normalised to standardise the value of the frequency \correctText{$B_1$}{$B_{b,1}$}. 
This choice then fixes $\mathcal{S}_{b,2}(\vec{r})$ to be
\begin{equation}
    \label{eq:f2}
    \mathcal{S}_{b,2}(\vec{r}) = \frac{A_{\correctText{}{b},x} A_z x + A_{\correctText{}{b},y} A_z y - (A_{\correctText{}{b},x}^2+A_{\correctText{}{b},y}^2)z}{\sqrt{A_{\correctText{}{b},x}^4 + A_{\correctText{}{b},y}^4 + A_{\correctText{}{b},x}^2A_z^2 + A_{\correctText{}{b},y}^2A_z^2 + 2A_{\correctText{}{b},x}^2A_{\correctText{}{b},y}^2}} .
\end{equation}
As previously stated, \correctText{$B_1$ and $B_2$}{$B_{b,1}$ and $B_{b,2}$} specify the frequency of the waves. 
If they are equal, the pattern in \fig{fig:eggCarton} emerges with even waves in both directions. 
However, if they are not equal, oblong `bumps' appear.
The corrugated plane emerges if \correctText{$B_1=0$, or $B_2=0$}{$B_{b,2}=0$} for waves in the other direction. 
In those cases, the other \correctText{$B_i$}{$B_{b,1}$} then defines the frequency of the waves. 
Although this study will focus on wavy planar walls, wavy colloids and cylinders are also possible, as detailed in \appndx{app:wavyspheres}. 

While the wavy surfaces are \correctText{defined}{generalized} to work in 3D, they work without modification in 2D. 
For simplicity, it is assumed that 2D simulations occur in the $xy$-plane, such that \correctText{$A_z=0$}{$A_{b,z}=0$ for all boundaries $b$}. 
Additionally, only one cosine wave is needed and so we set \correctText{$B_2=0$}{$B_{b,2}=0$} for all cases. 
\correctText{}{Thus, $B_{b,2}=0$ is not discussed in the main text.}

\subsection{Boundary Rules}
\label{sctn:boundaryRules}
The $b^\text{th}$ surface boundary exists at $\mathcal{S}_{b}(\vec{r})=0$, thus a particle is defined as being within the simulation domain if $\mathcal{S}(\vec{r}_i) \geq 0$ for particle position $\vec{r}_i$. 
If a particle is found to be outside the boundary, a set of boundary rules define how the generalized coordinates (position $\vec{r}_i(t)$, orientation $\vec{u}_i(t)$ and velocity $\vec{v}_i(t)$) are transformed. 
The boundary rules are applied to the components normal $\vec{\nu}_b$ and perpendicular $\vec{t}_b$ to the boundary (\appndx{app:norm}). 

Keeping in mind the goal of creating periodic boundary conditions, the position $\vec{r}_i(t)$ can be transformed by shifts in the normal $\mathcal{D}_{b,\nu}$ and/or tangential $\mathcal{D}_{b,t}$ directions. 
After crossing the surface $b$, particle $i$'s position is transformed as $\vec{r}_i \to \vec{r}_i + \mathcal{D}_{b,\nu} \vec{\nu}_b + \mathcal{D}_{b,t} \vec{t}_b$. 
Likewise, the particle's velocity may be transformed. 
The velocity is altered by multiplicative parameters $\mathcal{M}_{b,\nu}$ and $\mathcal{M}_{b,t}$, such that upon colliding with the surface the velocity becomes $\vec{v}_i \to \mathcal{M}_{b,\nu}\tens{\Pi}_b\cdot\vec{v}_i + \mathcal{M}_{b,t}\tens{T}_b \cdot \vec{v}_i$, where $\tens{\Pi}_b=\vec{\nu}_b\otimes\vec{\nu}_b$ and  and $\tens{T}_b=\tens{1}-\tens{\Pi}_b$ are the normal and tangent projection operators. 
Lastly, the orientation is also multiplicatively transformed $\vec{u}_i \to \mu_{b,\nu}\tens{\Pi}_b\cdot\vec{u}_i + \mu_{b,t}\tens{T}_b\cdot\vec{u}_i$, which is then rescaled back to a unit vector. 

The simplest boundaries employed in this study are periodic boundary conditions. 
Particles that cross a periodic boundary are simply shifted by the length of the simulation domain in the direction normal to the boundary, without change to their velocity or orientation. 
So $\mathcal{M}_{b,\nu} = \mathcal{M}_{b,t} = \mu_{b,\nu} = \mu_{b,t}=1$, while $\mathcal{D}_{b,t}=0$ and $\mathcal{D}_{b,\nu} = L_j$, where $L_j$ is the system size in the direction $j$. 

Solid walls are boundaries that do not apply a shift ($\mathcal{D}_{b,t} = \mathcal{D}_{b,\nu} = 0$) but do transform the direction of the velocity. 
For perfect-slip boundaries, the particles are reflected and $\mathcal{M}_{b,\nu} = -1$ and $\mathcal{M}_{b,t}=1$; whereas, for no-slip conditions bounce-back rules are applied with $\mathcal{M}_{b,\nu} = \mathcal{M}_{b,t} = -1$. 
In order for the particle to not violate the boundary, the particle is rewound to the location where it crossed the boundary, and from there it streams for the rest of the time step using its new velocity. 

Controlling the orientation when a particle crosses a boundary allows for different nematic anchoring conditions for solid boundaries. 
If both $\mu_{b,\nu}$ and $\mu_{b,t}$ are set to $1$ the particle's orientation does not change and no anchoring conditions are enforced. 
However, by setting the tangential component to $\mu_{b,t}=0$ homeotropic anchoring is achieved. 
Similarly, if $\mu_{b,\nu}=0$ planar anchoring is enforced on the boundary. 

\subsection{Strong Anchoring}\label{app:strongAnchoring}
For solid boundaries, the above boundary rules do not fully enforce perfect no-slip~\cite{Bolintineanu2012}. 
This is because when the boundaries intersect cells, they have a lower than average density since part of the cell is excluded, which lowers the viscosity of the cell. 
To combat this, `phantom' particles are added to make up the correct density. 
These particles are only used when generating the new velocities during the collision stage. 

In a similar fashion, the boundary rules for anchoring generate weak anchoring because only particles that have crossed the boundary have the anchoring condition applied, and are then mixed with particles that have not~\cite{Louise}. 
In order to generate strong anchoring conditions, all the particles in cells that intersect $\mathcal{S}_b$ are subjected to the boundary rules described in \sctn{sctn:boundaryRules}. 
Because the entire cell now essentially has the anchoring condition applied, this generates stronger anchoring. 

\subsection{Colloids as Moving Boundaries}\label{app:moving}
Hard colloids can be simulated as spherical (or \correctText{circular}{discoidal} in 2D) boundaries. 
However, the boundaries must be able to move as a response to the surrounding fluid, $\mathcal{S}_{b,0}(\vec{r}) \to \mathcal{S}_{b,0}(\vec{r},t)$. 
Firstly, when a particle collides with the boundary, the impulse of the collision must be applied to the moving boundary to conserve both linear and angular momentum \cite{Shendruk2013}. 
Additionally, when anchoring is used, the change in particle orientation applies a torque on the colloid. 
In this study, the anchoring condition on the colloids are always strong homeotropic anchoring. 
Moving boundaries also needs to be able to collide with other boundaries, both moving and fixed. 
For spherical \correctText{}{or discoidal} colloids of radius $R_b$, the algorithm checks if it has collided with another solid boundary and applies the same rules as when colliding with an MPCD particle. 

\section{Wavy Spheres, Ellipsoids and Cylinders}\label{app:wavyspheres}
In addition to \correctText{corrigated}{corrugated} and egg-shell structured planar walls, wavy spheres, ellipsoids and cylinders are possible via \eq{eq:surface_original}-\eqref{eq:f2}. 

\begin{figure}[tb!]
    \centering
    \begin{subfigure}[b]{0.15\textwidth}
        \centering
        \includegraphics[width=\textwidth]{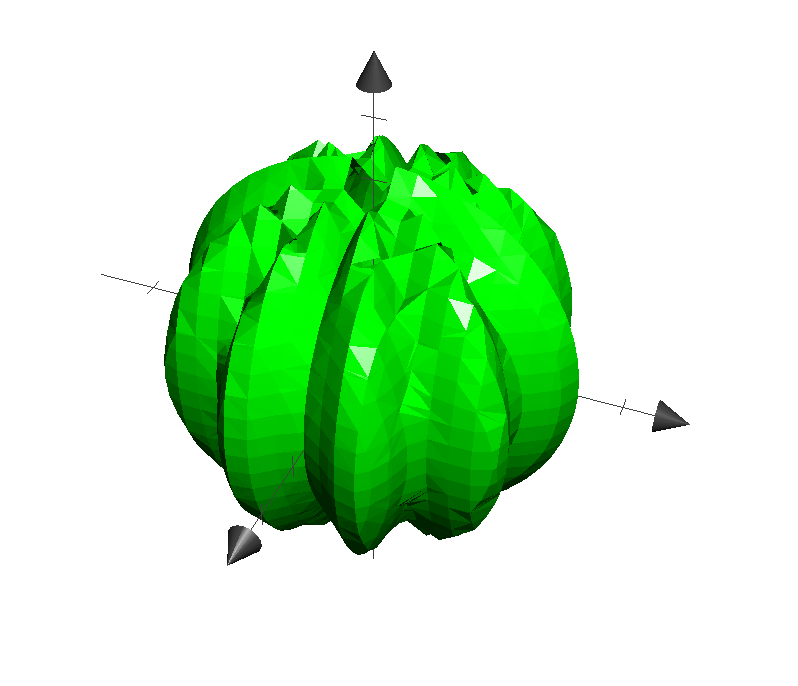}
        \caption{Longitudinal.}
        \label{fig:longitudeWaves}
    \end{subfigure}
    \begin{subfigure}[b]{0.15\textwidth}
        \centering
        \includegraphics[width=\textwidth]{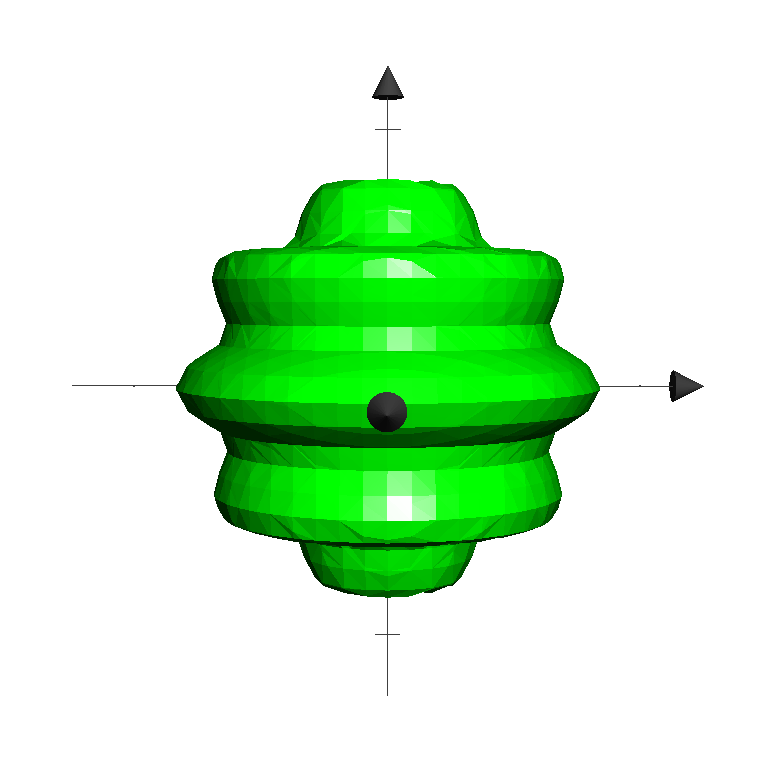}
        \caption{Latitudinal.}
        \label{fig:latidudeWaves}
    \end{subfigure}
    \begin{subfigure}[b]{0.15\textwidth}
        \centering
        \includegraphics[width=\textwidth]{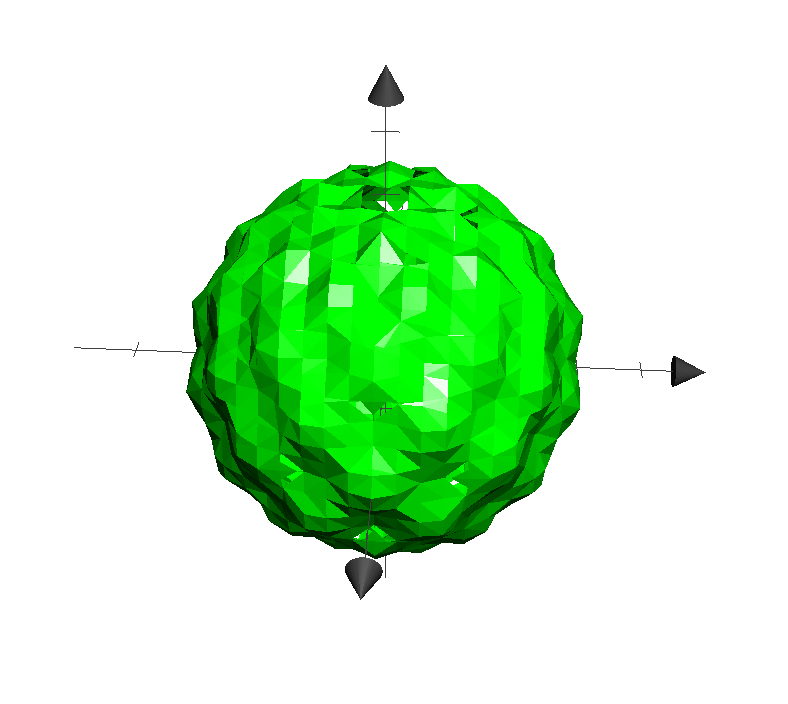}
        \caption{Egg Carton.}
        \label{fig:eggCartonSphere}
    \end{subfigure}
    \caption{Three types of wavy spheres. }
    \label{fig:wavySphere}
\end{figure}

\subsection{Spheres and Ellipsoids}\label{app:spheres}
For spheres, $\vec{p}_b=\vec{2}$, $\vec{A}_b=\vec{1}$ and $R_b$ is the radius of the $b^\text{th}$ sphere. 
For ellipsoids, $\vec{A}_b$ is the inverse of the $b^\text{th}$ ellipsoid's focii. 
There are two different directions in which a sphere can be wavy: longitudinal and latitudinal (\fig{fig:wavySphere}a-b). 
These can also be combined to create a sphere with egg carton waves on it (\fig{fig:wavySphere}c). 
The longitudinal and the latitudinal waves correspond to $\mathcal{S}_{b,1}(\vec{r})$ and $\mathcal{S}_{b,2}(\vec{r})$ respectively, which take the forms
\begin{align}
    \tan{\mathcal{S}_{b,1}(\vec{r})} &= \frac{A_{b,y}(y-q_{b,y})}{A_{b,x}(x-q_{b,x})} \label{eq:sphereS1} \\
    \tan{\mathcal{S}_{b,2}(\vec{r})} &= \frac{ \sqrt{ A_{b,x}^2(x-q_{b,x})^2 + A_{b,y}^2(y-q_{b,y})^2 } }{A_{b,z}(z-q_{b,z})} \label{eq:sphereS2}. 
\end{align}
To create longitudinal waves on a sphere $B_{b,2}=0$, and conversely $B_{b,1}=0$ for latitudinal waves. Neither is $0$ for the egg carton shapes on the sphere. The $B_{b,i}$ correspond to the number of waves around the sphere. Thus, due to the closed surface nature of spheres, all $B_{b,i}$ must be even numbers in order to have complete waves. The surface will still be closed for non-integer and odd $B_{b,i}$ but they will result in sharp points. These equations work for ellipsoids as well, however the waves are not perfectly even. 

\subsection{Cylinders}\label{app:cylinders}
\begin{figure}[tb!]
    \centering
    \begin{subfigure}[b]{0.15\textwidth}
        \centering
        \includegraphics[width=\textwidth]{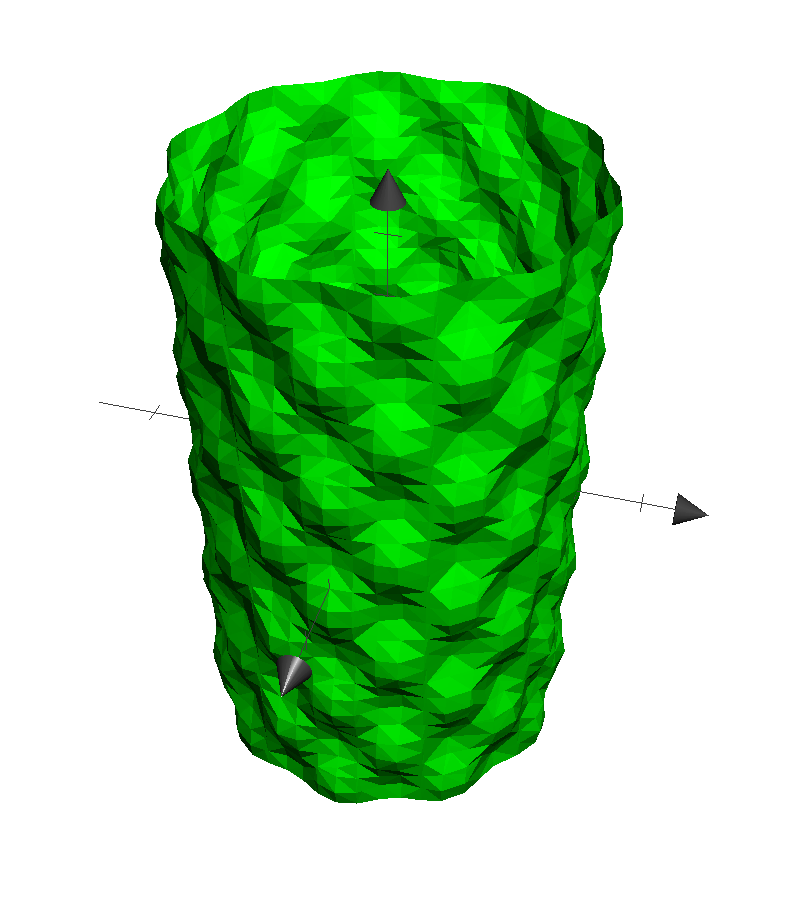}
        \caption{Egg carton.}
        \label{fig:eggCartonCyl}
    \end{subfigure}
    \begin{subfigure}[b]{0.15\textwidth}
        \centering
        \includegraphics[width=\textwidth]{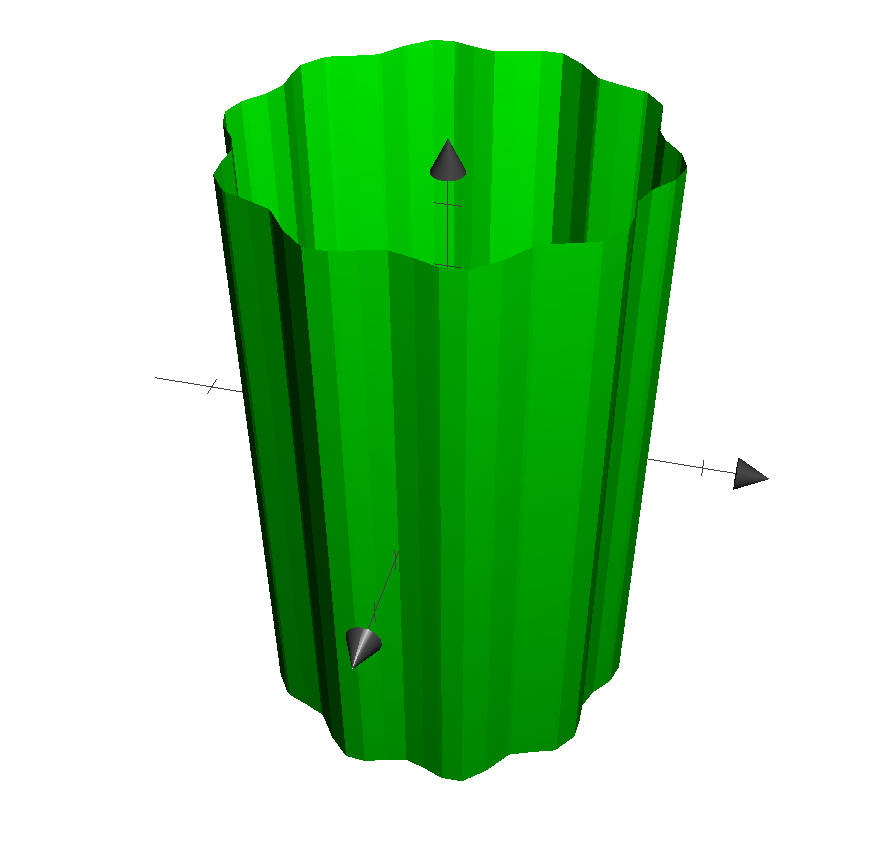}
        \caption{Polar undulations.}
        \label{fig:corrCyl}
    \end{subfigure}
    \begin{subfigure}[b]{0.15\textwidth}
        \centering
        \includegraphics[width=\textwidth]{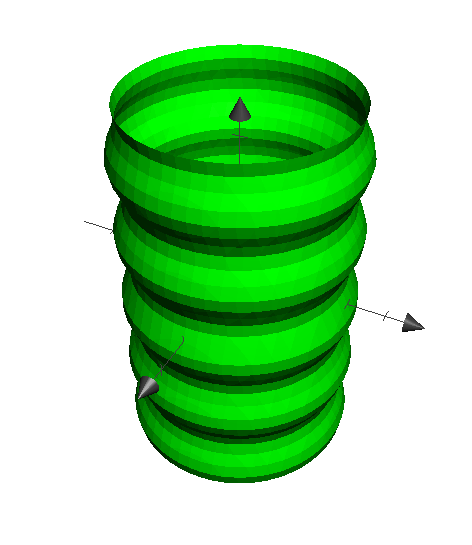}
        \caption{Axial undulations.}
        \label{fig:corrCyl2}
    \end{subfigure}
    \caption{The three types of wavy cylinders. }
    \label{fig:wavycylinders}
\end{figure}

Cylinders are a straightforward extension to spheres, where the surface is independent of one of the directions. 
Thus, either $A_{b,x}$, $A_{b,y}$ or $A_{b,x}$ is $0$ but otherwise follows the same form as spheres and ellipsoids. 
There are three types of wavy cylinders: waves around the cylinder, waves along the cylinder, and an egg carton pattern on the cylinder which combines both (\fig{fig:wavycylinders}).
The two functions $\mathcal{S}_{b,1}(\vec{r})$ and $\mathcal{S}_{b,2}(\vec{r})$ for cylinders have a cyclic property so if $A_{b,x} = 0$, then $(i,j,k) = (x,y,z)$, if $A_{b,y} = 0$: $(i,j,k) = (y,z,x)$; and if $A_{b,z} = 0$ then $(i,j,k) = (z,x,y)$. 
So then for any choice $A_{b,i} = 0$ the wavy surfaces are given by
\begin{align}
    \tan{\mathcal{S}_{b,1}(\vec{r})} &= \frac{A_{b,k}(r_k-q_{b,k})}{A_{b,j}(r_j-q_{b,j})} \label{eq:cylinderS1}\\
    \tan{\mathcal{S}_{b,2}(\vec{r})} &= \frac{\sqrt{ A_{b,j}^2(r_j-q_{b,j})^2 + A_k^2(r_k-q_{b,k})^2}}{r_i}  \label{eq:cylinderS2}.
\end{align}
Here, $\mathcal{S}_{b,1}(\vec{r})$ defines the waves around the cylinder (\fig{fig:corrCyl}) and $\mathcal{S}_{b,2}(\vec{r})$ defines waves along the direction of the cylinder (\fig{fig:corrCyl2}). 
By setting the opposite $B_{b,i}$ to 0, these can be achieved. 
If both $B_{b,i}\neq0$, the egg carton pattern emerges on the cylinder (\fig{fig:eggCartonCyl}).

\subsection{Calculating the Normal of the Boundary}
\label{app:norm}
For some of the boundary rules, the normal and tangential directions of boundary $b$ are needed. 
As the component of a vector tangential to the surface $\mathcal{S}_b(\vec{r})$ is just the vector minus the normal component, only the normal direction $\vec{\nu}_b(\vec{r})$ needs to be calculated as
\begin{equation}
    \label{eq:surfNorm}
    \vec{\nu}_b = \frac{\vec{\nabla}\mathcal{S}_b}{\left|\vec{\nabla}\mathcal{S}_b\right|} .
\end{equation}
Taking the derivative of \eq{eq:surface} gives
\begin{widetext}
\begin{align}
    \frac{\partial \mathcal{S}_b}{\partial r_i} &= 
        \frac{\partial \mathcal{S}_{b,0}}{\partial r_i}
        - A_c \left[ B_{b,1} \frac{\partial \mathcal{S}_{b,1}}{\partial r_i} \sin{(B_{b,1} \mathcal{S}_{b,1})} \cos{(B_{b,2} \mathcal{S}_{b,2})} 
        + B_{b,2} \frac{\partial \mathcal{S}_{b,2}}{\partial r_i} \cos{(B_{b,1} \mathcal{S}_{b,1})} \sin{(B_{b,2} \mathcal{S}_{b,2})} \right] . 
    \label{eq:derivSurf}
\end{align}
\end{widetext}
For all surfaces, the derivative of $\mathcal{S}_{b,0}(\vec{r})$ is given by
\begin{equation}
    \label{eq:S0/xi}
    \frac{\partial \mathcal{S}_{b,0}}{\partial r_i} = p_{b,i} A_{b,i}\left[A_{b,i}(x_i-q_{b,i})\right]^{p_{b,i}-1} .
\end{equation}
However, the derivatives of $\mathcal{S}_{b,1}(\vec{r})$ and $\mathcal{S}_{b,2}(\vec{r})$ vary depending on the surface type.

\subsection{Planes}\label{app:planes}
The normals for wavy planes can be found with the derivatives of $\mathcal{S}_{b,1}$
\begin{subequations}
    \begin{align}
        \frac{\partial \mathcal{S}_{b,1}}{\partial x} &= \frac{A_{b,y}}{\mathcal{A}} \nn \\
        \frac{\partial \mathcal{S}_{b,1}}{\partial y} &= -\frac{A_{b,x}}{\mathcal{A}} \label{eq:derivSb1_plane} \\
        \frac{\partial \mathcal{S}_{b,1}}{\partial z} &= 0 \nn
    \end{align}
\end{subequations}
and the derivatives of $\mathcal{S}_{b,2}$
\begin{subequations}
    \begin{align}
        \frac{\partial \mathcal{S}_{b,2}}{\partial x} &= \frac{A_{b,x} A_{b,z}}{\mathcal{R}} \nn\\
        \frac{\partial \mathcal{S}_{b,2}}{\partial y} &= \frac{A_{b,y} A_{b,z}}{\mathcal{R}} \label{eq:derivSb2_plane} \\
        \frac{\partial \mathcal{S}_{b,2}}{\partial z} &= \frac{\mathcal{A}^2}{\mathcal{R}} , \nn
    \end{align}
\end{subequations}
where we conveniently define 
\begin{subequations}
    \begin{align}
        \mathcal{A} &= \sqrt{A_{b,x}^2+A_{b,y}^2} \label{eq:A}\\
        \mathcal{R} &= \sqrt{A_{b,x}^4+A_{b,y}^4+A_{b,x}^2A_{b,z}^2+A_{b,y}^2A_{b,z}^2 + 2A_{b,x}^2A_{b,y}^2} . \label{eq:R}
    \end{align}
\end{subequations}

\subsection{Spheres and Ellipsoids}\label{app:spheres2}
The surface functions for spheres and ellipsoids include arctan functions, which make the derivatives of $\mathcal{S}_{b,1}$
\begin{subequations}
    \begin{align} 
        \frac{\partial \mathcal{S}_{b,1}}{\partial z} &= 0 \nn\\
        \frac{\partial \mathcal{S}_{b,1}}{\partial x} &= -\frac{A_{b,x} A_{b,y}}{\mathcal{B}} \left(y-q_y\right) \label{eq:derivSb1_sphere}\\
        \frac{\partial \mathcal{S}_{b,1}}{\partial y} &= \frac{A_{b,x} A_{b,y}}{\mathcal{B}} \left(x-q_{b,x}\right) \nn
    \end{align}
\end{subequations}
and the derivatives of $\mathcal{S}_{b,2}$
\begin{align} 
    \frac{\partial \mathcal{S}_{b,2}}{\partial z} &= - \frac{\sqrt{\mathcal{B}}}{\mathcal{C}} \nn\\
    \frac{\partial \mathcal{S}_{b,2}}{\partial x} &= \frac{A_{\correctText{}{b},x}^2(x-q_{b,x})(z-q_{b,z})}{\mathcal{C}\sqrt{\mathcal{B}}} \label{eq:derivSb2_sphere}
    \\
    \frac{\partial \mathcal{S}_{b,2}}{\partial y} &= \frac{A_{\correctText{}{b},y}^2(y-q_{b,y})(z-q_{b,z})}{\mathcal{C}\sqrt{\mathcal{B}}} , \nn 
\end{align}
where we define 
\begin{subequations}
    \begin{align} 
        \mathcal{B} &= \left[A_{b,x}(x-q_{b,x})\right]^2 + \left[A_{b,y}(y-q_{b,y})\right]^2 \label{eq:B}\\
        \mathcal{C} &= \frac{ \mathcal{S}_{b,0}+R_b^{p_{b,r}} }{ A_{b,z} }. \label{eq:C}
    \end{align}
\end{subequations}

\subsection{Cylinders}\label{app:cylinders2}
Cylinders are the most complicated boundary considered here due to the possibility of several orientations. 
The derivatives have a cyclic property so if $A_{b,x}=0$, then $(j, k, l) = (x, y, z)$, if $A_{b,y}=0$: $(j, k, l) = (y, z, x)$; and if $A_{b,z}=0$ then $(j, k, l) = (z, x, y)$. 
Letting $\mathcal{C}^\prime = \mathcal{C}A_{b,z}$, the derivatives are
\begin{align}
    \frac{\partial \mathcal{S}_{b,1}}{\partial r_i} &= \frac{A_{b,k} A_{b,l}}{\mathcal{C}^\prime}
    \times
    \begin{cases}
        0 & \text{for } r_i = r_j \\
        - (r_l-q_{b,l}) & \text{for } r_i = r_k \\
        r_k-q_{b,k} & \text{for } r_i = r_l \\
    \end{cases}  \label{eq:derivSb1_cylinder} \\
    \frac{\partial \mathcal{S}_{b,2}}{\partial r_i} &= \frac{1}{\mathcal{C}^\prime + r_j^2} 
    \times
    \begin{cases}
        -\sqrt{\mathcal{C}^\prime} & \text{for } r_i = r_j \\
        \frac{x_j}{2\sqrt{\mathcal{C}^\prime}} \frac{\partial \mathcal{S}_{b,0}}{\partial r_i} & \text{for } r_i = r_k \text{ or } r_l . 
    \end{cases} \label{eq:derivSb2_cylinder}
\end{align}

\section{Flow through Plane Channels}\label{app:planeChannel}

\begin{figure}[tb!]
    \centering
    \begin{subfigure}[b]{0.475\textwidth}
        \centering
        \includegraphics[width=\textwidth]{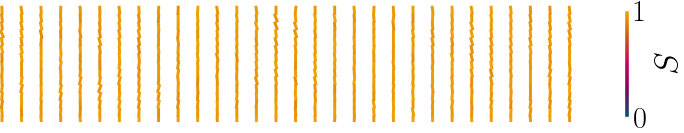}
        \caption{Bowser.}
        \label{fig:bowser}
    \end{subfigure}
    \begin{subfigure}[b]{0.475\textwidth}
        \centering
        \includegraphics[width=\textwidth]{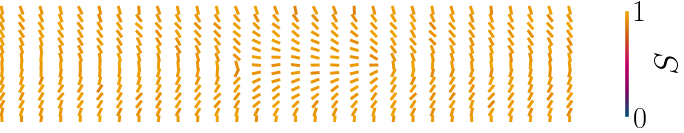}
        \caption{Transition.}
        \label{fig:bowserTransDowser}
    \end{subfigure}
    \begin{subfigure}[b]{0.475\textwidth}
        \centering
        \includegraphics[width=\textwidth]{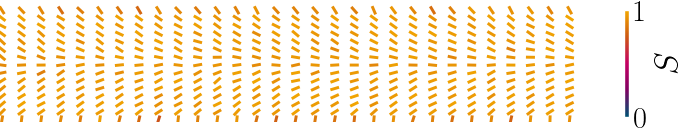}
        \caption{Dowser.}
        \label{fig:dowser}
    \end{subfigure}
    \begin{subfigure}[b]{0.475\textwidth}
        \centering
        \includegraphics[width=\textwidth]{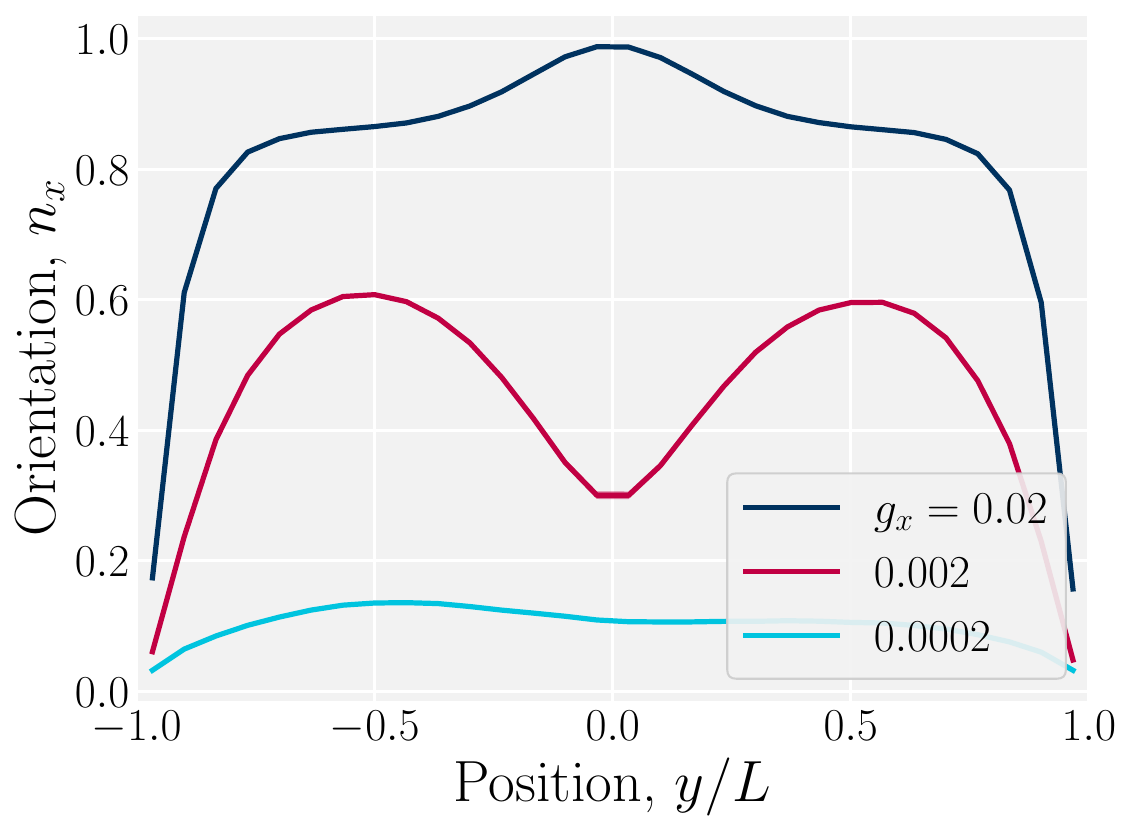}
        \caption{Director profile.}
        \label{fig:dirProfile}
    \end{subfigure}
    \caption{
        Nematic director profiles for different flows in a plane channel of height $h=30a$ (characteristic length $L=h/2$). 
        (a) Bowser state for weak flow ($g_x=0.0002a/\tau^2$). 
        (b) Transition state  ($g_x=0.002a/\tau^2$). 
        (c) Dowser state for strong flows ($g_x=0.02a/\tau^2$). 
        (d) Director profile $n_x=\hat{x}\cdot\vec{n}$ across the channel for the Bowser, transition and Dowser states. 
    }
    \label{fig:directorDeformation}
\end{figure}

Flowing isotropic fluids at low Reynolds number obey plane Poiseuille flow with parabolic flow profiles~\cite{bruus2007}, which is a direct consequence of the the non-slip condition at the walls of the channel. 
Though the coupling between flow profile, local director and anisotropic dissipation can induce complications\cite{Brochard1973}, N-MPCD simulates linearized fluctuating nematodynamics with isotropic viscous dissipation~\cite{Hijar2019}. 

The director can discontinuously transition from a homogenous alignment to deformed states~\cite{Copar2020}. 
For strong homeotropic anchoring and sufficiently small pressure gradients, the director remains uniformly homeotropic across the channel; however, for sufficent flow the director field bows into the Bowser state~\cite{Copar2020}, in which the field curves slightly in the direction of the flow (\fig{fig:bowser}). 
The Bowser state transitions (\fig{fig:bowserTransDowser}) to the the Dowser state (\fig{fig:dowser}) at sufficiently strong flows, for which the director fully aligns with the flow at the center  (\fig{fig:dirProfile}). 
The Bowser and Dowser states arise from the competition between nematoelastic free energy and flow strength~\cite{Anderson2015}. 
For the pressure gradient and channel height used in this study, the director is found to be in the bowed Bowser state. 
The planar anchoring case is not as interesting since the boundary conditions and strain rate coalign, though at sufficiently stronger flow the near-wall director may tilt inward as expected~\cite{Batista2015}. 

\begin{figure}[tb]
    \centering
    \begin{subfigure}[b]{0.495\textwidth}
        \includegraphics[width=0.95\textwidth]{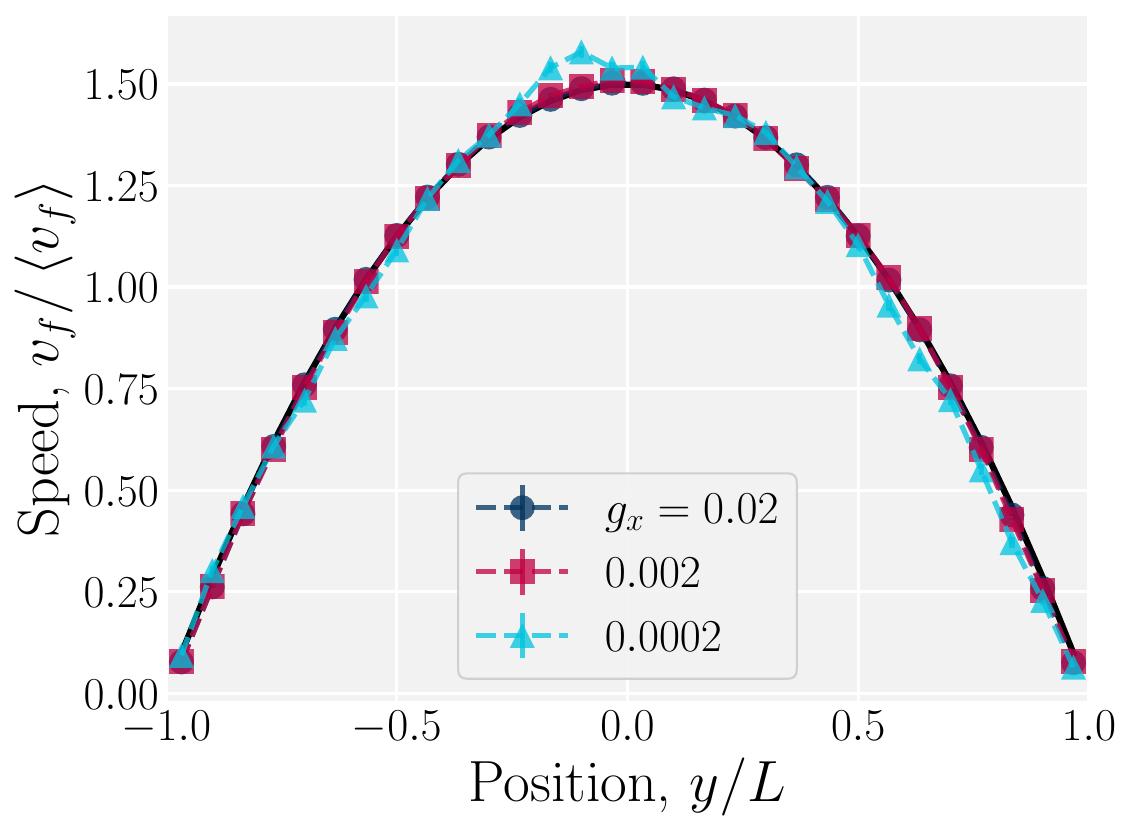}
        \caption{Flow profiles.}
        \label{fig:nemflowprof}
    \end{subfigure}
    \begin{subfigure}[b]{0.495\textwidth}
        \includegraphics[width=0.95\textwidth]{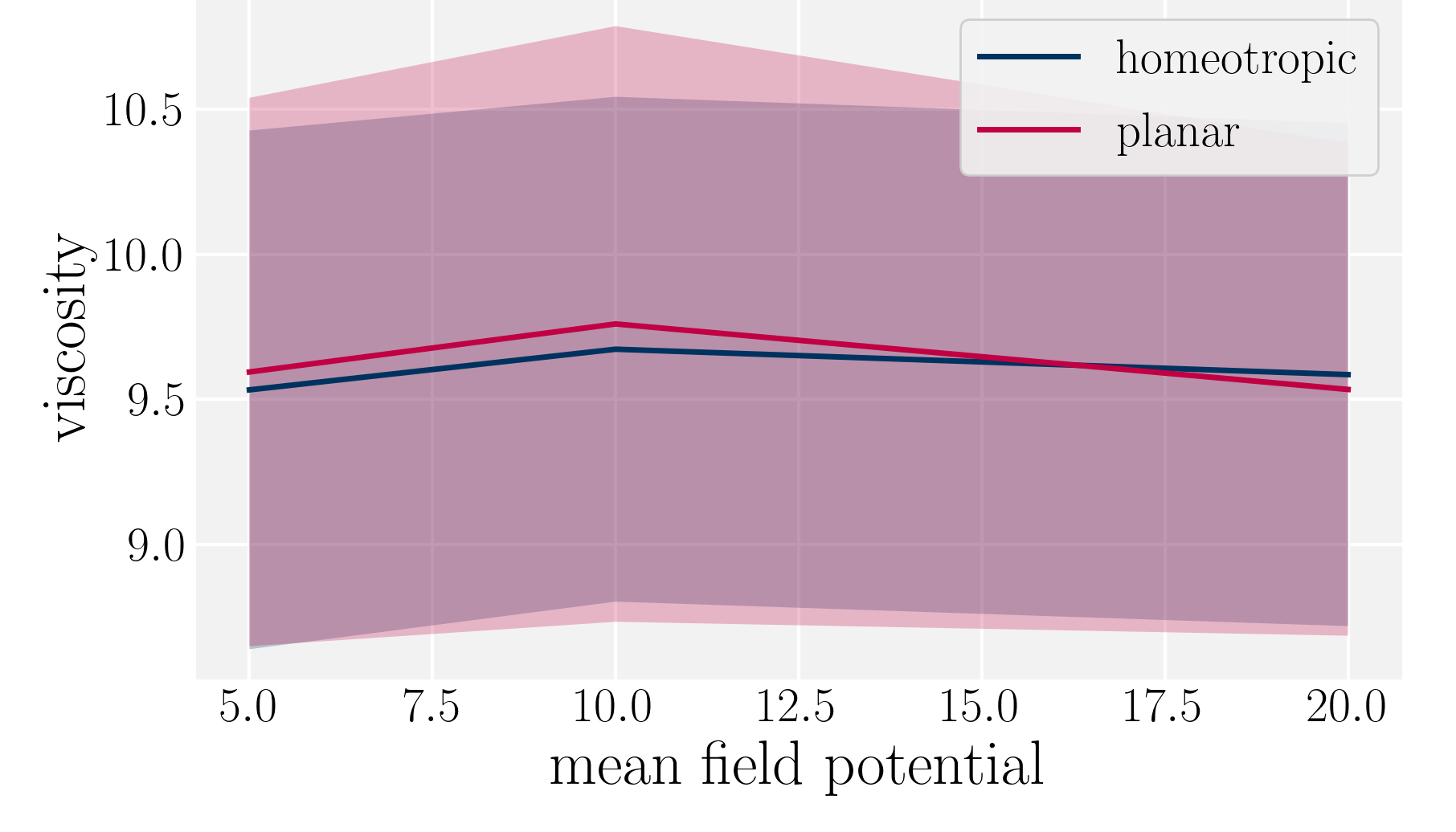}
        \caption{Viscosity from parabolic Poiseuille fit.}
        \label{fig:viscosity}
    \end{subfigure}
    \caption{
        Flow profiles in planar channels. 
        (a) The flow profiles for different pressure gradients $\rho g_x$, where $g_x$ is the component of the external acceleration in the $\hat{x}$ direction. 
        The flow obeys parabolic Poiseuille flow for all pressure gradients considered here in a plane channel of height $h=30a$ (with $L=h/2$). 
        (b) The viscosity of the fluid for different mean field potentials and wall anchoring in a plane channel of height $h=20a$. 
    }
    \label{fig:visc}
\end{figure}

Since the flow profiles remain parabolic (\fig{fig:nemflowprof}), the viscosity can be calculated directly from the average velocity of the nematic fluid $\av{v_\text{f}}$ and the plane Hagen–Poiseuille equation to be
\begin{equation}
    \label{eq:visc}
    \mu = \frac{h^2 \rho \vec{g}\cdot\hat{x}}{12 \av{v_\text{f}}} . 
\end{equation}
To measure the viscosity, the pressure gradient is kept sufficiently low to be in the Bowser state and viscosity is found to be independent of the values of mean field potential (for $U=\{5, 10, 20\}\kbt$) and anchoring conditions (\fig{fig:viscosity}), as expected since the nematogens are modelled as point-particles with a direction, rather than actual rod-like particles. 
An isotropic viscosity of $\mu=\left[9.6\pm0.4\right] \mu_0$ is found. 
This measurement gives the characteristic velocities due to nematicity and pressure to be $\tilde{V} = K/\mu L = \left[11.8\pm0.8\right] a/\tau$ and $V=L^2 \rho g_x /\mu = \left[0.208\pm0.009 \right]a/\tau$, respectively. 
This establishes the dimensionless Ericksen number for this study to be
\begin{equation}
    \label{eq:er}
    \text{Er} = \frac{\av{v_\text{f}}}{\tilde{V}} \sim 3\times10^{-3},
\end{equation}
which indicates elastic forces dominate viscous forces. 
Likewise, the dimensionless P\'{e}clet number is approximately $\text{Pe} = v_\text{f} L / D \sim 10^{2}$, which reflects that advection is more significant than diffusion in plane channels. 
The small $\text{Er}$/large $\text{Pe}$ regime is where we expect topological microfluidics to be most interesting. 
\correctText{}{In these simulations, elastic forces dominate over both velocity and diffusion, $\text{Er}/\text{Pe} \sim 3\times 10^{-5}$.
This can be seen explicitly by comparing trajectories in a nematic solvent, which are near the middle line, to trajectories in an isotropic solvent, which can diffuse away from the center (see \appndx{app:isotropic}).}

\section{Absence of stick-slip in isotropic fluid flows}\label{app:isotropic}
\begin{figure}[tb!]
    \centering
    \begin{subfigure}[b]{0.475\textwidth}
        \centering
        \includegraphics[width=\textwidth]{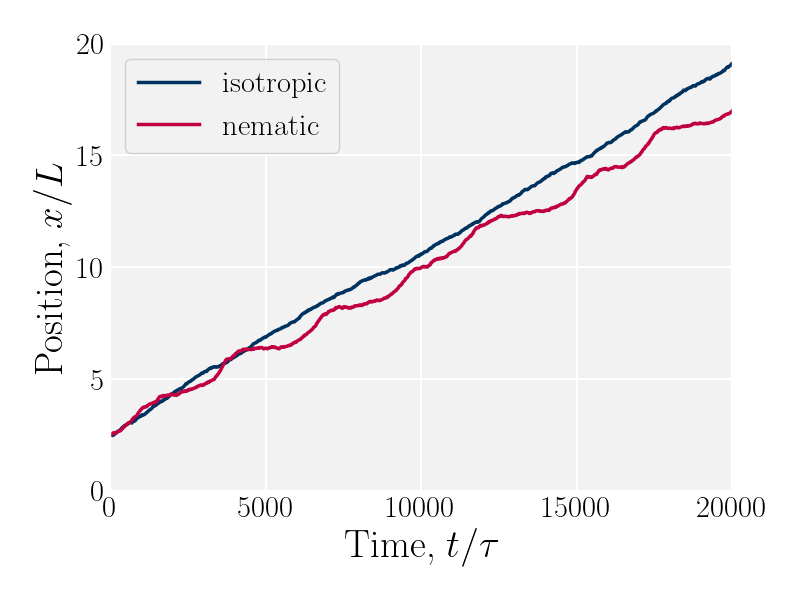}
        \caption{\correctText{}{$\hat{x}$-component of the colloid trajectories as a function of time.}}
        \label{fig:isoVSnemX}
    \end{subfigure}
    \begin{subfigure}[b]{0.475\textwidth}
        \centering
        \includegraphics[width=\textwidth]{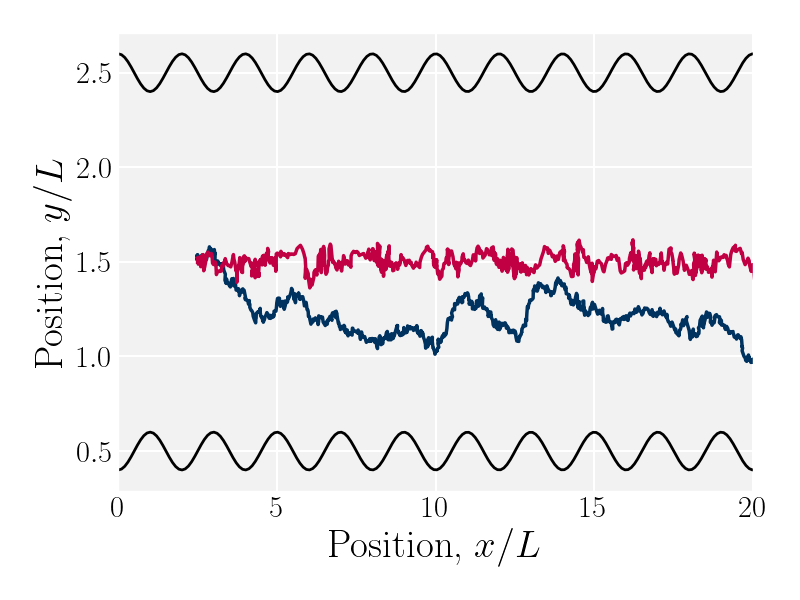}
        \caption{\correctText{}{Colloid trajectories.}}
        \label{fig:isoVSnemXY}
    \end{subfigure}
    \caption{
        \correctText{}{Example trajectories for a colloid suspended in a nematic compared to in an isotropic fluid. 
        The wall amplitude in both cases is $B_0/L=1$. 
        The nematic case has homeotropic anchoring on the colloid and planar anchoring on the walls.} 
    }
    \label{fig:isoVSnem}
\end{figure}

\correctText{}{Simulations of a colloid in an isotropic fluid advecting through a wavy channel do not exhibit stick-slip dynamics (\fig{fig:isoVSnem}). 
While the nematic case exhibits clear periods of movement (slip) and cessation (stick), the isotropic case exhibits a steady elution rate (\fig{fig:isoVSnemX}). 
Furthermore, nematic repulsion between the colloids and the walls keeps the colloid close to the channel centerline (\fig{fig:isoVSnemXY}). 
On the other hand, the colloid in an isotropic fluid is free to diffuse away from the centerline, sampling different local velocities and exhibiting greater Taylor dispersion. 
In this way, the lock-key dynamics investigated here differ from the Brownian dynamics of a driven diffusive colloid moving through a periodic landscape.
}

\section{Impact of Colloid Dynamics on Fluid Flow}\label{app:feedback}
\begin{figure}[tb!]
    \centering
    \includegraphics[width=0.5\textwidth]{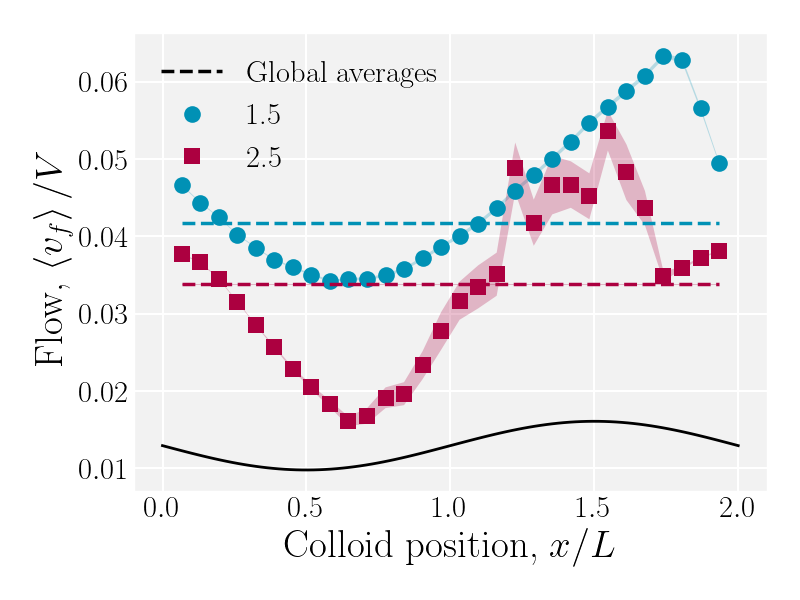}
    \caption{Average fluid velocity $\av{v_f}$ as a function of instantaneous colloid position $x$ for wall amplitudes $B_0=\left\{ 1.5, 2.5 \right\}a$.}
    \label{fig:flow}
\end{figure}

\correctText{}{The colloid elutes due to the drag force that the fluid exerts on it, but it also moves in response to the nematoelastic forces. 
Therefore, at any instance the colloid may resist or assist the fluid flow. 
The spatially averaged flow of the fluid is not constant with time during the stick-slip motion. 
When the colloid sticks, the combination of colloid drag and nematoelastic forces slow the fluid. 
When the colloid slips, the nematoelastic forces drive fluid flow. 
We measure the total average fluid velocity $\av{v_f}$ as a function of the colloid position (\fig{fig:flow}). 
A minimum average fluid velocity occurs just past the trough, where defects oppose elution. 
Here, nematic interactions work against the pressure gradient and the fluid slows. 
During the slip phase, the colloid speeds up due to nematic forcing and the fluid flow reaches a maximum as the colloid is forced over the crest by elastic interactions with the walls. 
As the wall amplitude is increased, crossing events become rarer and faster, causing the statistics to worsen for $x/L>1$.} 

\end{document}